\documentclass{pasa}%
\newcommand{\msolar} {$\rm{M_{\odot}}~$}
\newcommand{\msolarc} {$\rm{M_{\odot}}$}
\newcommand{\Lsolc} {$\rm{L_{\odot}}$}

\usepackage{graphicx}
\usepackage{etoolbox}
\pretocmd{\abstractname}{\newpage}{}{}
\usepackage{ulem}

\title[Titans of the Early Universe]{Titans of the Early Universe: The Prato Statement on the Origin of the First Supermassive Black Holes}

\author[Woods et al.]{Tyrone~E.\ Woods$^{1,2}$\thanks{tyrone.woods@monash.edu}, Bhaskar Agarwal$^{3}$, Volker Bromm$^{4}$, Andrew Bunker$^{5,6}$, Ke-Jung Chen$^{7}$, Sunmyon Chon$^{8}$, Andrea Ferrara$^{9,10}$, Simon C.\ O.\ Glover$^{3}$, Lionel Haemmerl\'e$^{11}$, Zolt\'an Haiman$^{12}$, Tilman Hartwig$^{13,14}$, Alexander\ Heger$^{1,15}$, Shingo Hirano$^{16}$, Takashi Hosokawa$^{17}$, Kohei Inayoshi$^{18}$, Ralf\ S.\ Klessen$^{3,19}$, Chiaki Kobayashi$^{20}$, 
Filippos Koliopanos$^{21,22}$, Muhammad A.\ Latif$^{23}$, Yuexing Li$^{24,25}$, Lucio Mayer$^{26}$, Mar Mezcua$^{27,28}$, Priyamvada Natarajan$^{29,30}$, Fabio Pacucci$^{30}$, Martin J.\ Rees$^{31}$, John A.\ Regan$^{32}$, Yuya Sakurai$^{10}$, Stefania\ Salvadori$^{33,34,35}$, Raffaella Schneider$^{36}$, Marco Surace$^{37}$, Takamitsu L. Tanaka$^{38}$, Daniel J. Whalen$^{37}$, Naoki Yoshida$^{12,13}$
}

\jid{PASA}
\doi{10.1017/pas.\the\year.xxx}
\jyear{\the\year}

\usepackage{aas_macros}
\usepackage{hyperref} 
\hypersetup{colorlinks,citecolor=blue,linkcolor=blue,urlcolor=blue}

\hypersetup{draft}

\begin{document}

\begin{frontmatter}
\maketitle

\begin{abstract}
In recent years, the discovery of massive quasars at $z\sim7$ has provided a striking challenge to our understanding of the origin and growth of supermassive black holes in the early Universe. Mounting observational and theoretical evidence indicates the viability of massive seeds, formed by the collapse of supermassive stars, as a progenitor model for such early, massive accreting black holes. Although considerable progress has been made in our theoretical understanding, many questions remain regarding how (and how often) such objects may form, how they live and die, and how next generation observatories may yield new insight into the origin of these primordial titans. This review focusses on our present understanding of this remarkable formation scenario, based on discussions held at the Monash Prato Centre from November 20--24, 2017, during the workshop ``\textsl{Titans of the Early Universe: The Origin of the First Supermassive Black Holes\/}.'' 
\end{abstract}

\end{frontmatter}

\section{INTRODUCTION }
\label{sec:intro}

\begin{figure}[t]
\begin{center}
\includegraphics[width=\columnwidth]{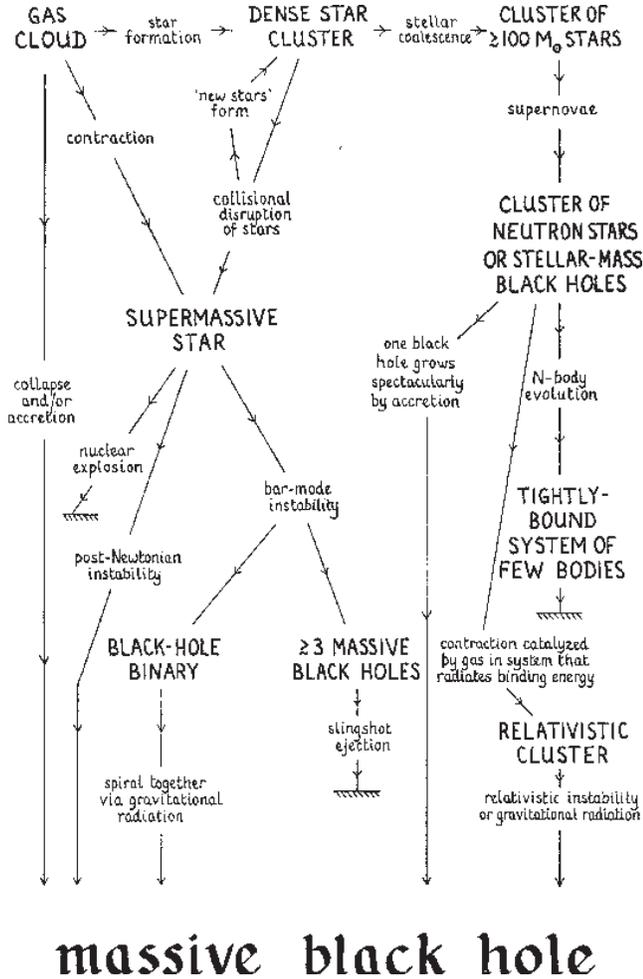}
\caption{Original diagram from \cite{Rees78,Rees84}, outlining the possible formation pathways for supermassive black holes. In this review, as in the conference, our focus is on the left side of the diagram.}\label{Rees}
\end{center}
\end{figure}

Most massive galaxies host a central supermassive black hole (SMBH), however the origin of these objects remains uncertain.  The canonical summary of possible formation pathways was first laid out in the Halley Lecture of 1978 at Oxford University \citep{Rees78}.  Within the so-called Rees diagram (see also Fig. \ref{Rees}), supermassive stars \citep{HF63,Iben63}, dense stellar clusters \citep[e.g.,][]{BR78}, and a host of other objects were laid out as possible intermediaries.  Many of these hypothesized progenitor objects were initially suggested to actually be the sources powering the emission seen in active galactic nuclei, before mounting observational evidence made it clear that these were in fact accreting SMBHs \citep{Rees84}. Common to all of the SMBH progenitor channels in Fig. \ref{Rees} is the concentration of a large quantity of gas in a sufficiently small volume, leading to runaway black hole growth. How often each channel may be realized in nature, however, remains an outstanding problem. The majority of these scenarios remain plausible but unproven today.

By far the greatest challenge to any theory of SMBH formation has been the discovery of luminous ($\gtrsim 10^{13}\,\mathrm{L}_{\odot}$) quasars at $z\sim7$, when the Universe was only $\sim800\,\mathrm{Myr}$ old \citep[e.g.,][]{Mortlock11,Wu15,Banados18}. The masses of these objects are all $\gtrsim 10^{9}\,\mathrm{M}_{\odot}$, inferred from the breadth of the observed Mg II 2798\AA\ line
\citep[e.g.,][]{MD04}, and consistent with their luminosities being near the Eddington limit. Among the most troubling examples, SDSS J010013.02+280225.8 is a known redshift $\sim$6.3 quasar that is already 1.2$\times10^{10}\mathrm{M}_{\odot}$ \citep{Wu15}, while even earlier in the Universe, ULAS J134208.10+092838.61 is a $0.8\times10^{9}\mathrm{M}_{\odot}$ quasar at $z=7.54$. How did these black holes reach of order 1--10 billion solar masses in the first billion years of the Universe? 

The problem is best illustrated if we consider the optimistic case of persistently Eddington-limited accretion for the entire prior lifetimes of these objects. A black hole may only grow in this way from an initial ``seed'' mass $M_{0}$ to a given mass $M_{\rm{BH}}$ in a time:
\begin{equation}
t_{\rm{growth}} \approx 0.45 \frac{\epsilon}{1 - \epsilon}\ln\left(\frac{M_{\rm{BH}}}{M_{0}}\right) \rm{Gyr}
\label{eq:t_growth}
\end{equation}
where $\epsilon\sim 0.1$ is the typical radiative efficiency for thin-disk accretion \citep[see e.g.,][for discussion]{Shakura_1973}. Even in this most favourable scenario, producing a $>10^{9}\,\mathrm{M}_{\odot}$ quasar from a typical $\sim 10$--$100\,\mathrm{M}_{\odot}$ Pop III remnant would require an accretion time greater than the age of the Universe at $z\sim7$, unless significantly lower radiative efficiencies may be invoked \citep[i.e., strongly ``super-Eddington,'' accretion, see e.g.,][for further discussion]{2014GReGr..46.1702N, inayoshi16}. Even so, numerical simulations suggest most such stellar-mass Pop III black holes were likely to have been ``born starving,'' unable to grow substantially via accretion early in the Universe, particularly due to their strong ionizing feedback and possible ejection from their halos via dynamical 3-body interactions \citep[e.g.,][]{JB07,WF12,smith18}. 

These simple considerations provide a strong hint that very rapid accretion rates and a massive ``seed'' are necessary ingredients in the origin of the most massive high-z quasars, although relatively lower-mass progenitors may yet be plausible for these objects \citep[see e.g.,][and references therein]{2012MNRAS.422.2051N, Pezzulli17}, and can reproduce the observed M-$\sigma$ relation in the local Universe \citep[see e.g.,][]{TK14,TK15}.  The relative abundance of light and massive seeds, and their role in the origin of all SMBHs, depends sensitively on the prevalence of halos where the formation of massive black hole seeds is possible \citep[e.g.,][]{Lodato_Natarajan_2007, Agarwal12,Dijkstra14,Habouzit16}. 

The first question then becomes what massive seed formation channels are viable. The collapse of dense stellar clusters to form a massive protostar has been considered for decades \citep[e.g.,][]{BR78}. The necessary inefficiency of this process however, in which much of the original energy and angular momentum of the system are shed with stellar mass ejected in 3-body interactions, strongly limits its ability to produce extremely massive seeds \citep[see e.g.,][and references therein]{Latif_2016}. More exotic channels, such as the growth of primordial black holes \citep{primordial}, or intermediate-mass black holes formed from dissipative dark matter \citep{dissipative}, remain somewhat speculative, and require further study. A promising model for producing both large seed masses and rapid accretion rates has emerged in the atomically-cooled halo scenario \citep[e.g.,][]{Dijkstra08}, in which exposure to a strong Lyman-Werner flux from an adjacent Pop III halo destroys the molecular hydrogen in a primordial halo which has not yet undergone fragmentation \citep{Regan17}.
This prevents ``normal'' Pop III star formation, and allows infall rates of up to $\sim0.1$--$1.0\,{\rm M}_{\odot}/\mathrm{yr}$.  Numerical simulations have consistently shown such high accretion rates lead to the formation of a nuclear-burning, supermassive star, which will later undergo collapse through a relativistic instability (referred to in the literature as a ``direct collapse black hole,'' DCBH), leaving a massive black hole remnant \citep{hos13,umeda16,woods17,haemmerle18a}. 

High baryonic streaming velocities relative to dark matter within some regions in the early Universe may provide another means to suppress star formation in low mass halos \citep{TH10} and promote the growth of a direct collapse black hole seed \citep{TL14,Hirano17}, although such a mechanism may primarily act as a catalyst within the atomically-cooled halo model \citep[e.g.,][]{Schauer17}. Other alternative channels for producing massive seeds, including massive high-z galaxy mergers \citep{Mayer10,MB18} and the disruption of dense stellar clusters \citep[e.g.,][]{BR78,Katz15,Sakurai17} have also been the subject of significant investigation, although questions remain regarding the nature of the remnants they produce \citep[see e.g.,][for a thorough discussion]{LF16}.

The upcoming launch of next-generation observatories such as the {\it James Webb Space Telescope} ({\it JWST}) and {\it Euclid}, as well as ground-based extremely large telescopes, promise to yield great insight into the nature and number of the most luminous objects in the early Universe. Already, observational case studies such as the Lyman-$\alpha$ emitter CR7 have provided invaluable tests for our ability to distinguish between models for luminous high-z objects \citep{Sobral15,Pallottini15,Hartwig16,Agarwal17,Bowler17,Pacucci17}. The ongoing search for intermediate-mass black holes (IMBHs) in the nearby Universe may soon allow us to constrain the distribution of black hole seed masses \citep{Mezcua17}. With this progress in mind, the workshop \textsl{Titans of the Early Universe: The Origin of the First Supermassive Black Holes} was held in late November of 2017 in Prato, Italy, with the goal of determining what theoretical and observational progress can now be made in understanding the plausibility, origin, and nature of supermassive stars and their direct collapse as massive black hole seeds, as well as the role of alternative formation channels. With respect to Fig. \ref{Rees}, then, our focus is on the left side of the diagram, and in particular all paths which pass through a ``Supermassive Star'' phase.

In the following, we begin in $\S2$ by providing a summary of some of the latest theoretical developments in understanding how the extremely rapid accretion rates ($\sim 0.1$--$1 \mathrm{M}_{\odot}$/yr) needed to fuel the growth of massive black hole seeds and very high-z quasars may be achieved, what effects may moderate or enhance this growth, and how progress can be made in future simulations. Then, in $\S$3, we outline recent numerical results which have reached a consensus on the initial product of these extreme accretion rates: high-entropy, hydrostatic nuclear-burning supermassive stars, which collapse through a relativistic instability to produce massive ($\sim10^{5} \mathrm{M}_{\odot}$) black hole seeds. In $\S4$, we discuss the relative formation rates of light and massive seeds, the influence of still-uncertain parameterized physics, and future prospects for encapsulating the underlying physics of massive seed/DCBH formation in simplified prescriptions for use in population modelling. Finally, in $\S5$ we discuss observational prospects for detecting evidence of black hole seed formation channels (and, eventually, determining the initial black hole ``seed mass function'') using next-generation observatories, before concluding in $\S6$.

\section{How could supermassive stars form?}
\label{supermassivestars}

Independent of whether the seeds of the most massive high-z quasars are light or heavy, there is a growing consensus that sustained extremely high accretion rates are a vital ingredient in their origin. The central problem is that the growth of black holes via mergers alone is unlikely to produce $\sim 10^{9}\,\mathrm{M}_{\odot}$ black holes by $z\sim7$ \citep[e.g.,][see \S4 for further discussion]{sesana07,TH09,2011BASI...39..145N}.  Rapid accretion rates ($\gtrsim 0.1\,{\rm M}_{\odot}\,\mathrm{yr}^{-1}$) are also essential if any direct collapse black hole seeds are to pass through a supermassive ($\sim 10^{5}\,\mathrm{M}_{\odot}$) stellar evolutionary phase, as the Eddington limit imposes a maximum nuclear burning lifetime of $\sim2\,\mathrm{Myr}$ for these objects.

Over the last 2 decades, numerical simulations of the collapse of primordial halos have grown increasingly sophisticated. Early numerical work focussed on investigating the collapse of low angular momentum halos \citep[e.g.,][]{LR94,Begelman06,Lodato_Natarajan_2006}.  It was soon understood that the dissociation of molecular hydrogen by Lyman-Werner radiation from the first stars could produce primordial halos in which collisionally-excited atomic line transitions dominate cooling, allowing the isothermal collapse of such halos at $\sim 10^{4}\,\mathrm{K}$ \citep{HRL97}.  Since both the Jeans mass and the infall rate scale with the cube of the sound speed, and hence as $T^{3/2}$, where $T$ is the temperature, such ``atomically-cooled halos'' provide a natural means for producing the conditions necessary for the formation of early quasars from massive seeds \citep[e.g.,][and references therein]{Volonteri10}.  Initial efforts to simulate the collapse of atomically-cooled halos presumed the prior destruction of $\rm{H}_{2}$ \citep[e.g.,][]{BL03,WTA08,RH09_a,RH09_b}. Subsequent work, however, has focussed on determining the initial conditions under which such a halo may arise, including the critical intensity for Lyman-Werner radiation, required to completely suppress H$_{2}$ cooling within a halo \citep[e.g.,][]{Dijkstra08,SBH10}. This quantity (``$J_{\rm{crit}}$'') is conventionally parameterised in the literature in terms of the specific intensity at the Lyman limit, $J_{\rm LW}$, expressed in units of $10^{-21}\ \rm erg/s/cm^2/Hz/sr$ \citep[e.g.,][]{Omukai:2001p128}. The overall picture which has emerged is that a strong Lyman-Werner flux, provided by a nearby halo which has recently undergone Pop III fragmentation, can indeed produce an atomically-cooled halo \citep[e.g.,][]{Dijkstra08,Regan17}. The number density of DCBHs which may be formed in this way is understood to vary as $J_{\rm LW}^{-4}$ \citep[e.g.,][]{Dijkstra08,IT15,Chon16}. Varying estimates for $J_{\rm{ crit}}$ correspond to the $\sim$8 orders of magnitude variation in existing estimates for the number density of DCBHs (see discussion in $\S$4.2); whether a strong Lyman-Werner flux alone is primarily responsible for triggering DCBH formation is increasingly in doubt however, and remains an area of active study.

This ``synchronised pairs'' model \citep{VHB14_a} implies very strict constraints on the timing, evolution, and separation of any pair of primordial halos which may provide a site for the formation of a massive black hole seed \citep{VHB14_b,Dijkstra14,Regan17}. Many questions remain, however, regarding the precise conditions under which atomically-cooled halos could form and plausibly reproduce the masses and space density of high-z quasars \citep{LF16}. To what extent does fragmentation pose a problem for the formation of supermassive stars \citep[e.g.,][]{Chon16, Regan18}? Can supermassive stars form without an atomically-cooled halo, e.g., with lower-mass star formation suppressed via high baryonic streaming velocities relative to dark matter in the early Universe, or is this only complementary to the standard scenario \citep{Schauer17}? These questions, as well as the next steps needed to make further progress in numerical simulations of atomically-cooled halos, are discussed in the following section.

\subsection{H$_{2}$ dissociation in primordial halos}

\subsubsection{Important Reactions}

The birth of the first stars (Pop III) in the Universe also marked the onset of stellar radiative feedback. This must then be accounted for in any self-consistent treatment of subsequent star formation. In addition to hydrogen-ionizing photons (with $E>13.6\,\mathrm{eV}$), stellar populations emit photons in two other energy bands which are of vital importance in determining the thermal and chemical state of primordial star-forming halos. Following \cite{Agarwal:2018jg}, we describe each of these in turn, before returning to address the treatment of irradiating stellar spectra in assessing the conditions necessary for the formation of atomically-cooled halos (and, perhaps, massive black hole seeds).

Photons in the Lyman-Werner band, with energies $11.2\,\mathrm{eV}<E<13.6\,\mathrm{eV}$, can dissociate molecular hydrogen (H$_{2}$). They have insufficient energy, however, to ionize atomic hydrogen, allowing Lyman-Werner photons which escape from nascent halos to travel substantial path lengths in the early Universe, destroying H$_{2}$ in other primordial minihalos and potentially delaying further Pop III star formation \citep{hrl96,Ciardi:2000p82}. Radiative de-excitations of excited H$_{2}$ provide a powerful source of cooling in primordial halos. In its absence, the next available strong cooling term becomes radiative transitions of collisionally-excited atomic hydrogen, requiring gas temperatures of $\sim 8\,000\,\mathrm{K}$.  For the subsequent isothermal collapse, such high temperatures imply a Jeans mass of $\sim 10^6\,\mathrm{M}_\odot$ at $n\sim10^3\,{\rm cm}^{-3}$. Therefore, atomically-cooled halos undergoing isothermal collapse permit the birth of truly supermassive objects, potentially leading to the formation of, e.g., DCBHs and/or quasi-stars (see subsequent sections).

Previous efforts to model the effect of a photo-dissociating background on star formation in the early Universe have focussed on the strength of this LW radiation, ($J_{\rm LW}$, see above). Numerical studies in this vein \citep[e.g.,][]{SBH10,2014MNRAS.443.1979L} suggest it is indeed possible to almost completely dissociate H$_2$ (and therefore suppress ``normal'' Pop III star formation) in primordial halos, requiring however very high values of the LW specific intensity (with the extent of the problem particularly dependent on the assumed spectrum, see below).

In addition to near-UV photons, stellar radiation in the infra-red (IR) can also provide an important contribution to feedback in the early Universe \citep{WolcottGreen:2012p3854}. In particular, photons in the energy range $h\nu \geq 0.76\,\mathrm{eV}$ can unbind the extra electron in H$^-$, which otherwise catalyses H$_2$ formation at low densities via the reactions:

\begin{eqnarray}
\nonumber
\rm H + e \rightarrow H^- + \gamma \\ 
\rm H^- + H \rightarrow H_2 + e  \nonumber
\end{eqnarray}

\noindent thus playing a key role in determining the equilibirum H$_2$ fraction. Therefore, cooler low-mass stars (in particular, second generation ``Pop II'' stars) can also play an important role in regulating early star formation. H$_{2}$ formation can also be catalysed via the reaction chain
   \begin{eqnarray}
   {\rm H^{+} + H} & \rightarrow & {\rm H_{2}^{+} + \gamma}, \\
   {\rm H_{2}^{+} + H} & \rightarrow & {\rm H_{2} + H^{+}}.\nonumber
   \end{eqnarray}

\noindent However, this is considerably less effective than forming H$_{2}$ via the H$^{-}$ ion, and so these reactions play an important role in supermassive star formation only in unusual circumstances \citep{2016MNRAS.456..270S}.

Taking all this together, the photodetachment of H$^-$ and the photo-dissociation of H$_{2}$ may be parametrised as:

\begin{eqnarray}
k_{\rm di} = C_{\rm di}\,\alpha\, J_{\rm LW} \\
k_{\rm de} = C_{\rm de}\,\beta\, J_{\rm LW}
\end{eqnarray}\\

\noindent where $C_{\rm di} = 1.38\times10^{-12}\ \rm s^{-1}$ and $C_{\rm de} = 1.1\times 10^{-10}\ \rm s^{-1}$ are rate constants, $\alpha$ and $\beta$ are rate parameters that depend on the spectral shape of the irradiating source, and J$_ {LW}$ is the LW specific intensity, as defined earlier.  Therefore, it is essential that both the spectral shape of the irradiating source(s) and their specific intensity be accounted for in modelling the equilibrium H$_{2}$ fraction in primordial halos.

\subsubsection{Is there a single critical Lyman-Werner flux?}

Much of the work that has been done to investigate the impact of radiation from the earliest star forming systems on the later collapse of primordial gas has adopted a highly simplified description of the spectral energy distribution of the radiation, often describing it in terms of a single temperature black body or a power-law spectrum. In particular, the use of a $T = 10^{5}\,\mathrm{K}$ black body spectrum (hereafter a T5 spectrum) to model emission from Pop III dominated systems and a $T = 10^{4}\,\mathrm{K}$ black body spectrum (hereafter a T4 spectrum) to model emission from Pop II dominated systems is particularly common, although some studies have also considered black-body spectra with intermediate temperatures \citep[e.g.,][]{Sugimura14,Latif2015}.

The earliest study to examine the impact of extremely large levels of incident Lyman-Werner radiation on the gravitational collapse of primordial gas was carried out by \cite{Omukai:2001p128}.\footnote{Earlier work by e.g.\ \cite{Haiman:2000p87} had typically focussed on much weaker radiation fields.} He showed that for a sufficiently large incident flux, H$_{2}$ cooling could be completely suppressed in the collapsing gas. In its absence, cooling from Lyman-$\alpha$ and Balmer series emission and from processes such as H$^{-}$ formation results in a quasi-isothermal collapse of the gas at a temperature $T \sim 7\,000\,\mathrm{K}$, dropping only gradually with increasing density, thereby leading to a very high Jeans mass of $\sim 10^{6}\,{\rm M}_{\odot}$ at $n \sim 10^{3}{\rm cm}^{-3}$.  In a low-mass dark matter minihalo, where the total gas reservoir often does not exceed this value, this would be enough to completely suppress the collapse of the gas. However, in an atomic cooling halo with a virial temperature $T_{\rm vir} \sim 10^{4}\,\mathrm{K}$ and with a much larger gas reservoir, collapse would not be suppressed. However, this high temperature would inhibit fragmentation, which leads to the idea that in such systems the gas may collapse directly without fragmenting to form a single supermassive star (or quasi-star) with a mass of $10^{4}$--$10^{5}\,{\rm M}_{\odot}$.

A key parameter in this scenario is the strength of the Lyman-Werner radiation field required in order to suppress H$_{2}$ cooling, commonly referred to as $J_{\rm{ crit}}$. The earliest studies \citep{Omukai:2001p128,BL03} found that extremely high values of $J_{\rm {crit}}$ were required, but \cite{SBH10} later showed that this was a consequence of the simplified treatment of H$_{2}$ collisional dissociation adopted in these studies, and that simulations with a more accurate treatment of this process yielded significantly lower values for $J_{\rm {crit}}$. The issue of the correct value of $J_{\rm{crit}}$ has subsequently been addressed by a number of other authors, but despite this considerable uncertainties remain. Single-zone models of the thermal and chemical evolution of the gas, similar to those studied by \cite{Omukai:2001p128}, typically find $J_{\rm{crit}} \sim 10$ for a T4 spectrum and $J_{\rm{crit}} \sim 1\,000$ for a T5 spectrum (see, e.g., \cite{Sugimura14} and \cite{Glover15a}).\footnote{The large difference between the value of $J_{\rm{crit}}$ for a T4 spectrum and that for a T5 spectrum is a  consequence of the fact that with a T4 spectrum, the destruction rate of H$^{-}$ ions for a given value of $J_{\rm LW}$ is much larger than for a T5 spectrum with the same $J_{\rm LW}$, owing to the difference in spectral shapes.} However, there is a factor of 2--3 uncertainty in these values that arises from uncertainties in the input rate coefficient data \citep{Glover15b}. In addition, there may be as much as an order of magnitude uncertainty arising from how H$_{2}$ self-shielding is treated \citep{Sugimura14,Glover15b}, with different self-shielding prescriptions yielding substantially different results. 

In addition, three-dimensional simulations of the collapse of highly-irradiated primordial gas typically find values of $J_{\rm{crit}}$ that are much larger than those found in one-zone models, with $J_{\rm{crit}} \sim 400$ for a T4 spectrum and $J_{\rm{crit}} \sim 10^{4}$ for a T5 spectrum \citep{SBH10,Regan_14,Latif2015}. The significant difference between one-zone and 3D model predictions for the value of $J_{\rm{crit}}$ is not completely understood, but may be a consequence of the large number of weak shocks that one finds in the 3D models, which produce a pervasive low level of ionization in the collapsing gas that is missing in the one-zone models.

Additional uncertainties arise once one considers other sources of radiation. For example, if a high redshift X-ray background is present, then this can significantly increase $J_{\rm{crit}}$ if the strength of the X-ray background is high enough \citep{IO11,IT15,Glover2016}, although the effect is much greater in one-zone models than in 3D models \citep{Latif2015}. Accounting for H$^{-}$ detachment by the Lyman-$\alpha$ photons produced during the collapse of the gas can also lead to a factor of a few reduction in $J_{\rm{crit}}$, potentially corresponding to up to a $\sim$100-fold increase in the anticipated number density of DCBHs \citep{JD17}.

Using a single black body spectrum to represent the integrated light from all of the stars in a galaxy is, however, a very crude approximation. A more realistic treatment necessitates incorporating more detailed models for spectral energy distributions (SEDs) from either single \citep[Starburst99,][]{Leitherer:1999p112}, or more recently-available binary \citep[BPASS,][]{2016MNRAS.462.3302E} stellar population and spectral synthesis codes. This allows one to compute the rate parameters $\alpha$ and $\beta$ for models of early stellar populations as a function of stellar mass, star formation rate (SFR), metallicity (Z), and age of the stellar population. Carrying out this analysis, \cite{Agarwal15a} and \cite{Agarwal_binary} found that $\alpha$ and $\beta$, and therefore $k_{\rm de}$ and $k_{\rm di}$, vary over several orders of magnitude during the lifetime of an instantaneous burst of star formation, though less than an order of magnitude for a continuous SFR. This implies that for galaxies with realistic spectra, one cannot summarize the effects of their radiation with a single $J_{\rm{crit}}$ parameter, since the value of this will depend on $\alpha$ and $\beta$, and hence on the properties of the individual stellar populations. Instead, the radiation constraint for DCBH formation can be better understood in terms of the interplay between $k_{\rm de}$ and $k_{\rm di}$.

\begin{figure}
\centering
\includegraphics[width=\columnwidth]{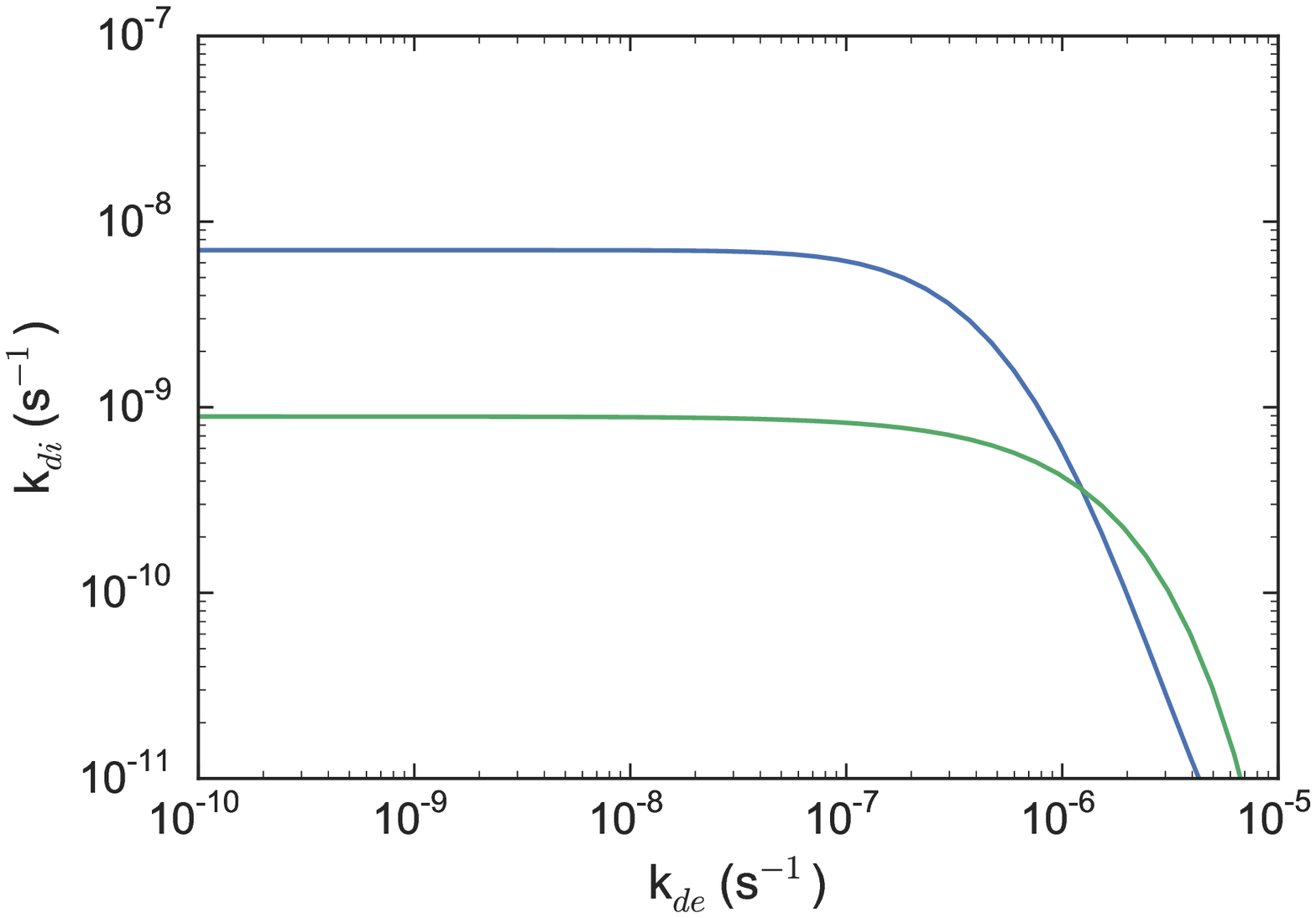}
\caption[Critical curve for direct collapse]{The critical curve(s) in k$_{de}$ - k$_{di}$ phase space, characterizing tthe photo-detachment of H$^{-}$ and the photo-dissociation of H$_{2}$ in an irradiated halo \citep{Sugimura14,Agarwal16,Wolcott2017}. Each curve divides the parameter space into two regions, depending on the equilibrium H$_{2}$ fraction resulting from the given values of k$_{de}$ and k$_{di}$. Rates above the curve lead to collapse into a DCBH, and below result in fragmentation into Pop III stars. The difference between the curves is due to the difference in the Jeans length and self-shielding of H$_2$. The blue curve \citep{Agarwal16} is obtained with a Jeans length that is twice that of the one used in the green curve \citep{Wolcott2017}.}
\label{fig.critcurve}
\end{figure}

A particularly useful way to express the radiation constraint \citep{Sugimura14,Agarwal16,Wolcott2017} is in the form of a critical curve in the $k_{\rm de}$--$k_{\rm di}$ parameter space (see Fig. \ref{fig.critcurve}) which separates those sets of reaction rates that lead to efficient H$_{2}$ cooling and fragmentation from those that lead to the suppression of H$_{2}$ cooling and supermassive star formation. This curve is a consequence of the chemo-thermal evolution of the irradiated gas, and although it is affected by many of the same uncertainties that hamper our efforts to determine $J_{\rm{crit}}$ for idealized spectra, it has the significant advantage that it does not depend on the spectral properties of the source of the radiation. We can therefore study what types of sources are likely to lead to the suppression of H$_{2}$ cooling by examining where the photochemical rates produced by different sources lie with respect to this critical curve.

In summary, no single-valued critical Lyman-Werner intensity $J_{\rm{crit}}$ appears to provide a suitable criterion for the formation of an atomically-cooled halo under arbitrary circumstances. Alternative prescriptions based on the availability and photo-destruction of relevant species, such as that formulated by \cite{Sugimura14}, \cite{Agarwal16}, and \cite{Wolcott2017} are more useful for understanding the equilibrium H$_{2}$ fraction in primordial halos irradiated by a external Lyman-Werner flux. Whether this irradiation provides the primary catalyst for DCBH formation remains an open question, with recent results adding further evidence that e.g., structure formation dynamics may in fact play a more significant role \citep{Wise2019}. This and other complicating factors are addressed in the following sections.

\subsection{Fragmentation}

One of the key requirements for the formation of a supermassive primordial star (SMS) with $M_\star \gtrsim 10^5~{\rm M}_\odot$ is to suppress vigorous fragmentation of its parent gas cloud. It is anticipated that for a nearly isothermal cloud in an atomic-cooling halo, where H$_2$ is dissociated by intense LW radiation and/or collisions with atomic hydrogen (see the previous section), the gas undergoes monolithic collapse without a major episode of fragmentation. Significant progress has been made in the past decade to understand the details of isothermal collapse of such an H$_2$-free gas in atomic cooling halos, both via analytical frameworks and numerical simulations. Three-dimensional hydrodynamical simulations show that the gas cloud is unlikely to fragment into gravitationally bound clumps during the runaway collapse phase, and forms a central protostar surrounding a massive accretion disk \citep{Latif_2013,Inayoshi_14,Becerra_15}.

\begin{figure}[t]
\begin{center}
\includegraphics[width=0.45\textwidth]{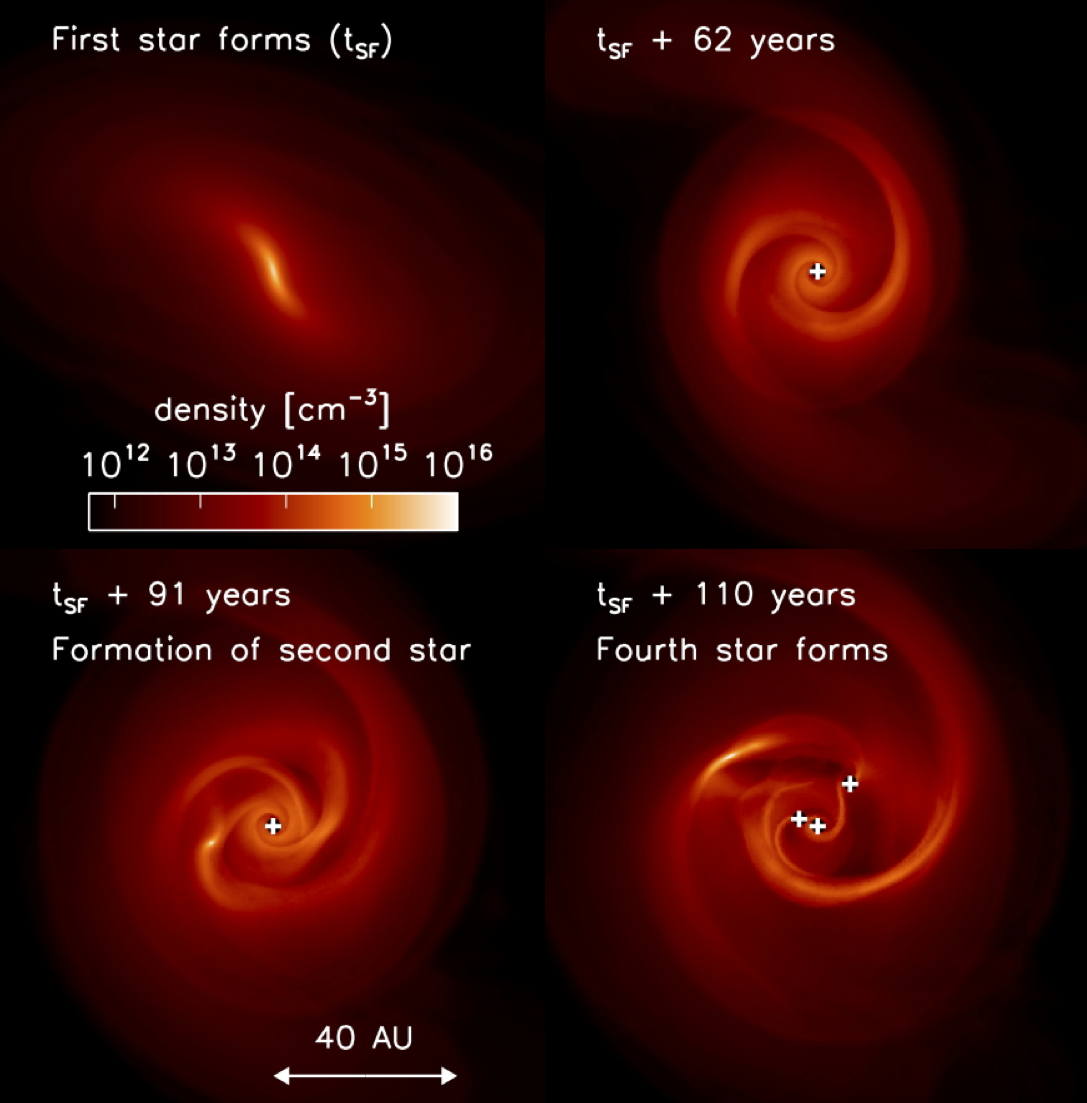}
\end{center}
\caption{Fragmentation of the accretion disk in the center of a primordial halo for conditions leading to the formation of normal-mass Pop III stars. The image is adopted from \citet{Clark11}.}\label{fig:frag}
\end{figure}

However, as material falls in, conservation of angular momentum leads to the build up of a centrifugally supported accretion disk. As the disk grows in mass it may become gravitational unstable, fragment, and form multiple objects \citep{Clark11,Clark2011b,Smith11,Greif12,Regan_14,Sakurai_2016}. This is illustrated in Figure \ref{fig:frag} for the case of an H$_2$ dominated disk which will form normal-mass Pop III stars. High resolution simulations suggest that fragmentation mostly occurs within the central parsec, but may also happen on larger scales in some cases, depending on the larger environment, merger history, spin, and so forth. Even if fragmentation occurs within the disk on small scales, many clumps may eventually migrate inward and merge with the central protostar. This is because the migration timescale in a self-gravitating disk is as short as the orbital timescale \citep{Clark11, Smith2012, Inayoshi_Haiman_14,Latif_Schleicher_15}. This leads to intermittent mass accretion onto the central object, with mass inflow rates up to  $0.1~{\rm M}_{\odot}~{\rm yr}^{-1}$ or more \citep{Sakurai_2016,Latif_16,Regan18,Becerra18}. In some cases, tidal force caused by nearby galaxies strongly influences the morphology of a massive H$_2$ cloud that potentially forms a SMS, leading to the potential formation of a smaller cluster of SMSs under a strong tidal field \citep{Chon16,Chon_2018}. Clearly, this area requires further attention. 

The efficiency of fragmentation depends strongly on the thermal evolution of the gas \citep{Klessen2016}, and so an adequate description requires one to take into account all relevant processes regarding radiative cooling and chemical reactions. For this purpose, recent simulations have employed detailed chemical models in order to explore the impact of H$^-$ cooling (${\rm H}+{\rm e}^- \rightarrow {\rm H}^- + \gamma$; ${\rm H}+{\rm e}^- \rightarrow {\rm H} + {\rm e}^- +\gamma$), and opacities due to H$^-$ bound-free/free-free transition and Rayleigh scattering of hydrogen atoms. In fact, H$^-$ cooling decreases the temperature with increasing density, i.e., $\gamma_{\rm eff} <1$, where $\gamma_{\rm eff}$ is the specific heat index ($P\propto \rho^{\gamma_{\rm eff}}$). Based on linear stability analysis, a runaway-collapsing cloud, theso-called the Larson-Penston self-similar solution \citep{Larson_1969}, is unstable against non-spherical perturbations when $\gamma_{\rm eff}\lesssim 1.1$. As a result, the collapsing region is elongated with the density and fragmenting into clumps \citep{Lai_2000, Hanawa_Matsumoto_2000,Sugimura_2017}. However, since the growth timescale is sufficiently long, turbulence in the collapsing gas would suppress the unstable modes. On the other hand, when the density increases to $\simeq 10^6-10^{13}~{\rm cm}^{-3}$, cold clumps with $T\sim 10^3\,\mathrm{K}$ are produced by the thermal instability induced by the combination of the adiabatic cooling due to turbulent expansion and the subsequent H$^-$ and H$_2$ cooling \citep{Inayoshi_14}. Since the cold clumps are not massive enough to be gravitationally bound, the evolution of the central collapsing region is not affected. We note that the thermal instability would play an important role after a rotationally supported disk forms at the center because the gravitational collapse is significantly delayed. A similar outcome is observed in less conventional
direct collape scenarios, such as in the simulations of multi-scale inflows driven by mergers between massive galaxies at $z \sim 8-10$ \citep{MB18}. In the latter case, gas is already
metal-enriched but the extremely high optical depth in the inner regions relegates rapid cooling and sporadic fragmentation into transient clumps just to the outer rims of the disk.

At densities above $\rm 10^{16} ~cm^{-3}$, the central region becomes opaque to all radiative processes and the temperature rises above $\rm 10^4$ K \citep{Inayoshi_14, VanBorm_14, Latif_16, Becerra18}. The opaque core (i.e., protostar) further increases the density and temperature adiabatically via mass accretion from the surrounding material. Due to the rapid input of high-entropy material, the central protostar evolves with an expanding envelope during the earliest stages of gas accretion \citep{Inayoshi_14} (see also \S 3). In the late phase, gas supply onto the protostar is dominated by episodic bursts because the surrounding accretion disk is unstable against its self-gravity, forming multiple fragments. In the quiescent phases of the episodic accretion, the protostar contracts via radiative diffusion and starts to emit ionizing photons.  \citet{Chon_2018} found that even though the ionizing photons produced from the proto-SMS heats the surrounding gas, the feedback effect cannot reverse the gas inflow to the center. This is in line with results from radiation hydrodynamic simulations of present-day star formation \cite[see, e.g.,][]{Krumholz2009,Peters2010, Peters2011, Kuiper2011}. However, radiative heating can reduce the level of disk fragmentation, and so further numerical simulations with a high spacial resolution and multi-frequency radiative transfer are required to fully quantify this effect.  In addition, studies with a larger sample of halos in a cosmological volume will allow us to address the fate of gas collapse, fragmentation, and formation of SMSs in atomic-cooling halos.

So far, the main focus of these studies have been on understanding the isothermal collapse and its implications for the formation of a SMS. This scenario demands rather special conditions and mandates that the halo should be metal free and the formation of $\rm H_2$ remains suppressed (see previous section for details). These requirements make the sites of SMSs less abundant if not rare but this is still an open question. However, less idealized conditions such as halos polluted with trace amount of metals $Z/Z_{\odot} \leq 10^{-5}$, halos irradiated by a moderate strength UV flux below the critical limit, or halos with large baryonic streaming velocities (see the following, dedicated subsection on this topic) may still form a SMS \citep{Latif15,Latif_16,Hirano17,Schauer17, Inayoshi18}.

In fact, halos irradiated by a moderate UV flux of strength $\sim 1000 J_{21}$ can still provide large inflow rates of $\sim 0.1\,\mathrm{M}_{\odot}\,\mathrm{yr}^{-1}$, sufficient to produce a SMS \citep{hos12, Latif14, Latif15,Regan18}. Under these conditions, gas remains hot with a temperature $\sim8\,000\,\mathrm{K}$ at scales above $\sim1$ parsec, similar to the isothermal case, while in the core of the halo sufficient $\rm H_2$ can form to allow the temperature to fall to a few hundred K. Similar conditions are observed for halos irradiated by an intense UV flux, but polluted with trace amounts of metals ($Z/\mathrm{Z}_{\odot} \leq 10^{-5}$). For further discussion regarding the role of streaming motions in the birth of SMS, see the following subsection. 

Less idealized conditions are more prone to fragmentation on scales below 1 pc, but clump migration in combination with dynamical/viscous heating at these scales stabilizes the disk. Therefore, conditions still seem conducive for the formation of a massive central object. Unfortunately, a better understanding of these conditions is still limited due to the computational constraints, however they may provide a potential pathway for the formation of SMSs at $z > 10$.

\subsection{Streaming Velocities}

Before recombination, baryons and photons are tightly coupled. This leads to  acoustic waves  \citep{Sunyaev1970} that also result in oscillations between baryons and dark matter. At $z \approx 1000$ the relative velocities are $\sim 30\,$km$\,$s$^{-1}$, with coherence lengths of several $10\,$Mpc to $100\,$Mpc \citep{Silk1986,Tseliakhovich2010}. This streaming velocity decays linearly with $z$, and reaches values of about  $1$km$\,$s$^{-1}$ at a redshift of $z \approx 30$. It is thus comparable to the virial velocity of the first halos to cool and collapse, and has the potential to strongly influence gas dynamics and star formation within these halos \citep{Tseliakhovich2011, Fialkov2012}. 

Early studies of this process, based on numerical simulations, indeed suggest that the kinetic energy provided by the streaming velocity provides additional stability. It reduces the baryon overdensity in low-mass halos and delays the onset of cooling, altogether resulting in a larger critical mass for collapse \citep{Greif2011, Stacy2011, Maio2011, Naoz2012, Naoz2013, Oleary2012, Latif2014Stream, Schauer17} and different angular momentum evolution of dark matter and dense gas \citep{chiou2018, druschke2018}. They may also have substantial impact on the resulting $21\,$cm emission \citep{Fialkov2012, McQuinn12,Visbal2012}. Specific to the subject of this review, it has been suggested that the presence of large streaming velocities may create the ideal conditions for the formation of supermassive black holes by preventing fragmentation and the formation of normal Population III stars \citep{tanakali2014}. Indeed, \citet{VHB14_a} and \citet{Schauer17} indicate that streaming velocities may suppress the formation of H$_2$, reduce the amount of cooling, and allow halos to reach the mass limit for atomic cooling to set in. However, their results also demonstrate that once collapse has set in, H$_2$ will rapidly form in the center of the halo and lead to fragmentation and the onset of a more normal mode of star formation. Therefore, we may conclude that external irradiation is still needed in the presence of streaming velocities to enable the formation of supermassive black holes by direct collapse. Indeed, \citet{Schauer17} argue that streaming velocities can facilitate the formation of synchronized halo pairs. In particular, if two halos form in close proximity to each other, then they can reach the atomic cooling phase at roughly the same time and start to go into collapse separated only by a few million years or less. The one forming stars earlier can then provide enough ultraviolet radiation to suppress H$_2$ cooling in the other. If this second halo goes into collapse before the stars in the first halo explode as supernovae or before the metals from these explosions have reached it, (i.e., if strong cooling and fragmentation can be prevented), then this collapse may lead directly to the build up of a supermassive black hole \citep{Agarwal17,agarwal2018}.

\subsection{Magnetic fields}
\label{subsec:magnetic-fields}

The presence of dynamically important magnetic fields has the potential to  significantly alter the picture presented so far. The current Universe is highly magnetized \citep{Beck1996}, but our knowledge of magnetic fields at high redshifts is very sparse. Possible sources of magnetic fields in the early Universe are battery processes \citep{Biermann1950}, the Weibel instability \citep{Lazar2009, Medvedev2004}, or thermal plasma fluctuations \citep{Schlickeiser2003}. Other theories resort to phase transitions occurring during cosmic inflation \citep{Sigl1997, Grasso2001,Banerjee2003, Widrow2012}. The resulting fields are typically too weak to have any dynamical impact, and so, magnetohydrodynamic effects are usually neglected when studying supermassive black hole formation. 

\begin{figure}
\begin{center}
\includegraphics[width=0.45\textwidth]{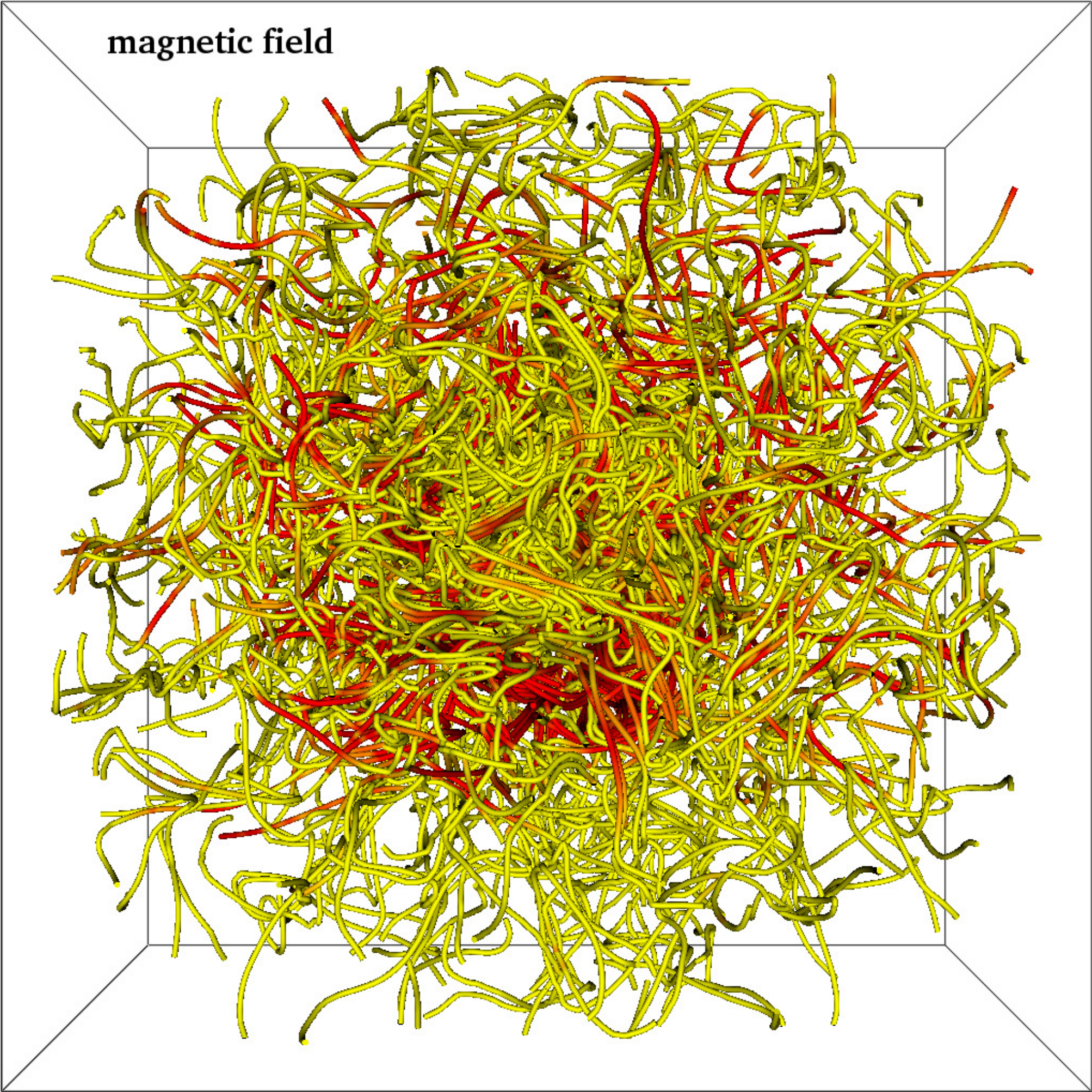}
\end{center}
\caption{Magnetic field lines (yellow: 0.5$\mu$G, red: 1$\mu$G) illustrating the complex magnetic field structure in the center of a contracting primordial halo resulting from the small-scale dynamo. The image is adopted from \citet{federrath2011b}.}\label{fig:mag}
\end{figure}
This simplifying assumption could be incorrect, however, due to the small-scale turbulent dynamo that is active during all phases of structure formation. This is able to efficiently amplify even extremely small primordial seed fields to the saturation level.  The amplification timescale is comparable to the eddy-turnover time on viscous or resistive length scales \citep{Kazantsev1968, Subramanian1998,Sur2010,Sur2012,Schober12a,Schober12b} depending on the Prandtl numbers, and in either case is much shorter than the dynamical or collapse timescale. The resulting magnetic field configuration is extremely complex, as illustrated in Figure \ref{fig:mag}, with the energy density being dominated by fluctuations on the smallest scales accessible to the system. Once  backreactions become important, the growth rate slows down, and saturation is reached within a few large-scale eddy-turnover times \citep{Schekochihin2002, Schekochihin2004, Schober2015}, which again is shorter than the typical lifetimes of protostellar accretion disks. The field in this phase is dominated by the more coherent larger-scale rotational structure of the accretion disk. The magnetic energy density at saturation is thought to reach levels of  0.1\% up to a few $\times$ 10\% of the kinetic energy density \citep{Federrath2011, Federrath2014}. 

Dynamically significant magnetic fields can strongly affect the evolution of protostellar accretion disks. They remove angular momentum from the star-forming gas \citep{Machida2008,Machida2013, Bovino2013NJP,Latif2013a,LatifMag2014}, drive protostellar jets and outflows \citep{Machida2006}, and most importantly for our discussion, they reduce the level of fragmentation in the disk \citep{Turk2011, Peters2014} and thus may facilitate the formation of supermassive black holes by direct collapse. To better understand this process, and in particular, to quantitatively determine how magnetic fields influence the dynamical evolution of the gas, remains an area of very active research.

\subsection{Black Hole Seed Growth: Simulation Requirements}
   
\cite{Chon_2018} and \cite{Regan18b} recently modelled the formation of SMSs in cosmological environments with semi-analytical recipes for the evolution of the SMS using the \texttt{Gadget} and \texttt{Enzo} codes, respectively \citep{Springel:2005p667,Enzo_2014}. Both simulations included radiative transfer to approximate stellar feedback from the SMS and thus make progress towards modelling the correct thermal state of the gas surrounding the SMS. The SMS is thought in most cases to collapse to a black hole at the end of its life because it either encounters the GR instability above a few hundred thousand \msolar \citep{woods17} or it exhausts its nuclear fuel.

The DCBH is likely born in the high gas densities in which the SMS had grown, with infall rates of $10^{-2}$ - $10^{-1}$ \msolar yr$^{-1}$. Accretion onto the black hole is regulated by a number of processes on scales down to nearly its event horizon, which cannot be resolved in current cosmological simulations. Therefore, the growth of the black hole is usually approximated by Bondi-Hoyle (B-H) accretion.  B-H accretion assumes that flows onto the black hole are spherical, and it is characterized by the B-H radius, $R_{\mathrm{B-H}}$:
\begin{equation}
R_{\mathrm{B-H}} = 2 G M_{\mathrm{BH}}/c^2_\mathrm{s},
\end{equation}
where $M_{\mathrm{BH}}$ is the black hole mass and $c_\mathrm{s}$ is the sound speed in the surrounding medium. Although flows onto the black hole are not spherical, the B-H model provides a reasonable benchmark for accretion rates. In numerical simulations, the accretion rate is usually just taken to be the inflow rate through $R_{\mathrm{B-H}}$ and the luminosity of the black hole is computed as $\epsilon \dot{m}c^2$, where $\epsilon$ is the radiative efficiency, the fraction of the mass that is converted to energy upon being swallowed up by the black hole.  Past simulations have simply deposited this luminosity locally as heat \citep[e.g.,][]{dm12,yu14,costa14,hir14} but the newest models now perform multi-frequency radiation transport from the black hole \citep{aycin14,smidt16,smidt18} as well as mechanical 
feedback in the form of jets (Regan et al. 2018c).

If the medium is not heated by the SMS, a 50,000 \msolar DCBH will have a B-H radius of $\sim$ $10^{-2}$ pc at birth. However, this radius can drop by up to three orders of magnitude when the black hole begins to accrete because X-rays and outflows heat the gas and raise its sound speed by a similar factor.  The B-H radius of a SMS can be resolved in  cosmological simulations, but doing so for a DCBH would restrict the simulation to such small time steps that it would be difficult to evolve the black hole or study its effect on its own subsequent growth.  Consequently, resolving the B-H sphere of a SMBH seed will remain a challenge for the foreseeable future.

\subsubsection{Modelling Accretion and Feedback from a Seed Black Hole below the Eddington Limit}
\begin{figure*}
\begin{center}
\hskip-0.67cm\includegraphics[width=0.75\textwidth]{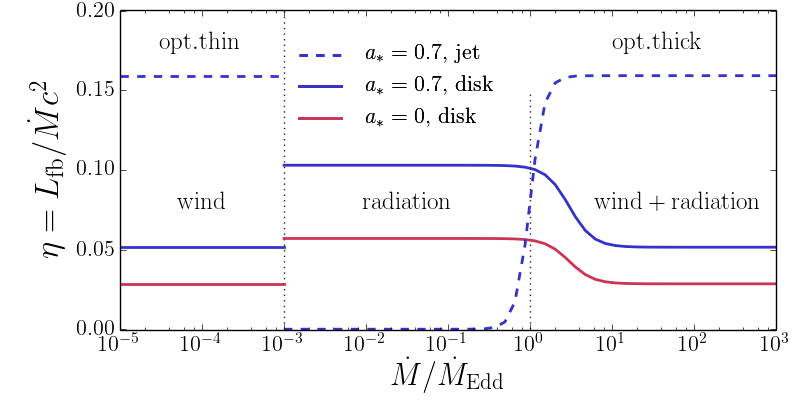}
\end{center}
\caption{Figure taken from \cite{Sadowski_2016} showing the regimes where both radiative and mechanical feedback are expected to dominate. $a_*$ is the normalised spin of the black hole. Radiative efficiency is plotted on the y-axis against accretion rate on the x-axis.}\label{sadowski_fig}
\end{figure*}

As already noted above, a rapidly accreting black hole seed will generate significant feedback, which must be accurately modelled by the simulation in order to correctly estimate the resulting growth. Feedback from a black hole can be broken down into two different components - radiative and mechanical. Furthermore, the regimes of black hole accretion are thought to lead to different feedback characteristics. 
Below an accretion rate of $\dot{M} \sim 10^{-3} \dot{M}_\mathrm{Edd}$ radiative feedback is expected to be minimal as the material surrounding the 
disk is optically thin and radiative feedback ineffective. In this case, mechanical feedback in the form of radio jets is thought to 
be the dominant feedback mechanism. Above this limit but below the Eddington limit (i.e., $10^{-3} \dot{M}_\mathrm{Edd} \lesssim \dot{M} \lesssim \dot{M}_\mathrm{Edd}$), radiative feedback is expected to be the dominant feedback mechanism. Then for super-Eddington flows, the 
radiative feedback component may become trapped by the rapid inflow; radiative feedback may then once again become inefficient, though jets may again dominate (see Figure \ref{sadowski_fig}).
However, it should be noted that these regimes are contentious and jets may in fact be active at all phases of the accretion scale \citep[e.g.][]{Sadowski_2017}.\\
\indent Radiative feedback from the accretion disk surrounding a black hole is the dominant form of feedback in the regime between radio-mode accretion and super-Eddington accretion.
The accretion onto a black hole and the radiative feedback resulting from this accretion is usually captured by the radiative efficiency parameter, $\epsilon$. For a geometically thin but
optically thick \citep[Shakura-Sunyaev,][]{Shakura_1973} disk accreting close to, but below, the Eddington limit, the radiative efficiency is $\epsilon = 0.057$ for a non-spinning black hole. Radiative feedback from the accretion disk can 
be modelled in a straightforward manner, by determining the appropriate spectral energy distribution (SED) for a given black hole mass and accretion rate \citep[e.g.][]{Done_2012}. \\
\indent Thermal feedback is often used as an approximation to model the impact of both radiative and mechanical feedback. This is simpler and less computationally expensive to invoke, but may miss the essential physics of radiative and/or mechanical feedback. However, in the regime where a black hole is accreting above the Eddington limit, the SED is unknown and may not necessarily follow
that of the relatively well understood thin disk. In this case modelling the radiative feedback is problematic as we do not know the SED and the only recourse, at the present time, may be thermal
feedback modelling. 

\subsubsection{Super-Eddington Accretion - slim disk accretion}
The slim disk model of accretion onto a black hole surface is derived from numerical 
integration of the Navier-Stokes equations, where the point at which the gas velocity exceeds the local sound speed is set as a critical point \citep{Abramowicz_1988}. The slim disk solutions can be viewed as a generalisation of the thin disk solutions dervied by \cite{Shakura_1973}. Although different models for super-Eddington accretion exist \citep[e.g.][]{Jiang_2014} we focus here on a 
discussion of the slim disk model as it is widely applied and captures the essential physics of super critical accretion 
flows. \\
\indent The slim disk model is valid for accretion flows which exceed the Eddington limit, a regime in which the thin disk solutions break down (in fact the thin disk model breaks down at approximately $\dot{m} \sim 0.5 \dot{m}_\mathrm{Edd}$). 
Black holes accreting at rates above the Eddington limit have been shown to generate powerful jets \citep{McKinney_2013, Sadowski_2014, Sadowski_2016a, Jiang_2014} which may significantly disrupt the accretion flow. While the jets have been shown to be highly collimated at the scales of the high resolution general relativistic radiative magneto-hydrodynamic (GRRMHD) simulations used to model them - the scales at which these simulations operate ($\lesssim$ 1000 Schwartzchild radii) is
typically orders of magnitude separated from the scales at which astrophysical codes can operate. Furthermore, as the bow shock from the jet propagates outwards, it will broaden and disrupt gas perpendicular to the disk. Simulations of jets from accreting seed black holes which use the results of much higher resolution, much smaller scale GRRMHD simulations are only now being attempted.  \\
 \indent Furthermore, as the accretion flow onto a black hole exceeds the Eddington rate, the radiative efficiency of the feedback has been shown to reduce - in some cases, significantly. Thus in the super-Eddington regime, the dominant form of feedback can be mechanical. For black holes accreting above the Eddington limit, \cite{Madau_2014} provide a empirical fit to the GRRMHD simulations of \cite{Sadowski_2009}, where the radiative efficiency of a black hole accreting at rates above the critical Eddington limit has indeed been shown to decrease, thus potentially allowing more accreted material to reach the black hole surface compared to the Eddington limited case. Of course, the net effect of the jet needs to be accounted for to determine whether the jets positively or negatively impact the accretion onto the black hole. \\
\indent In accretion disk simulations, the degree to which the radiative efficiency can decrease is, however, not yet established, as it appears to depend on the specific way by which radiation transportis treated, as well as other aspects of the numerical implementation
\citep[see review chapter by][for a detailed discussion]{MB18}. In particular, the
recent RMHD simulations of \cite{Jiang_2017}, which are among the few to study
accretion on massive black holes rather than stellar mass black holes, exhibit radiative
efficiencies that are an order of magnitude higher than those in the slim disk model,
albeit being still almost an order of magnitude lower than standard viscous disk efficiencies. Interestingly, radiative efficiencies are also found to decrease with
increasing accretion rates, being as small as $1\,\%$ of $\dot M c^2$ for accretion rates
above $100\,\mathrm{M}_{\odot}\,\mathrm{yr}^{-1}$.

\indent Finally, assuming that rapid accretion onto a seed black hole can be achieved, how easily can the black hole accrete the inflowing material? \cite{Sugimura_2018} investigated the angular momentum barrier for black holes attempting to accrete at high-z. Using 2D axially symmetric radiation hydrodynamic simulations, they found that the interplay of radiation feedback and angular momentum transport significantly suppresses accretion. They concluded that in order for the black hole to efficiently accrete the incoming material, the angular momentum of the gas must be significantly suppressed within a centrifugal radius $\lesssim 0.01 \times R_{Bondi}$. At face value this appears to introduce yet another significant barrier to achieving rapid black hole growth onto seeds at high-z. However, the much smaller scale GRRMHD simulations which model the magneto-rotational instability (MRI) in a self-consistent manner can efficiently transport angular momentum outwards \citep[e.g.][]{Sadowski_2017} showing that rapid accretion is possible, at least at small scales. 
\cite{Lupi16} carry out 3D hydrodynamical simulations of massive circum-nuclear disks,
a few pc in size, which are likely present in gas-rich galaxies at high redshift. Using
both AMR grid-based codes and Lagrangian mesh-less particle codes, they show that such
systems become quickly clumpy and gravitoturbulent. The large scale turbulence arising from
gravitational instability drives efficient angular momentum transport through the disk
down to sub-pc scales. By initializing stellar mass black hole seeds and assuming a slim
disk model as a sub-grid recipe to treat black hole accretion, these authors suggest that there is indeed a bottleneck in the accretion flow from large to small scales, even when
radiative heating of the ambient medium originating from accretion is taken into account.
These simulations, however, still fail to resolve the gas flow at scales significantly below a few
parsecs. \\
\indent Further high resolution simulations, incorporating both the physics of SMS formation, evolution and transition into a black hole, as well as the feedback effects of the resulting accretion onto the black hole, will be required to determine how attractive this pathway to eventual SMBH formation may be. As with the case of growing Pop III remnants (so-called light seeds), accretion can be shut-off and it is 
as yet unclear what conditions are required to achieve optimal growth possibly including super-Eddington accretion.

\section{How do supermassive stars live and die?}

\begin{figure}
\includegraphics[width=0.5\textwidth]{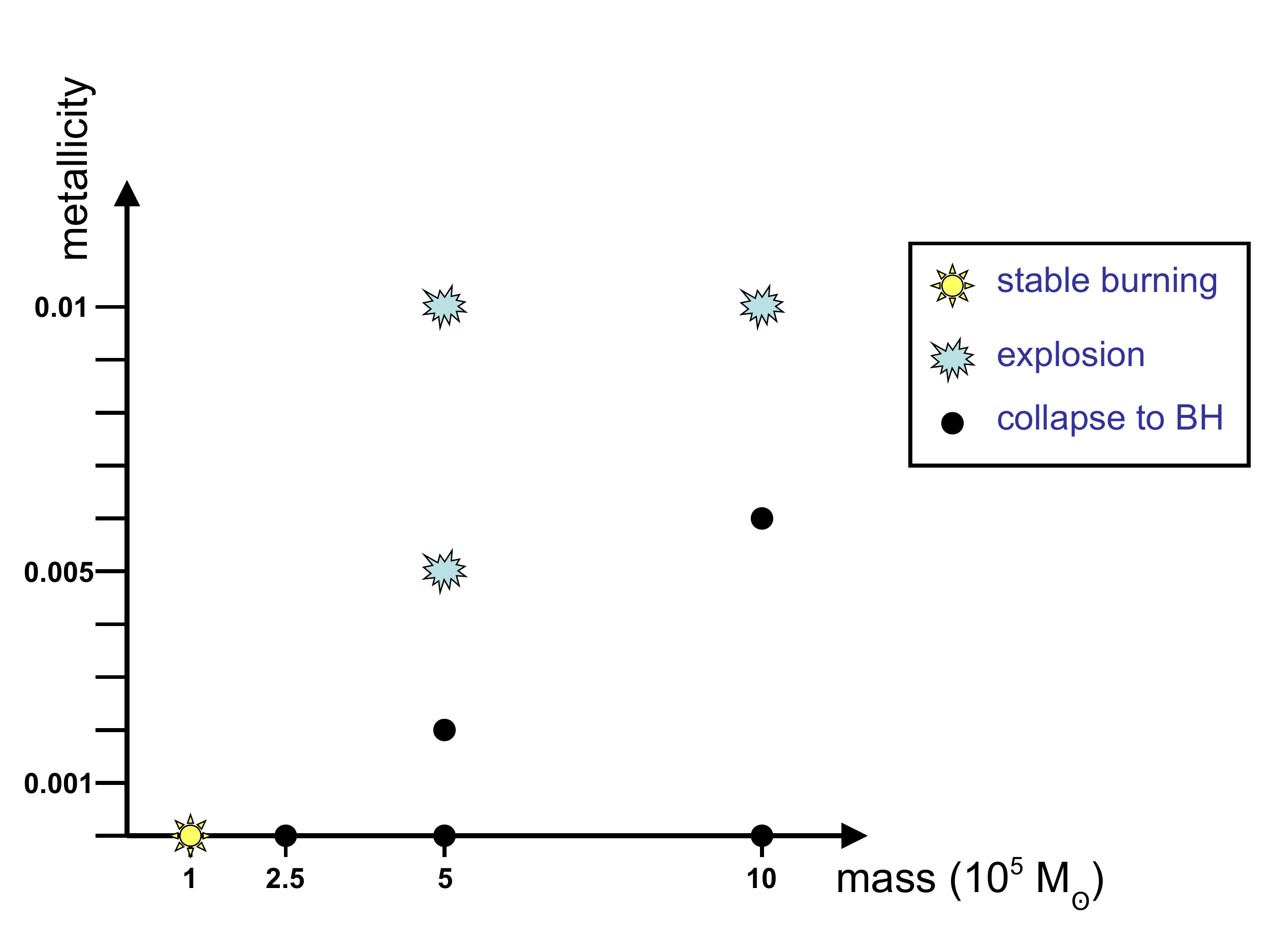}
\caption{Evolutionary outcomes for monolithically-formed supermassive stars as a function of mass and metallicity Z. Adapted from \cite{fuller86}, although see \cite{chen14}. Note that no rapidly-accreting models have been found to explode \citep[e.g.,][]{umeda16,woods17}.}\label{monolithic}
\end{figure}

Supermassive stars were originally conceived as a model for quasars themselves, rather than their progenitors \citep{HF63,Iben63,Fowler64}. Early analytic developments for the most part ignored how a SMS may have originated, assuming a single monolithically contracting cloud, and neglecting the impact of its formation history on its subsequent evolution. In this case, many of the physical characteristics of SMSs can be computed analytically, being well-approximated as polytropes of index 3. Such stars are, however, extremely prone to instability \citep{chandra39}.  It was soon understood that even in the relatively modest gravitational potential of a SMSs, post-Newtonian corrections to the equation of state would place a firm upper limit on its mass of $\sim 10^{5}\,\mathrm{M}_{\odot}$, above which the star would collapse on its dynamical timescale \citep{chandra64}. This led to a sustained interest in their evolution as potential progenitors of supermassive black holes \citep{Rees84}. Later, numerical experiments using the detailed stellar evolution code \textsc{Kepler} \citep{wzw78} suggested that for primordial composition SMSs, any collapse inevitably led to the formation of a black hole, preceded at lower masses by stable hydrogen burning \citep[][see also Fig. \ref{monolithic}]{fuller86}.  More recently, it has been suggested that there could be a narrow range of masses around $50,\!000\,\mathrm{M}_\odot$ for which primordial, ``monolithic'' SMSs may instead undergo a thermonuclear explosion, leaving no gravitationally bound remnant \citep{chen14}.  It remains uncertain, however, whether this result is robust against, e.g., details in assumptions and modelling of the underlying physics.  The narrowness of the mass window may make such explosions very rare and hence hard to exclude by direct observations, but see $\S$5 for further discussion.

Producing a single, supermassive, contracting cloud as a precursor to a supermassive star, e.g., through the collapse and coalescence of a dense stellar cluster, appears unlikely, however, to produce stellar masses greater than $\sim$ a few $10^{3}\,\mathrm{M}_{\odot}$ \citep[see][and references therein]{LF16}, though other channels may be viable \citep[see, e.g.,][]{Mayer10,MB18}.  In the atomically-cooled halo scenario, a supermassive proto-star instead grows (relatively) gradually at accretion rates of order $0.1$--$10$ $\,\mathrm{M}_{\odot}\,\mathrm{yr}^{-1}$.  This has a profound effect on the structure of the star: it never reaches thermal equilibrium \citep{Begelman10} and therefore detailed stellar evolutionary calculations are needed to predict the star's observational characteristics and fate \citep{hos12}.  In particular, although such stars still encounter the post-Newtonian instability and collapse, simple polytropic models underestimate the final mass \citep[see, e.g., discussion in][]{woods17}.  For non-rotating SMSs, the general properties of accreting supermassive stars are now well-understood \citep{hos13,haemmerle18a}.  In particular, the effective temperatures of rapidly-accreting SMSs are found to be significantly cooler than in the monolithic case, and no thermonuclear explosions are expected to reverse their collapse to black holes.  There is also a broad agreement on the final mass at collapse as a function of infall rate.  The largest discrepancies in final masses are found at the highest accretion rates, up to about a factor of $2$ \citep{umeda16,woods17,haemmerle18a}.

Beyond this qualitative agreement, however, significant differences remain in the predicted effective temperatures and final evolutionary states of SMSs found by the different numerical studies ($\S$\ref{sec-accr} below).  Furthermore, efforts to include rotation in detailed evolutionary calculations have clearly demonstrated a bottleneck in the growth of any SMS; an extremely efficient mechanism for angular momentum removal must be in place if an accreting proto-star is to avoid spinning-up to the mass-shedding limit before reaching the SMS regime \citep{haemmerle18b}.  How this may take place remains uncertain, although this problem is known to exist for massive star formation at all metallicities in general.  In the following, we discuss the latest progress on these issues, as well as benchmarking stellar evolutionary calculations from the participating groups. 

\subsection{The Structure and Evolution of Monolithic Supermassive Stars}

\label{sec-monolithic}

What do we mean when we say a star is supermassive?  Attempts at naming schemes for `big' stars based strictly on total mass have been made in the past \citep[e.g.,][]{heger02,heger03,heger10,heger12}; here, however, we use an approach based on the relevant physics.  In order to arrive at a clear definition, we must first consider the structure of massive stars in general.  For a gravitationally-bound sphere of ideal gas, hydrostatic equilibrium requires greater central temperatures at greater total masses.  This means that for higher and higher stellar masses, more and more of the total pressure in the star comes from radiation.  For very high masses, gas pressure is only a small perturbation to the total pressure.  In this case, it can be shown that, to a good approximation, the total pressure scales as $\propto \rho ^{4/3}$; more formally, the star is well-approximated as a polytrope of index $3$ \citep{KippenhahnWeigert}. The temperature gradient in a radiation-dominated $n\approx3$ polytrope is almost equal to the adiabatic gradient, and therefore nuclear burning easily drives convection throughout the star.  Consequently, convective mixing enforces a flat entropy profile; in the high-entropy limit relevant to massive, radiation-dominated stars, the adiabatic pressure gradient may then be well approximated as \citep{chandra39}:

\begin{equation}
\Gamma = \left(\frac{\partial\ln P}{\partial\ln \rho}\right)_\mathrm{\!\!ad} = 1 + \frac{1}{n} \approx \frac{4}{3} + \frac{\beta}{6}\label{Psupport}\;,
\end{equation}

\noindent where the ratio of gas pressure to total pressure, $\beta$, can be approximated in the limit $\beta \ll 1$ as

\begin{equation}
\beta = \frac{P_\mathrm{g}}{P_{\rm{tot}}} \approx \frac{4.3}{\mu}\left(\frac{\mathrm{M}_{\odot}}{M}\right)^{1/2}\;.
\end{equation}

\noindent where $\mu$ is the mean molecular weight of the gas, assuming the star is chemically homogeneous. For very large $M$, the pressure gradient $\Gamma \rightarrow 4/3$. This ``softness'' of the equation of state leaves the star very unstable against small perturbations \citep{Fowler64,GoldreichWeber,KippenhahnWeigert}. For this reason, a number of typically negligible effects must be taken into account in any detailed numerical modeling of extremely massive stars \citep{fuller86}, including, perhaps surprisingly, general relativistic corrections to the equations of stellar structure.

The critical role post-Newtonian effects play in the fate of extremely massive, quasi-statically contracting stars and proto-stars was first noted in the works of \cite{HF63} and \cite{Iben63}.  To summarize \citep[following][]{fuller86}, hydrostatic equilibrium requires that the total energy of a star, a function of its total mass, central density, and entropy, be at an extremum.  An initially low-density/high-entropy supermassive proto-star, produced by, e.g., the disruption and collapse of a dense stellar cluster \citep{BR78} will lose energy to radiation and contract, increasing its central energy and decreasing its equilibrium energy.  The effect of including general relativistic corrections to the equations of stellar structure, however, is that the equilibrium energy will have a minimum. As the star contracts, its central density may exceed a critical threshold $\rho _{\rm{crit}}$, above which energy must be added to achieve hydrostatic equilibrium.  Therefore, if a quasi-statically contracting polytrope reaches this point before nuclear burning can stabilize its evolution, it will undergo a dynamical collapse.  This instability was first formally derived by \cite{chandra64} and \cite{feynman}, and in the literature is variously called the Chandrasekhar, Chandrasekhar-Feynman, or post-Newtonian instability.  For a polytrope of index 3, the critical central density for the onset of a dynamical instability is

\begin{equation}
\rho _{\rm{crit}} \approx 1.994 \times 10^{18} \left(\frac{0.5}{\mu}\right)^{3}\left(\frac{\mathrm{M}_{\odot}}{M}\right)^{7/2} \mathrm{g}\,\mathrm{cm}^{-3}\label{rhocrit}\;,
\end{equation}

\noindent Another, perhaps more intuitive, way to express this criterion is as a condition on the critical pressure gradient needed to support the star against collapse; for a fully Newtonian star, this is $\frac{4}{3}$, whereas in the post-Newtonian limit, general relativity requires a slightly steeper gradient:

\begin{equation}
\Gamma_{\rm{crit}} \approx 4/3 + 1.12\,\frac{2\,G\,M}{R\,c^{2}}\label{GammaCrit}\;,
\end{equation}

\noindent where the coefficient $1.12$ is valid for the specific case of a polytrope with index $n=3$ \citep[c.f.,][]{chandra64}.  Comparing Eq.~\ref{GammaCrit} with Eq.~\ref{Psupport}, we see that the adiabatic pressure gradient in a fully convective, radiation-dominated polytrope of $n=3$ falls below the threshold needed to support it against collapse above a characteristic mass:

\begin{equation}
M_{\rm{SMS}} \approx \left( 0.32\; \frac{R\,c^{2}}{\mu\,G\,\mathrm{M}_{\odot}}\right)^{2/3}\,\mathrm{M}_{\odot} \sim 10^{5}\,\mathrm{M}_{\odot}
\end{equation}

\noindent This defines the realm of truly supermassive stars; very massive stars below this regime (roughly $10^{2}$--$10^{4}\,\mathrm{M}_{\odot}$) invariably survive to nuclear-burning and collapse on the electron-positron pair instability \citep[e.g.,][]{BAC84}. Understanding the boundary between these two regimes, however, and the role nuclear burning can play in supporting the star against collapse, or even reversing this and producing an explosion, requires detailed stellar evolution calculations.

Computational modeling efforts to follow the evolution and fate of supermassive stars in detail were carried out as early as the 1970s \citep[e.g.,][]{AF72a,AF72b}.  The first systematic effort in modelling these objects to include a comprehensive treatment of nuclear burning \citep[including the rp-process,][]{WW81}, post-Newtonian corrections, and hydrodynamics was that of \cite{fuller86}.  Beginning with a grid of highly-inflated (high-entropy) polytropes of index $n=3$, \citeauthor{fuller86} allowed their models to quasi-statically contract and followed the onset of nuclear burning.  They found that SMSs of primordial composition either collapsed directly to black holes within approximately the Kelvin-Helmholtz time, or survived to undergo the hydrogen-burning main sequence.  Modern simulations incorporating a complete, nuclear reaction network place the boundary between these two regimes at $\approx155,000\,\mathrm{M}_{\odot}$ \citep{woods19pre}.  It was found that more massive SMSs, however, with an initial metallicity, $Z\gtrsim1/4\,\rm{Z}_{\odot}$ could be completely disrupted in thermonuclear explosions, not seen in the primordial case \citep[though see][]{chen14}. 

This behaviour is easily understood.  For primordial SMSs, the absence of CNO elements necessitates that the star contract until the onset of He-burning through the triple-alpha reaction, as hydrogen-burning through the pp-chain alone cannot stabilize massive stars against further contraction \citep{heger10,KippenhahnWeigert}.  This means that vigorous nuclear burning begins in primordial SMSs only after most of their mass has fallen much deeper into their potential wells.  For SMSs above some threshold, nuclear burning is unable to reverse the collapse.  For SMSs of higher metallicity, with abundant CNO to catalyse nuclear burning, this is not the case, and thermonuclear explosions are possible in some cases (see Fig.~\ref{monolithic}). 

It remains uncertain, however, whether conditions under which non-zero metallicity SMSs could form ever arise in the Universe.  In particular, the disruption of dense stellar clusters alone does not appear to be able to produce sufficiently massive objects \citep[see][for a review]{LF16}.  The merger of massive, gas-rich proto-galaxies may be capable of producing SMSs of arbitrary metallicity, however, it is not clear at present whether the structure of such stars would in any way resemble the monolithic polytropes explored above.  It is clear, however, that the rapidly-accreting supermassive protostars produced in the atomically-cooled halo scenario, feasible only for primordial or extremely low metallicities, emerge with a radically different internal structure \citep[e.g.,][]{hos12, hos13, umeda16, woods17, haemmerle18a}.  This has profound consequences for their evolution and fate, which we outline in detail in the following subsection. 

\subsection{Rapidly-accreting Supermassive Stars}
\label{sec-accr}
\subsubsection{Non-rotating Case}
\label{sec-accr-nonrot}
Rather than forming ``all-at-once'' from the contraction of a single monolithic cloud, the prevailing theory for the formation of massive black hole seeds is now the ``synchronized pairs'' scenario, giving rise to an atomically-cooled halo (see $\S$2). In this picture, supermassive stars form under extreme accretion rates of $\sim 0.01$--$10.0\,\mathrm{M}_{\odot}\,\mathrm{yr}^{-1}$ onto a small central seed. This means that under typical conditions, the time needed for the star to reach the supermassive regime ($\gtrsim 10^{5}\,\mathrm{M}_{\odot}$) is comparable to its nuclear-burning lifetime. 

\begin{figure*}
\begin{center}
\includegraphics[width=0.75\textwidth]{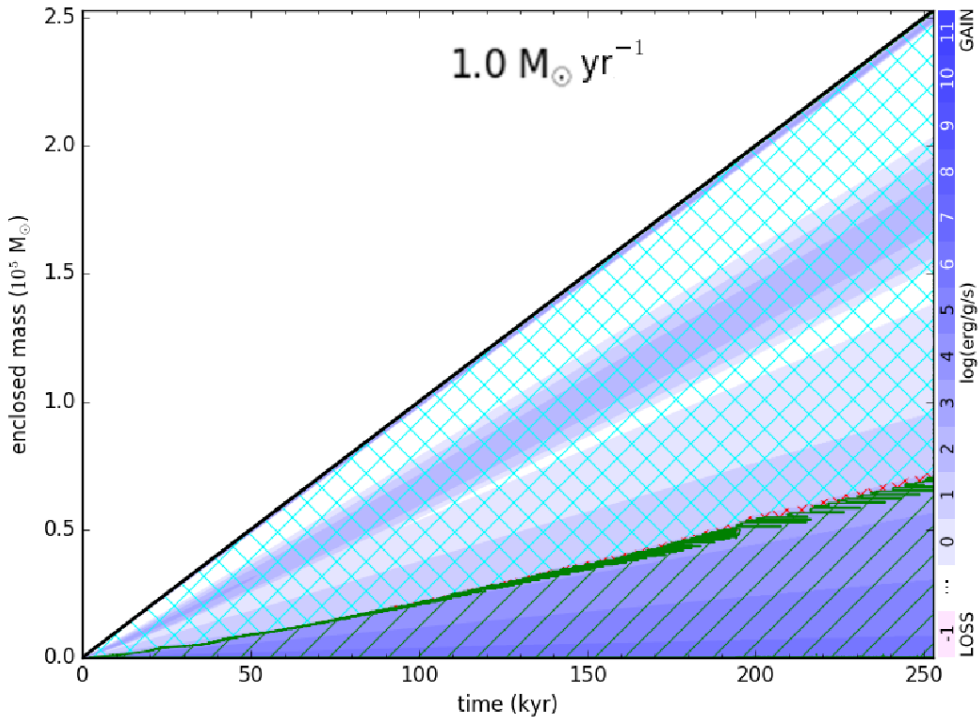}
\end{center}
\caption{Kippenhahn diagram for a supermassive star accreting $1\,\mathrm{M}_{\odot}\,\mathrm{yr}^{-1}$, showing the structure of the star at any given mass coordinate as a function of time. Green single-hashed regions are convective, blue double-hashed regions are radiative. The blue shading denotes energy generation, with the scale given on the right axis. From \cite{woods17}.}\label{Kippenhahn}
\end{figure*}

This has profound consequences for the evolution of the rapidly-accreting protostar \citep{Begelman10}. In particular, early in its lifetime the accretion timescale

\begin{equation}
t_{\rm{acc}} = \frac{M_{*}}{\dot M}
\end{equation}

\noindent is considerably shorter than the Kelvin-Helmholtz (thermal) timescale

\begin{equation}
t_{\rm{KH}} = \frac{GM^{2}_{*}}{R_{*}L_{\rm{int}}}
\end{equation}

\noindent with $L_{\rm{int}}$ the internally-generated luminosity of the star and $R_{*}$ its radius. Unable to cool to thermal equilibrium, these stars dramatically expand before reaching the main sequence \citep{OP01,OP03}. The internal structure of such stars was first investigated by \cite{Begelman10}, who modified the original polytropic formulation in order to incorporate a mass-dependence in the equation of state (so-called ``hylotropes''). Since the continuously-growing outer layers of a rapidly-accreting supermassive star can not generally relax within its nuclear-burning lifetime, Begelman showed that the structure of the star on the main-sequence would deviate dramatically from that of an n=3 polytrope, with an inner, nuclear-burning convective core surrounded by a high-entropy envelope (see also fig. \ref{Kippenhahn}). In the time it takes to become supermassive, the core will have already evolved significantly. Accretion continues to proceed at a rate faster than the envelope can cool, so that the envelope's entropy profile steeply increases outwards.

The first efforts to produce numerical simulations of rapidly-accreting supermassive stars followed shortly thereafter. \cite{hos12} followed the early evolution of such an object until reaching $\sim1000\,\mathrm{M}_{\odot}$. 
They found that the effective temperature of the inflated protostar fell to an almost constant $\approx5\,000\,\mathrm{K}$, approximately the Hayashi limit for primordial stars \citep{Hayashi61,Volonteri10}. The extreme temperature sensitivity of $H^{-}$ absorption, the primary source of opacity in this regime, prevents the star from cooling and its outermost layers from expanding further. Given that the luminosity of a massive, radiation-dominated SMS throughout its evolution will be very near the Eddington limit ($\propto M_{*}$), the Stefan-Boltzmann law provides a relation between the radius of a SMS and its mass:

\begin{equation}
R_{*} \approx 2.3\times 10^{3}\,\mathrm{R}_{\odot} \left(\frac{M}{100\,\mathrm{M}_{\odot}}\right)^{1/2}
\end{equation}

\noindent closely approximating numerical results \citep{hos12}. 

During the star's subsequent evolution, \cite{hos13} found that the radius continues to grow monotonically with the mass until the star reaches $\approx$~$10^{4}\,\mathrm{M}_{\odot}$. At this point, the ever-lower gas density in the outer envelope falls too low for $H^{-}$ absorption to be an effective source of opacity, and the SMS contracts somewhat \citep[see also][]{haemmerle18a}. Critically, however, the SMS never grows hot enough to become a significant source of ionizing photons, in sharp contrast to the monolithic case \citep{Johnson_2012}, reaching only $\sim 1-2 \times 10^{4}\,\mathrm{K}$ in detailed numerical simulations using different implementations \citep[e.g.,][]{hos13,umeda16,woods17,haemmerle18a}. Therefore, in contrast to the case for lower accretion rates ($\sim 10^{-3}\,\mathrm{M}_{\odot}\,\mathrm{yr}^{-1}$) typical of massive star formation in the present-day Universe \citep{OP01}, the growing luminosity of accreting SMSs never becomes self-limiting, since the star always remains too cool to ionize a surrounding H II region \citep{Johnson_2012,hos13}. This is essential to the viability of SMSs as the progenitors of the first black holes in the atomically-cooled halo scenario.

A potential caveat to this, however, is that all of the studies discussed above assumed constant accretion rates throughout the lifetimes of their model SMSs. Under more realistic conditions, however, accretion may be highly variable, e.g., if mediated by a disk which becomes gravitationally unstable and fragments \citep[e.g.,][]{SGB10}. Indeed, such disk fragmentation is expected in atomically-cooled halos \citep[see e.g.,][]{Regan_14}. The resulting ``clumps'' of matter would then migrate inward through the disk and merge with the central SMSs in bursts of rapid accretion. \cite{sak15} were the first to examine in detail the effect this process would have on the evolution of SMSs. They found that in the quiescent phases in between these bursts, the outer envelope of the SMS would begin to thermally relax and contract. Consequently, the effective temperature of the SMS may grow high enough to produce a substantial flux of ionizing photons. If left unchecked, this phenomenon could provide a significant source of feedback, potentially  halting the growth of the star. Subsequent bursts of accretion, however, will invariably re-inflate the SMS. \cite{sak15} found that for quiescence durations greater than

\begin{equation}
\Delta t_{\rm{quiescent}} \gtrsim t_{\rm{KH,env}} \sim \text{few}\times 10^{3}\,\mathrm{years}
\end{equation}

\noindent the ionizing photon luminosity of the contracting SMS may grow sufficient to produce an H II region (actually, this is the point in their calculation where 1 ionizing photon is produced by the SMS for every atom accreted, a simplifying assumption). Numerical simulations of the collapse of atomically-cooled halos suggest quiescence times in excess of a few thousand years may be expected \citep[e.g.,][]{Inayoshi_Haiman_14}. UV feedback may therefore play a role in limiting the growth of SMSs, although more detailed radiative transfer calculations are needed.

Perhaps the most vital question regarding the evolution of SMSs, at least in the context of the formation of high-redshift quasars, is the final mass and fate of such objects. Here again the radically different structure of rapidly-accreting SMSs sharply contrasts with the monolithic case. In particular, the lower densities of inflated, rapidly-accreting SMSs should allow them to avoid collapse at higher masses than otherwise expected \citep{Begelman10,umeda16,woods17}. Classically, a monolithic SMS undergoes the post-Newtonian instability once its core density satisfies eq. \ref{rhocrit}; the pressure support in an n=3 polytrope can no longer hold the star up against collapse. Another way of expressing this constraint is evident from comparison of eqns. \ref{Psupport} and \ref{GammaCrit}; both the critical pressure support needed to prevent collapse (including the post-Newtonian correction) and the adiabatic pressure gradient are very nearly 4/3, each including only a small perturbation. Equating the two gives us yet another formulation of the condition for instability:

\begin{equation}
\frac{\beta}{6} < 1.12\,\frac{2\,GM}{R\,c^{2}}
\end{equation}

\noindent which, if satisfied within a fully convective, radiation-dominated star, should signal the onset of collapse. 

\begin{figure}
\begin{center}
\includegraphics[width=\columnwidth]{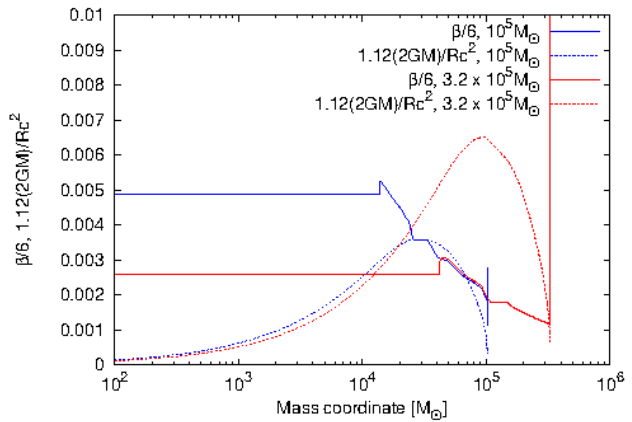}
\caption{Evaluation of the polytropic criterion for the onset of the post-Newtonian instability for a numerical simulation of the structure of a SMS accreting $10\,\mathrm{M}_{\odot}\,\mathrm{yr}^{-1}$.  This is shown for two moments in its evolution: once it has reached $10^{5}\,\mathrm{M}_{\odot}$ (blue lines), and at $\approx3.2\times10^5\,\mathrm{M}_{\odot}$ (red lines), the latter being shortly before the collapse of the star \citep{woods17}. Solid lines plot $\Gamma_1-4/3\approx\beta/6$, and dashed lines indicate the $n=3$ polytropic criterion for instability, both as a function of mass coordinate. For comparison, the solid line for a true n=3 polytrope would be flat with mass coordinate, and the instability would arise once the dashed line rose above it at any point in the star.}\label{criterion}
\end{center}
\end{figure}

A SMS, however, is emphatically not well-approximated by an n=3 polytrope, with the specific entropy rising steeply throughout its outer envelope. An illustrative example is given in Fig. \ref{criterion}, for the extreme case of a supermassive star accreting $10\,\mathrm{M}_{\odot}\,\mathrm{yr}^{-1}$.  Here we have plotted either side of the inequality above; for a fully convective star, the onset of collapse should arise when the post-Newtonian correction to the critical pressure support exceeds that provided  by gas pressure. For rapidly-accreting stars, however, the structure of the star at this point is not well-modeled as an $n=3$ polytrope, even when only the inner convective core is considered. Incorporating implicit hydrodynamics (and post-Newtonian corrections to the equations of stellar structure) allows one to follow the onset of collapse in detail using a detailed stellar evolutionary code \citep[e.g.,][]{fuller86,umeda16,woods17}. In the case presented in Fig. \ref{criterion}, the SMS clearly avoids collapse long after the polytropic criterion has been satisfied within the core; the accreting SMS only encounters the post-Newtonian instability and undergoes hydrodynamic collapse after reaching $\approx330\,000\,\mathrm{M}_{\odot}$.

Evidently, following the accretion history and evolution of a SMS, with the ability to follow its hydrodynamic response in detail at the onset of instability, is essential in order to accurately predict the final mass at collapse of massive black hole seeds. Several studies have now investigated this question for the case of non-rotating supermassive stars accreting zero-metallicity gas. For $>0.1\,\mathrm{M}_{\odot}\,\mathrm{yr}^{-1}$, the final mass at collapse is well-approximated as a logarithmic function of the accretion rate:

\begin{equation}
\rm{M}_\mathrm{SMS,final}\approx\left[0.83\,\log_{10}\left(\frac{\dot{M}}{\mathrm{M}_{\odot}\,\mathrm{yr}^{-1}}\right)+2.48\right]\times10^5\,\mathrm{M}_{\odot}
\end{equation}

\noindent although there is some divergence at higher accretion rates between different codes, rising to a factor of $\sim 2$ for $\sim 10\,\mathrm{M}_{\odot}\,\mathrm{yr}^{-1}$ (see \S\ref{benchmark} for further discussion). For the first time then, the seed masses of the first quasars may be described as a function of their initial conditions, albeit only for the non-rotating case. In principle, a rapidly-accreting protostar should gain not only mass but angular momentum, quickly reaching very high spin velocities at the surface. In the following section, we turn to the evolution of such rotating supermassive stars.

\subsubsection{Rotation \& Mass Loss}

Star formation always requires mechanisms to extract angular momentum from the collapsing pre-stellar gas \citep{spitzer1978}.
The existence of stars proves that nature finds ways to circumvent this angular momentum problem, through magnetic fields, gravitational torques or viscosity in accretion disks.
The efficiency of these mechanisms, however, depends on the properties of the forming star, mainly on its mass and metallicity.
SMSs are not immune to this angular momentum problem.
Thus their rotation is not only of interest in itself, but is also another mechanism which could prevent SMS formation by rapid accretion.
Furthermore, rotation could impact the life and death of SMSs by stabilizing the star against radial pulsations \citep{fowler1966} and by allowing a small fraction of the stellar mass to stay outside the horizon after the collapse \citep{baumgarte1999b}.

As a consequence of this angular momentum problem, rotating monolithic models (Sect.~\ref{sec-monolithic}), which did not address the formation process, assumed that SMSs rotate at the Keplerian velocity, losing mass as they contract.
This limit is thus called the `mass-shedding' limit \citep{baumgarte1999a}.
In particular, it implies a strong deformation of the star by rotation.
Moreover, since monolithic models are fully convective, the star is assumed to rotate as a solid-body.

\begin{figure}
\includegraphics[width=.95\columnwidth]{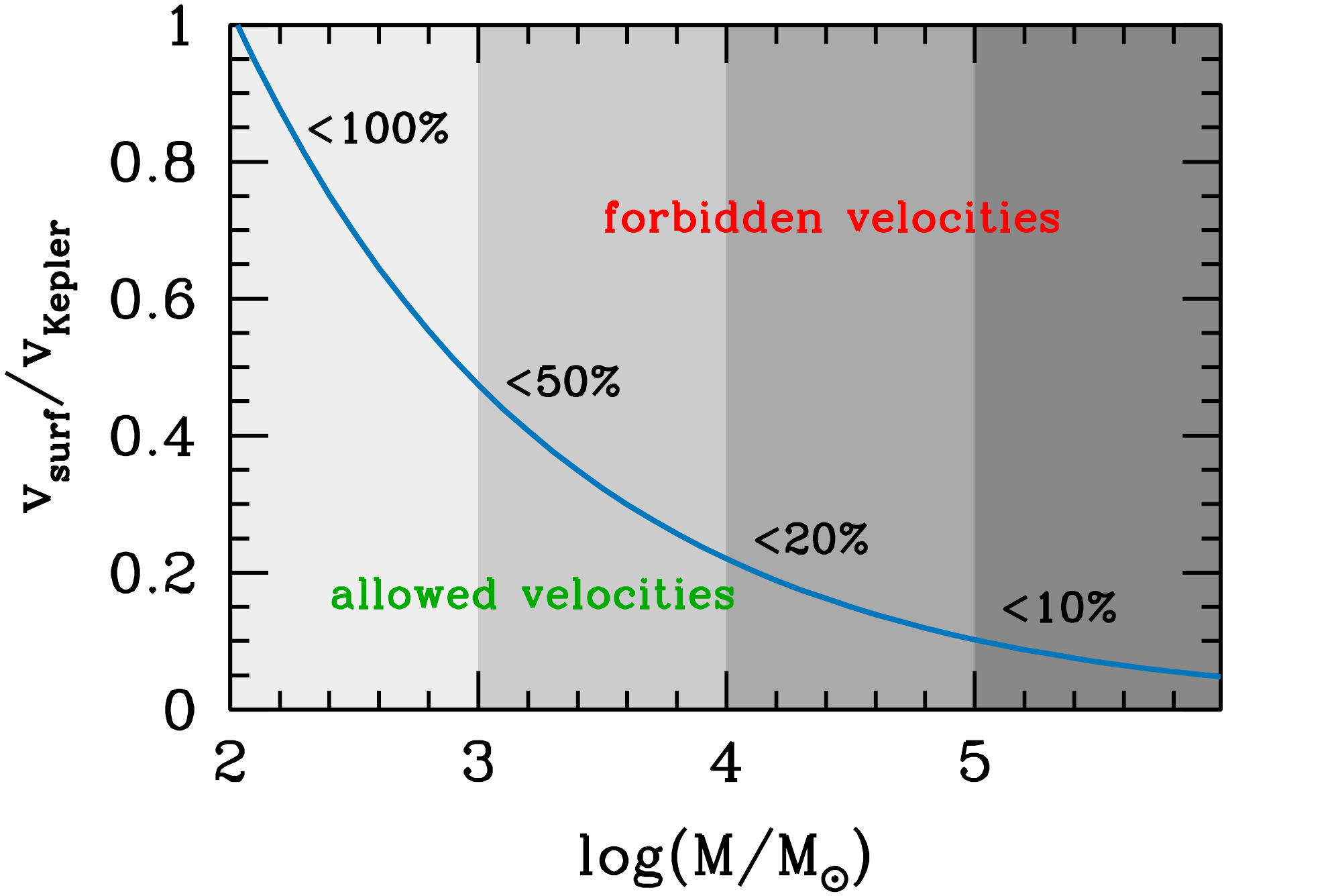}
\caption{Constraint from the $\Omega\Gamma$-limit on the surface rotation velocity of accreting SMSs, as a function of their mass.
The blue curve indicates the upper velocity limit imposed by the $\Omega\Gamma$-limit \citep{haemmerle18b}.}\label{omgam}
\end{figure}

This picture changes completely when we include the accretion process.
SMSs forming by accretion must be slow rotators, with surface velocities smaller than 10~--~20\% of their Keplerian velocity \citep{haemmerle18b}.
At such low velocities, the deformation of the star by rotation is negligible.
This constraint is a consequence of the $\Omega\Gamma$-limit \citep{maeder2000}: for stars near the Eddington limit, like SMSs, the critical velocity at which the effective gravity vanishes at the equator is reduced compared to the Keplerian limit, due to the contribution of radiation pressure to hydrostatic equilibrium.
The constraint from the $\Omega\Gamma$-limit on the surface velocity of accreting SMSs is illustrated in Fig.~\ref{omgam}.
If at a given stage the surface velocity violates this constraint, accretion stops and the star cannot reach higher masses.
Thus SMSs formed by accretion never reach the mass-shedding limit.

Moreover, in contrast to the monolithic models, accreting SMSs rotate highly differentially, with a frequency in the core 4 -- 5 orders of magnitude higher than that at the surface \citep{haemmerle18b}.
Indeed, accreting models are mostly radiative and contract in a highly non-homologous way (Sect.~\ref{sec-accr}).
Angular momentum transport in radiative regions turns out to be negligible on such short timescales, and non-homology enhances differential rotation in the case of local angular momentum conservation.

For the constraint on surface velocity to be satisfied, the angular momentum of the accreted material must be smaller than about $1\,\%$ of the Keplerian angular momentum \citep{haemmerle18b}.
If we assume accretion to proceed via a Keplerian disk, it implies that 99\% of the angular momentum in the inner disk must be extracted.
The mechanism by which the accretor gets rid of this angular momentum excess is still unknown.
Three of the most common mechanisms invoked in understanding this problem for modern star-formation have also been studied in the context of PopIII SMSs, and are thought to be sufficiently efficient:
\begin{itemize}[label=-]
\item magnetic fields \citep{latif2016a};
\item gravitational torques \citep{Begelman06,WTA08};
\item viscosity in the disk \citep{takahashi2017}.
\end{itemize}
In particular, the efficiency of gravitational torques and viscosity grows as the mass density in the disk increases.
Thus if at any stage the constraint from the $\Omega\Gamma$-limit is violated, accretion stops and mass accumulates in the disk, until angular momentum transport becomes efficient enough for accretion to start again.
In this picture, the upper velocity limit shown in Fig.~\ref{omgam} is not only an upper limit, but the actual velocity expected for SMSs of various masses.
Nevertheless, understanding the specific role of each of these mechanisms in the context of SMS formation requires progress in numerical simulations of the accretion process, including a self-consistent treatment of the inner accretion disk, which currently limits our knowledge on this question.

Due to their low surface velocities, accreting SMSs do not lose mass through rotational mass-loss.
As well, radiative mass-loss is inefficient because of the low opacity on the Hayashi limit, which results in winds that never reach the escape velocity \citep{nakauchi2017}.
The only mechanism allowing for mass-loss is pulsational instability \citep{Inayoshi13, hos13}.
However, the mass-loss rate due to pulsations is not expected to exceed about a few 0.001 M$_\odot$ yr$^{-1}$, i.e., several orders of magnitude smaller than the accretion rate.
Thus mass-loss does not prevent SMS formation by accretion.

\subsection{Benchmarking Results}
\label{benchmark}

Numerical studies of the evolution of rapidly-accreting SMSs (Sect.~\ref{sec-accr}) agree on several conclusions:
\begin{itemize}[label=-]
\item SMSs accreting at the rates predicted for atomically cooled halos ($0.1$--$10\,\mathrm{M}_\odot\,\mathrm{yr}^{-1}$) evolve as red supergiant protostars, along the Hayashi limit. As a consequence, their ionizing feedback has negligible impact on accretion.
\item They collapse at masses of several $10^5\,\mathrm{M}_\odot$, during central H-burning, through the GR instability. The higher the rate, the larger the final mass.
\item The structure of accreting SMSs is mostly radiative, with a convective core due to H-burning.
\end{itemize}

Aside from this general agreement, however, questions remain regarding
\begin{itemize}[label=-]
\item the exact value of the final mass (and especially reducing the uncertainty at high accretion rates);
\item the size of the convective core at given mass (in particular at collapse);
\item the central hydrogen abundance at a given mass.
\end{itemize}

\begin{figure}
\includegraphics[width=1.05\columnwidth]{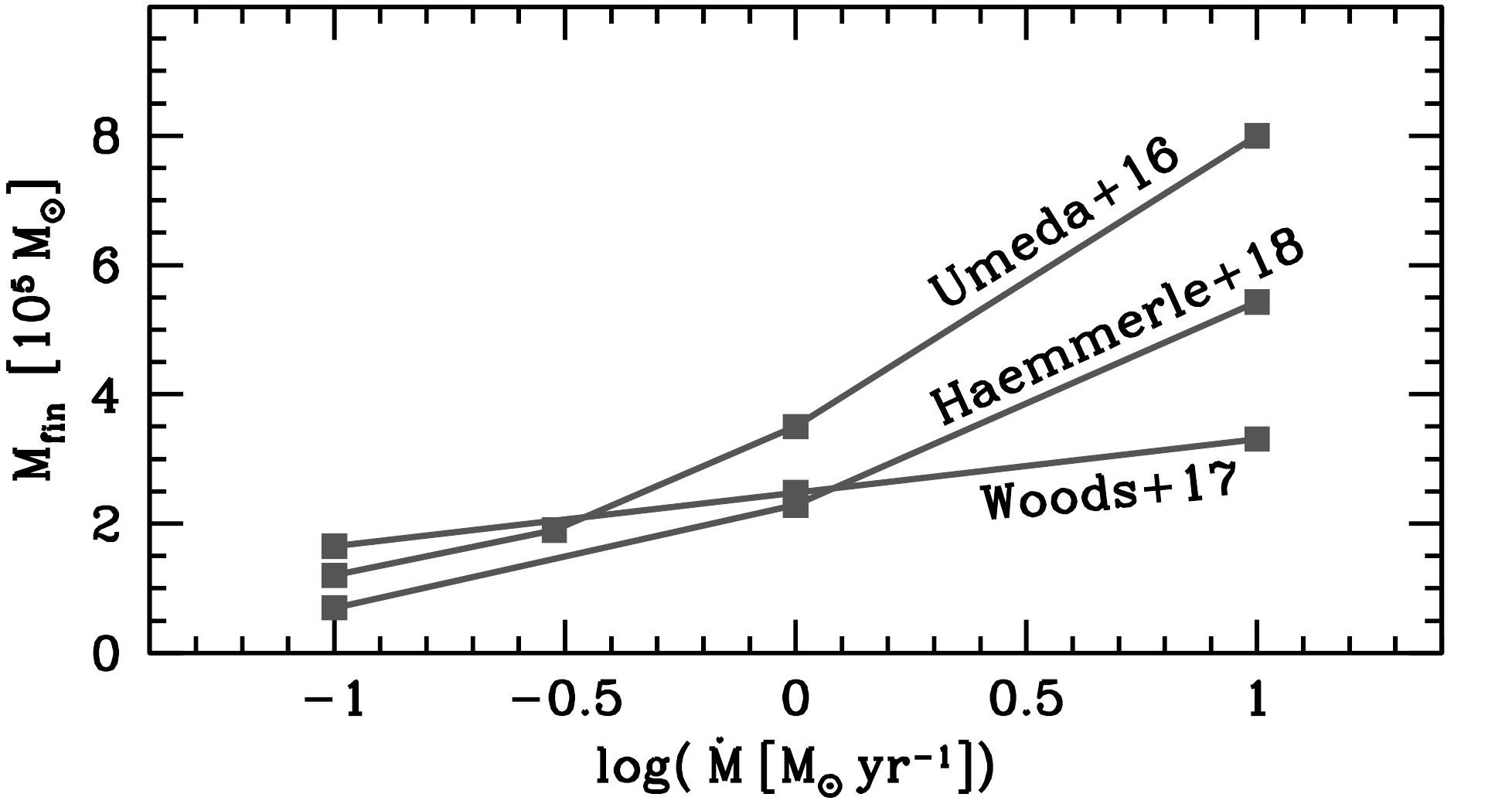}
\caption{Final masses of accreting SMSs as a function of their accretion rate, from \cite{umeda16}, \cite{woods17} and \cite{haemmerle18a}}\label{mfin}
\end{figure}

\noindent Quantitative comparisons are given in Table~\ref{tab-bench} and Fig.~\ref{mfin}.

\begin{table*}\begin{center}\begin{tabular}{ccccccc}\hline
$\dot M=$			&	0.001	&	0.01	&	0.1		&	1		&	10		&	M$_\odot$ yr$^{-1}$\\\hline\hline
collapse:	&	fuel exhaustion	&	fuel exhaustion		&	GR		&	GR		&	GR		&	\\\hline
$M_\mathrm{fin}=$	&	0.1-0.2	&	0.3		&	1-2		&	2-4		&	3-8		&	$\times10^5$ M$_\odot$\\\hline
$M_\mathrm{core}/M_\mathrm{fin}=$
					&			&	50-90\%	&	50-75\%	&	30\%	&	20\%&	(W+17)\\
					&	80\%	&	80\%	&	50\%	&	20\%	&	5\%	&	(H+18)\\\hline
$X_c=$				&	0		&	0		&	0		&	0		&	0.5	&	(U+16)\\
					&   		& 	0		&	0.2		&   0.5		&	0.5	&	(W+17)\\
					&	0		&	0		&	0.6		&	0.7		&	0.75&	(H+18)\\\hline
$T_{\rm eff}=$				&	$\sim10^5$		&	$10^4-10^5$		&	$\sim10^4$		&	$\sim10^4$		&	$\sim10^4$	&	K\\\hline
\end{tabular}\end{center}
\caption{Properties of SMSs at collapse for various accretion rates,
according to \cite{umeda16} (U+16), \cite[][W+17]{woods17} and \cite[][H+18]{haemmerle18a}}\label{tab-bench}\end{table*}

The origin of these divergences is not clear.
The differences in the final masses may arise from the differences in the size of the convective core, since the larger the convective core, the closer the structure is to an $n=3$ polytrope, and thus the earlier the instability arises.
The differences in the size of the convective core could be due to the treatment of convection and atmosphere, the accretion of entropy, or the implemented opacities.
In particular, the various studies used different prescriptions for convection \citep[e.g., use of Schwarzschild vs. Ledoux criterion, inclusion of overshooting, see][]{KippenhahnWeigert,Ledoux} relying on the free parameters intrinsic to the mixing-length theory.
Thus a precise knowledge of the structure of rapidly-accreting SMSs would require one to go beyond mixing-length theory.

The accretion of entropy (i.e., the thermal boundary condition) could also impact the structure of the star.
The simplest prescription is `cold-disc accretion', where the entropy of the accreted material matches continuously that of outermost layer of the star.
This is the lowest limit for accretion of entropy, corresponding to a disk-like accretion geometry, where any entropy excess can be radiated in the polar directions before being advected.
On the other hand, some codes include advection of entropy through various prescriptions \citep{hos13,woods17}.
The advection of entropy, however, is thought to be small compared to the energy budget of the star, due to the high luminosity at the Eddington limit \citep{hos13}.

\subsection{Discussion and Outstanding Problems}

From the preceding discussion, it is clear that the last decade has seen accelerating progress in understanding the structure and fate of massive black hole seeds; however, many questions remain. Aside from the outstanding discrepancies between numerical solutions outlined in the previous section, a problem of particular interest is the maximum mass attainable for DCBH seeds prior to collapse.  Does such a limit exist? From $\S$\ref{sec-accr-nonrot}, it is clear that the answer depends directly on the maximum accretion rate attainable within a DCBH-forming halo. Given the logarithmic dependence of the final mass at collapse on the accretion rate, however, it appears to be difficult to construct DCBH seeds with initial masses greater than $\sim10^{6}\,\mathrm{M}_{\odot}$.  A related, and perhaps more worrying question, is the extent to which a massive seed black hole can continue growing without merger events in an isolated halo \citep[e.g.,][]{TH09} Though still uncertain, progress on this front may come soon from the search for IMBH relics in the low-redshift Universe (see $\S$\ref{sec:imbhs}).

Closely related to the problem of SMS evolution and collapse is the viability and nature of quasi-stars \citep{Begelman10}. These objects form from the collapse to a black hole of a relatively smaller seed mass, surrounded by a massive, giant-like envelope. In this case, the pressure needed in order to support this envelope is provided by the radiation released by accretion at the base of the envelope, in a manner analogous to Thorne-$\dot Z$ytkow objects \citep{TZ75}. \cite{Ball11} \& \cite{Ball12} were the first to produce detailed numerical models for the evolution of quasi-stars, however these early efforts neglected hydrodynamics and nuclear-burning. Including these effects will be vital in any complete model for the evolution of quasi-stars. \cite{FR16} and \cite{FR17} investigated the structure of rotating quasi-stars, and found no stable solutions for truly supermassive objects. At the moment, it is also unclear what accretion history would naturally produce a quasi-star.  These questions should be investigated further.

Another outstanding issue emerging from numerical investigations of SMS evolution is the possibility of ``supermassive supernovae.'' While a relativistic collapse may be reversed for Z$\gtrsim$0.005 \citep{fuller86}, a SMS cannot be formed through the atomically-cooled halo scenario at such high metallicity (though recall the discussion in $\S$2 of other possibilities). Conversely, for primordial metallicities, only an extremely narrow range of masses were found to explode in the monolithic case \citep{chen14}, and further studies will be necessary in order to verify whether this result is robust against small changes to the input physics. If SMSs can indeed undergo thermonuclear explosions, they would provide a remarkable nucleosynthetic signature in the early Universe. Note, however, that the elemental abundances of extremely metal-poor stars in the Milky Way Galaxy \citep{UN03,ITKN14} and damped Lyman $\alpha$ systems (DLAs) at high redshifts \citep{KTN11} suggest that chemical enrichment there was driven more predominantly by core-collapse supernovae from $\sim 10-40 M_\odot$ stars \citep{ITKN18}.
Observational prospects for investigating the deaths of SMSs and their possible mass return are discussed further in $\S$5.

\section{Formation Rates of Black Hole Seeds}

New pathways that can lead to the formation of massive seeds are now emerging. If DCBHs can indeed form in the early Universe, the obvious question becomes how often? And where? Modelling the statistics of early quasar formation necessarily requires a careful understanding of the physics of star formation in primordial halos, with the first stars' contribution to the build-up of Lyman-Werner radiation and metals in the early Universe setting bounds on the formation of atomically-cooled halos (recall \S2). For those black hole seeds which do form (``massive'' or not), an immense supply of infalling matter must also then be available within a very short time frame in order to produce $\sim 10^{9}\,\mathrm{M}_{\odot}$ quasars by z$\sim$6, particularly as even under the most optimistic assumptions the growth of black hole seeds via mergers appears to be insufficient to account for the observations \citep[see e.g.,][]{TH09}.

Increasingly, the need for available gas to feed the growth of the first quasars points to massive seeds as the likely progenitors of the most massive high-redshift quasars \citep[e.g.,][]{Agarwal16, Valiante_2017, Gallerani_2017, Pacucci_2017, Pacucci_2018}. Massive seeds have the advantage of boosting further growth, and being born in more gas rich environments \citep{Latif_2013, Pacucci_2015,Volonteri_2016}. Low-mass seeds may still play a role, however many questions remain. How may gas be brought in from cosmological scales to the accreting black hole? And how can feedback during bursts of accretion and the response of the gas to that feedback be mitigated, allowing the seed to continue to grow? 

Concurrent with the growth in our understanding of their formation, there has been an explosion of new efforts to construct semi-analytic models (SAMs) that can predict the abundance and growth of DCBH seeds \citep[e.g.,][]{Dijkstra08,Agarwal12,Dijkstra14,Agarwal14,Habouzit16hSAM,V16}, with halo properties derived either from Monte Carlo algorithms based on the extended Press-Schechter formalism \citep{1991ApJ...379..440B} or from cosmological N-body simulations. Theoretical models now agree on the expected number density of DCBHs/SMSs, so long as the same recipes (i.e., $J_{\rm{crit}}$ threshold and metal pollution) are implemented. At the same time, however, there has been increasing pressure to move away from simplified recipes, particularly the idea of a single value of $J_{\rm{crit}}$ existing in nature for atomically-cooled halos. Rather, the shape of the incident spectra and the importance of H$^-$ photo-detachment, along with H$_2$ photo-dissociation, must be accounted for in any realistic model (see discussion in $\S$2). The importance of other key criteria, such as avoiding external metal pollution, has also likely been overestimated in the past (c.f. $\S$4.2). Going forward, further progress may require new observational constraints, such as the mass function and abundance of IMBHs (see $\S$~5). There is now a consensus that the number of seeds in a theoretical model should be larger than the number of $z=6$ quasars, as simulations indicate that not all seeds formed will later grow into $10^9\,\mathrm{M}_{\odot}$ BHs. If mergers between seeds are unimportant for their growth, then IMBHs found in the local Universe could trace the evolution of host galaxies that underwent only a limited number of mergers. In the following, we focus on addressing these and other questions regarding the likely abundance of black hole seeds in the Universe, whether massive or not, and summarise a number of recent results.

\subsection{Light and Heavy Seeds}
The very recent possible detection of first star formation occurring $\sim 180 \, \mathrm{Myr}$ after the Big Bang \citep{Bowman_2018}, combined with the detection of massive quasars already formed and shining by $z \sim 7$ \citep{Fan_2006,Mortlock11,Wu15}, indicate that the growth process to form the first super-massive black holes (SMBHs) occurred in $\lesssim 600 \, \mathrm{Myr}$. As discussed above, these time constraints profoundly challenge our view of how black holes grow: how massive must the ``seeds'' of the first SMBHs have been? In this context, the word ``seed'' refers to the original black hole that, growing in mass by accretion and mergers, generates a SMBH of order $\sim 10^{9-10} \, \mathrm{M}_{\odot}$. In the following, we expand upon the previous discussion of the ``seeding'' problem in order to place the ``heavy'' seed scenario in context, by contrasting it with the viability of ``light'' seeds.

Assuming that accretion is the main driver of growth, the standard scenario predicts that the black hole mass $M_{\bullet}$ increases exponentially, with an e-folding time scale $t_\mathrm{S} \sim 0.045\,\epsilon_{0.1} \, \mathrm{Gyr}$, named Salpeter time (see also Eq. \ref{eq:t_growth}) . Here, $\epsilon_{0.1}$ is the matter-radiation conversion factor normalized to the standard value of $10\%$, valid if the growth occurs at the Eddington rate. Another implicit assumption of this treatment is that there is a constant gas inflow to feed the growth. If this is not the case, a duty cycle ${\cal D}$ can be used to express the fraction of cosmic time during which the black hole is actually accreting. The relevant equation to describe the time evolution of the mass of the seed is thus:
\begin{equation}
M_{\bullet}(t) = M_{\bullet ,0} \exp\left({\frac{t}{t_\mathrm{S}}}\right) \, ,
\label{eq:growth}
\end{equation}
where $M_{\bullet ,0}$ is the mass of the black hole seed at $z \sim 20$, corresponding to a cosmic age of $\sim 180 \, \mathrm{Myr}$.
A common terminology in the field differentiates ``light seeds'' ($\lesssim 10^{2} \, \mathrm{M}_{\odot}$) from ``heavy seeds'' ($\gtrsim 10^{4} \, \mathrm{M}_\odot$) (e.g., see the recent review \citealt{Valiante_2017}). Light seeds are formed at the endpoint of the evolution of Pop III stars, while heavy seeds are formed by a number of different processes active in the early Universe and described below.

It is very challenging to envision that the z $\sim 7$ SMBHs were formed from light seeds, if their accretion rates are indeed restricted by the Eddington limit. 
In fact, this growth process would require constant accretion at the Eddington rate until $z \sim 7$ to allow the formation of $\sim 10^9 \, \mathrm{M}_\odot$ quasars. This process, while not strictly impossible, is very unlikely, due to the stringent requirement of a steady gas reservoir with low angular momentum.
In order to solve the problem, two pathways can be devised directly from the evolution equation (Eq.~\ref{eq:growth}): either decreasing the time-scale $t_S$ or increasing the initial mass $M_{\bullet ,0}$.

To decrease the e-folding time scale, while keeping $M_{\bullet ,0}$ in the light seed regime, it is necessary to reach super-Eddington accretion rates \citep{Volonteri_2005}, often indicated with $\lambda_{\rm Edd} \equiv \dot{M}/\dot{M}_{\rm Edd} > 1$. Note that, at least in a spherically symmetric accretion scenario, this is equivalent to assuming that the radiative efficiency $\epsilon$ is lower than the standard value adopted for Eddington-limited accretion.
Several works (\citealt{Alexander_2014, Madau_2014, Volonteri_2014}) have predicted the occurrence of largely super-Eddington accretion episodes at high redshift.

The second possibility assumes that the environmental conditions (i.e., pristine or very low-metallicity gas) of the early Universe ($z \gtrsim 10-15$) allowed alternative pathways to form massive black hole seeds: (i) the direct collapse of un-enriched and self-gravitating pre-galactic disks (\citealt{Begelman06, Lodato_Natarajan_2006,Lodato_Natarajan_2007}), (ii) the collapse of a primordial atomic-cooling halo into a direct-collapse black hole (DCBH, \citealt{BL03,SBH10, Johnson_2012}), or (iii) the formation of a very massive star from runaway stellar mergers in a dense cluster (\citealt{Omukai:2008p113, Devecchi_2009, Davies_2011}). 
Alternatives to these two scenarios include the possibility that at least a fraction of $z>6$ quasars might be lensed (e.g., \citealt{Wyithe_Loeb_2002, Fan_2019, Pacucci_2019}). A lensing effect would in fact decrease the mass of the SMBH powering the quasar.

Currently, the general consensus is that massive seeds are the most likely progenitors of early SMBHs (e.g., \citealt{Agarwal16, Valiante_2017, Gallerani_2017, Pacucci_2017, Pacucci_2017_limits, Pacucci_2018}), due to a combination of two factors.
First, as can be clearly seen in Eq. \ref{eq:growth}, larger initial masses $M_{\bullet ,0}$ boost the growth early on \citep{Pacucci_2015, Volonteri_2016}; since the process follows an exponential growth law (assuming that the black hole grows constantly), this results in a very relevant boost.
Second, the environment where heavy seeds formed tend to be characterized by higher gas densities (e.g., \citealt{Latif_2013, Pacucci_2015}), at the cusp of the density distribution of primordial halos; this, in turn, increases the chances of feeding the black hole seed with large gas inflows early on.
Several works (e.g., \citealt{Pelupessy_2007, Alvarez_2009}) have shown that light seeds, instead, are formed in low-density regions, mainly due to the effect of stellar feedback that sweeps the surrounding gas away, causing a delay before the central density is rebuilt.

Thus, the combination of large seed mass and high density leads to a very efficient and rapid growth of heavy seeds.
On the contrary, light seeds will struggle to grow, at least at very early times, given the low seed mass and the low-density environments in which they are born.

From a purely analytical point of view, it is possible to show that even light seeds can grow to the SMBH regime early on in the history of the Universe \citep{Volonteri_2005}.
The main concern for this scenario, however, is the availability of a steady inflow of gas at super-Eddington rates. 
Mass supply rates available in high-$z$ galaxies certainly allowed for episodes of super-Eddington accretion. Assuming a velocity dispersion $\sigma$ ($\sigma_{100}$ is expressed in units of $100 \, \mathrm{km \, s^{-1}}$) for the host, the characteristic free-fall rate of self-gravitating gas exceeds the Eddington rate by a factor of \citep{Begelman_2017, Pacucci_2017}:
\begin{equation}
\dot{M}_\mathrm{ff} \sim 10^5 \sigma_{100}^3 \dot{M}_{\rm Edd} \left( \frac{M_{\bullet	}}{10^6 \, \mathrm{M_{\odot}}}\right)^{-1} \, .
\end{equation}
Indeed, simulations (e.g., \citealt{Dubois_2014}) and analytical estimates (e.g., \citealt{WL_2012}) suggest that super-Eddington infall rates were quite common in the early Universe. \cite{Begelman_2017}, for example, show that the fraction of AGN accreting at super-Eddington rates could be as high as $\sim 10^{-3}$ at $z=1$ and $\sim 10^{-2}$ at $z=2$. Directly comparing these considerations with the growth of massive seeds, however, requires more detailed calculations of their rates of formation and growth, which we explore in the following subsection.

\subsection{Narrowing in on the Number of DCBHs}

Recent efforts to determine the abundance of DCBHs in the $z>6$ Universe have found number densities spanning several orders of magnitude, from $\sim 10^{-1}$ to $10^{-9}$ cMpc$^{-3}$. 
This is because these estimates have employed a large variety of models using various methods and simulation set-ups, which we outline below \citep[see also][]{2017PASA...34...31V}. 
It is necessary for models  to include at least three main ingredients which lead to the physical conditions needed for DCBH formation: self-consistent star formation in a halo right from when it first appeared as a minihalo, treating metal-pollution in said halo, and local build-up of the LW radiation field emanating from the first galaxies. 

As discussed earlier, exposure to strong LW radiation is one of the possible ways to achieve the low H$_2$ fraction needed for DCBH formation \citep{Omukai:2001p128,Ciardi:2000p82,Haiman:2000p87}. \citet{Ahn:2009p77} demonstrated that the spatial variation of LW radiation intensity was key for this scenario. Most of the models exploring DCBH abundances now include a spatially varying LW radiation flux from local irradiating sources. This radiation intensity is either computed directly from stellar particles depending on their age, distance, and redshift \citep{Agarwal14, Habouzit16hydro}, or from the stellar mass painted on DM halos \citep{Agarwal12, Dijkstra14,Habouzit16hSAM,Chon16}. Overall, studies agree on the fact that the maximum local specific intensity from Pop~III stars appears to be almost always below the critical intensity, whereas a majority of pristine halos exposed to radiation from Pop~II stars see a level of $J_{\rm{crit}}$ above the critical radiation threshold. The spatial distribution of radiation intensity is in good agreement between various studies, using either position based LW radiation modeling \citep{Agarwal12,Chon16}, or probabilistic determination of the radiation field \citep{Dijkstra08}. 

Another important requirement for the DCBH scenario is that the gas should be free of metals. This requirement can decrease the number of DCBH candidates by orders of magnitude (see, e.g., Fig.~5 in Ferrara et al. 2014). There is a general consensus that both prior episodes of star formation in halo progenitors, as well as galactic winds from nearby halos, may play a fundamental role in governing the fate of the collapsing gas. In principle, stellar winds from neighbouring galaxies could prevent DCBH formation on scales of $\leqslant 10\,\rm{kpc}$, thereby reducing their abundance \citep{Dijkstra14}. However, typical separation between direct collapse candidate halos and LW sources is of the same order \citep[$\sim10\, \rm{kpc}$, ][]{Agarwal12,Habouzit16hSAM}. Hydrodynamical studies of metal-mixing in halos due to external SNe winds have found, however, that the extent of the impact is often over-estimated in large simulations. In practice, metal-pollution in primordial gas is complicated by a number of considerations, including the rate at which winds expel metals from their original galaxies, the propagation of metals through the IGM, and subsequent mixing in gas of varying density \citep{Cen:2008p841,2015MNRAS.452.2822S,Maio2018}. Recently, \cite{Agarwal_metals} devised a semi-analytical model which explores the worst case scenario for DCBH formation, given metal-pollution driven by the same galaxies which provide the necessary LW flux. They report that even at an extremely short distance, $\sim300 \rm{pc}$, and with instantaneous metal mixing, the irradiating galaxy is not able to sufficiently pollute the DCBH candidate halo \citep{Omukai:2008p113,LatifDust} to prohibit DCBH formation.

In general, there are two approaches employed by the community in order to follow the evolution of any proto-galaxy, using analytic prescriptions depending on their dark matter halo:

\begin{itemize}[label=-]
\item pure semi-analytic models (pSAMs): which employ analytic algorithms (e.g., Monte Carlo) based on the extended Press-Schechter formalism \citep[EPS][]{Press:1974p19, Lacey:1993p59}. This approach allows one to statistically capture the abundance and mass assembly history of halos in the Universe, however, the trade-off is a lack of halo properties such as positions, spin etc. and thus, such an approach relies on a probabilistic determination and implementation of physical effects.
\item hybrid semi-analytic models (hSAMs): that employ cosmological N-body simulations \citep[e.g.][]{Springel:2005p667} to extract DM halo properties (e.g., mass and spatial distribution), and superimpose their analytic models. While such an approach provides physical properties such as positions, spins, virial radius (etc.) of halos at any given redshift, resolution limitations do not allow one to create a robust statistical sample of $z\sim 6$ quasar hosts while resolving the smallest Pop III host halos at the same time.
\end{itemize}

\noindent Differences between pSAMs employing PS merger trees and hSAMs using N-body dark matter only simulations (or hydrodynamical simulations), arise mainly due to the physical processes that can be included in the models. pSAMs  assign probabilities to physical properties, such as the probability for dark matter halos to have a given stellar mass,  of being star-forming at a given time or being metal-free. This is because pSAMs are designed to present a statistical view of the first galaxies and can not capture the inter-dependence of halo assembly histories. On the other hand, hSAMs are able to follow the individual histories of halos and galaxies, and capture the underlying interdependence.  Therefore whether a halo is star-forming is naturally captured, and the local radiation intensity or metal pollution can be consistently computed from the position based distribution of halos produced by the simulations.

\citet{Dijkstra08,Dijkstra14} and \cite{V16} are studies that use their own pSAMS. \citet{Dijkstra08} compute the probability distribution function of DM halos that are exposed to the expected LW flux on the basis of their clustering properties. They found that only a small fraction, $<10^{-6}$, of all atomic cooling halos are exposed to a sufficient LW flux ($J_{\rm LW}>10^3$), producing a number density of $<10^{-6}$ cMpc$^{-3}$ potential DCBHs hosts. In a later study, \citet{Dijkstra14} estimate the abundance to be $n_{\rm DCBH}\sim 10^{-9}-10^{-6}$ cMpc$^{-3}$ between $z=20$ and 7, and explore the effects of model assumptions on their results, including the LW photon escape fraction and $J_{\rm{crit}}$, underlying the role of galactic winds in decreasing the formation rate of DCBH sites. In their pSAM designed to explore the formation of a $z\sim 6$ SMBH J1148, \citet{V16} predict a mean number density of $\sim 10^{-7}$ cMpc$^{-3}$ DCBHs at $z>6$. Although several DCBH seeds form in the progenitors of the $10^{13}\,\mathrm{M}_\odot$ DM host halo of J1148, only a fraction of these seeds eventually end up contributing to the mass budget of the final SMBH.

\cite{Agarwal12}, \cite{Agarwal14}, and \cite{Habouzit16hSAM} are some examples of hSAMs. \citet{Agarwal12} report a higher number density in the range $10^{-2}-10^{-1}$ cMpc$^{-3}$ for $\rm{J_{\rm{crit}}}=30$. In their recent work \citet{Habouzit16hSAM} found that the number density of DCBH sites can lie anywhere in the range $10^{-7} - 10^{-2}$ cMpc$^{-3}$ depending on the value of $J_{\rm{crit}}$ imposed. If a higher critical flux is imposed for DCBH formation ($J_{\rm{crit}} > 100$), then \citet{Dijkstra14,Habouzit16hydro}  find a lower DCBH abundance that is just sufficient to reproduce the population of quasars. In other words, the abundance of DCBH sites is inversely proportional to the value of $J_{\rm{crit}}$ imposed in the models. The hSAMs self-consistently track each halo's mass assembly history (and thus the star formation) and compute the build up of the global and local LW radiation flux on the basis of the positions of halos at any given redshift. {Studies done using smaller simulation boxes that resolve minihalos and allow for a more developed chemical network, have led to higher DCBH site abundances. This can be particularly attributed to the lower $J_{\rm{crit}}=30$ \citep{Agarwal12,Agarwal14}} which in turn suggests that the DCBH scenario may be able to seed the SMBHs of more {{\textit{normal}}} galaxies. A recent study by \citet{Habouzit16hydro} argues that this strongly depends on the implementation of SN feedback; weak SN feedback may better explain BHs in normal galaxies, however simulations with strong SN feedback produce galaxies and BHs in better agreement
with observed statistics. 

Alternative direct collapse scenarios, such as the galaxy merger-driven
model, have also been studied with the aid of hSAMs applied to the Millenium simulation (Bonoli, Mayer \& Callegari 2014; MB18). These models, that apply several constraints on the properties of host galaxies of direct collapse seeds, still end up with a predicted number density of bright QSOs at $z \sim 6$ that is a factor of few higher than the observed one (MB18).

\citet{Habouzit16hydro} perform a comparison between the pSAM study of \citet{Dijkstra14} and a variety of hSAMs \citep{Agarwal12,Agarwal14,Habouzit16hSAM}, and find that \citet{Dijkstra14} overestimates the stellar mass in halos. \citet{Dijkstra14}, however, underestimate both the number of galaxies contributing to the LW flux, and the size of metal- polluted regions. Interestingly, the effect of these varying assumptions can sometimes be to compensate for each other, providing similar results for the number density of viable DCBH sites \citep{Habouzit16hydro}. The studies discussed in this section provide upper limits on the abundance of DCBHs, with a consensus emerging that the number density of DCBHs is greater than the observed number density of quasars at $z\sim 10 \ \rm{cGpc^{-3}}$ \citep[e.g.][]{2016ApJS..227...11B}. However, in their current formulation, they may yet be unable to explain the presence of central black holes in galaxies like the Milky Way and dwarf galaxies, which have much higher number densities, $\sim$0.1 $\rm{cMpc}^{-3}$.

\section{Observations Past and Future}

A significant motivation for the many theoretical developments outlined in the preceding sections has been the remarkable discovery of $\sim 10^9\,\mathrm{M}_{\odot}$ quasars in the early Universe, more of which continue to be found \citep[e.g.,][]{Mortlock11,Wu15,Matsuoka18,Banados18}. The challenge now is to find unique, unambiguous signatures for the formation of these objects which may be realistically sought after immediately or in the near future. In particular, the need now is to distinguish between the formation of massive black holes from low and high mass seeds, either by finding distinct signatures of the formation history of particular observed quasars at z$\sim$7, directly tracing the evolution of related progenitors at even higher redshift with next generation telescopes, or searching for evidence of either formation scenario in the local Universe.

The Luminosity Function (LF) of quasars at high redshift can provide an essential probe of the growth of SMBH \citep[e.g.,][]{TH09}. Ongoing efforts such as the VISTA surveys \citep{Sutherland09}, as well as future observations with the Large Synoptic Survey Telescope \citep{ivezic2008lsst} and Euclid \citep{Euclid} will probe the evolution of the quasar LF over cosmic timescales, eventually reaching out to z$\sim$7--8 \citep{Manti17, abell2009lsst}. Constraining the number density and accretion efficiencies of very high-z quasars in this way may soon make it possible to assess the viability of light and heavy seed models.

New diagnostics have also been proposed. For the massive seed scenario, X-ray ionization by very early massive black holes may leave a distinct imprint on the 21cm background at z$\gtrsim$20 \citep{Tanaka16}, and the sites of DCBH formation may reveal themselves by a unique 3cm (restframe) emission line \citep{dijkstra16}. The first direct, unambiguous evidence of any Pop III stars may come from observations of their explosive deaths as supernovae at high-redshift \citep{hummel2012,magg16,hartwig18c}. Owing to the high effective temperatures expected for Pop III stars, it has long been argued that strong He II recombination line emission can provide a clear signature for their presence within a stellar population \citep[e.g.,][]{Schaerer03,Visbal15,Cai15}. He II emission alone, however, is not a reliable diagnostic. Without additional line measurements, a He II nebula's ionizing source cannot be unambiguously distinguished from e.g., an AGN \citep{Bowler17}, and conversely, even if a stellar population produces a strong He II-ionizing flux, this may not produce a correspondingly high He II line luminosity if the ionization parameter is sufficiently low \citep[see e.g., discussion in][appendix 2]{Woods13}. 

The difficulty in interpreting high-z emission line spectra is clearly demonstrated by the case of CR7 \citep{Sobral15}. Initially thought to be the first definitely Pop III-like galaxy ever observed \citep[though for earlier claims see e.g.,][]{2002ApJ...565L..71M, 2008IAUS..255...75D}, this bright z = 6.6 Lyman-$\alpha$ emitter has provided a critical test of our ability to discriminate between theoretical interpretations of high-z galaxy observations. In the following, we begin with a focused overview of the rapid developments and lessons learned in studying CR7, before discussing alternative means to either directly or indirectly probe the formation and growth of black holes in the early Universe. We then discuss how, in the near future, the search for intermediate-mass black holes may yield new insight into the ``initial mass function'' of black hole seeds, while the {\it JWST} and other upcoming telescopes may reveal how they grow with time, or even reveal SMSs themselves (or their explosions!), while future gravitational wave observatories may reveal their growth via mergers.

\subsection{What have we learned from CR7?}
\label{subsec:learned_CR7}
The CR7 system was first identified as a Lyman-$\alpha$ emitter by \citet{matthee15} and spectroscopically confirmed by \citet{Sobral15}. It is the brightest Lyman-$\alpha$ emitter at $z=6.6$, with $L_{\mathrm{Ly}\alpha} \simeq 10^{44}\,\mathrm{erg}\,\mathrm{s}^{-1}$ and $L_\mathrm{HeII}=10^{43.3 \pm 0.4}\,\mathrm{erg}\,\mathrm{s}^{-1}$ \citep{Sobral15}. More remarkable than the sheer brightness of this source was the absence of any metal line detection and the remarkable equivalent width (EW) of He{\sc ii} (EW$_{\mathrm{HeII}} = 80 \pm 20$ \AA, \citealt{Sobral15}), indicating a hard ionizing spectrum, and suggesting the presence of Pop III stars \citep{2000ApJ...528L..65T, 2001ApJ...553...73O}. \citet{Sobral15} indeed found that the SED of the complete system could only be reproduced when including a young ($<5\,\mathrm{Myr}$) population of massive metal-free stars, with a total stellar mass of at least $10^7\,\mathrm{M}_\odot$ for one of the three clumps. This was one of the most compelling observations of a Pop~III-like galaxy and has therefore triggered many theoretical interpretations and further observations.

By re-analysing the VLT/X-Shooter data of \cite{Sobral15}, \citet{shibuya18} found only upper limits on the He{\sc ii} luminosity with EW$_\mathrm{HeII} < 60$ \AA\, and no other detected emission lines in the UV. In a more recent re-analysis of the X-Shooter data, \citet{sobral17} confirmed the detection of He{\sc ii} at lower significance than before ($\sim 3\sigma$) with $L_\mathrm{HeII} \simeq 10^{43.4 \pm 0.5}\,\mathrm{erg}\,\mathrm{s}^{-1}$. Using HST/WFC3 grism observations and a re-analysis of the VLT data, \citet{sobral17} find a EW$_{\mathrm{He}{\sc II}} < 10$ \AA, and a gas metallicity of $0.05 < Z/Z_\odot < 0.2$. They also report that there is no convincing evidence for strong variability ($>0.3$\,mag) of any of the components of CR7, which would be a possible smoking gun for an accreting BH.

\citet{Bowler17} use Spitzer/Infrared Array Camera observations and interpreted the blue IRAC color as a [O{\sc iii}] and H$\beta$ detection \citep[see also][]{bowler17b,pacucci16}. They advocate that CR7 could be a low-mass, narrow-line AGN or a young, low-metallicity starburst with a metallicity of $Z \sim 1/200 Z_{\rm{\odot}}$. As discussed in more details below, \citet{Pacucci17} showed that this updated photometry can still be explained by a DCBH model. In fact, while a zero-metalliticy environment is required at formation, subsequent star formation can certainly provide low levels of metals even in the early phases of DCBH accretion.

Based on recent ALMA observations, \citet{matthee17} reveal a clear $\mathrm{[C{\sc II}]}$ signature within multiple components of CR7. They do not find any FIR continuum emission due to dust, which may also disfavour an AGN. They argue that SFRs of $5-28\,\mathrm{M}_\odot\,\mathrm{yr}^{-1}$ in the individual clumps, together with a metallicity in the range $0.1 < Z/Z_\odot < 0.2$, can account for the observations.

In addition to the original interpretation of CR7 as a Pop III-like galaxy, \citet{Sobral15} also proposed the possibility of a DCBH. This scenario is favourable over other AGN formation channels because of its natural absence of metals during the formation process. Soon after the initial discovery, theoretical models employing hydrodynamical simulations \citep{Pallottini15} and a semi-analytical model \citep{Agarwal16_CR7} indicated that CR7 could indeed host a DCBH. In particular, clumps B and C could provide the required LW flux for clump A to have formed a DCBH, which would have the required observational features about 100\,Myr after its formation.

Separately, \citet{Hartwig16} showed that CR7 can not host the necessary $10^7\,\mathrm{M}_\odot$ of Pop~III stars. Chemical feedback from previous SNe prevents the accumulation of sufficient metal-free gas. An alternative scenario by \citet{visbal16} demonstrates how photoionisation feedback could delay the collapse of a pristine halo so that it can accumulate enough gas for a massive Pop~III burst, in agreement with the original observations of CR7.

The velocity offset between Lyman-$\alpha$ and He{\sc ii}, and the extent of the Lyman-$\alpha$ emitting region imply a source lifetime of at least 10\,Myr, in contradiction with a very young population of massive stars \citep{smith16}. \citet{smidt16} demonstrated, using Enzo \citep{Enzo_2014} and radiative transfer in post-processing, that an AGN powered by a black hole of mass $3 \times10^7\,\mathrm{M}_\odot$ accreting at $f_\mathrm{Edd}=0.25-0.9$ can account for the observations.
Additional radiative processes have been suggested to prevent pristine gas from collapsing for long enough to accumulate the required high masses, but it is challenging to reproduce the required $10^7\,\mathrm{M}_\odot$ of Pop~III stars \citep{xu16,visbal16,visbal17,yajima17}.

In the light of metal line detections in CR7, \citet{Agarwal17} present an improved analytical model. If a galaxy is externally polluted with metals after the formation of a central DCBH, such a system is within $3\sigma$ of the new observations. Also \citet{Pacucci17} present a new radiation-hydrodynamic model that can reproduce the observations by \citet{Bowler17} with a DCBH of mass $M_\mathrm{BH} = 7 \times 10^6\,\mathrm{M}_\odot$, a gas metallicity of $Z=5\times 10^{-3}$, and a line-of-sight column density of neutral hydrogen of $3 \times 10^{24}\,\mathrm{cm}^{-2}$.

Although CR7 has turned out to be less outstanding than initially thought, it has triggered many important discussions with results that will remain valid, regardless of the nature of CR7. Various mechanisms can delay the collapse of pristine gas and only strong photoionization feedback can result in a Pop~III stellar population more massive than $10^6\,\mathrm{M}_\odot$ \citep{visbal16}. Thanks in part to the intense scrutiny given to CR7, we now have a much greater understanding of the conditions under which a DCBH can form, such as the window of opportunity to provide enough LW flux without polluting the halos \citep{Hartwig16,Agarwal16_CR7,Agarwal17}, and what observational signatures can distinguish a DCBH from other energetic sources at high redshift, such as the 3\,cm fine structure line \citep{dijkstra16b}. A number of alternative diagnostics have now been proposed which may confirm or refute the DCBH scenario, both for individual high-z objects and as a fraction of all SMBHs formed, which we address in the following sections.

\subsection{What do IMBHs tell us about the seeds of the first SMBHs?}
\label{sec:imbhs}
Intermediate-mass BHs (IMBHs) of $10^{2}-10^{5}$ M$_{\odot}$ are thought to be
the local relics of those high-z seed BHs that did not become supermassive. 
They should be found in nearby dwarf galaxies with low mass, low metallicity, 
and which have not significantly grown through accretion and mergers, hence
resembling the first galaxies formed in the early Universe \citep{Volonteri10}.
As a direct trace of high-z black hole formation, then, which kind of IMBH masses should we expect for different SMBH seed 
scenarios, and how frequent are IMBHs in dwarf galaxies?

\begin{figure}[ht]
\includegraphics[width=1.1\columnwidth]{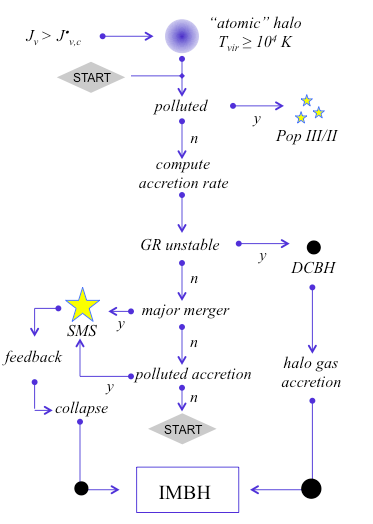}
\caption{From \cite{ferrara14}. Schematic view of the scenario for 
intermediate mass black hole formation and growth accounted for 
in the cosmological model by \cite{ferrara14}. The IMF of 
IMBHs derived by the authors model all of these different physical 
processes, thus taking into account both IMBH formation 
via truly direct collapse and via a SMS phase.}\label{formation_flowchart}
\end{figure}

\cite{ferrara14} investigated the initial mass function (IMF) of direct 
collapse BHs along with the mass distribution of their hosting halos. 
Fig.~\ref{formation_flowchart} illustrates the possible paths for direct collapse  formation and 
growth, which have been accounted in their model. To study the final outcome 
and frequency of such distinct paths the authors used a statistical merger tree 
approach, which includes metal-pollution from both the inhomogeneously 
enriched IGM and the star-forming progenitor halos \citep{salvadori14}. 
Their results show that the IMF of direct collapse black holes extends over a large range of masses, $M_{BH}\approx (0.2-20) \times 10^5\,\mathrm{M}_{\odot}$, and it is 
bi-modal. Further, they find that the hosting halos of direct collapse black holes have 
low masses, $M_\mathrm{h} \approx 10^{7.5}-10^8\,\mathrm{M}_\odot$ at $z\approx (7-15)$. In 
the local Universe, therefore, un-evolved metal-poor dwarf galaxies that have 
masses $M_\mathrm{h} < 10^9\,\,\mathrm{M}_{\odot}$ are thus the key objects to search for the living 
relics of such IMBHs. Yet, given the strict conditions required for IMBHs to form 
from direct collapse (see Fig.~\ref{formation_flowchart} and Sect.~\ref{supermassivestars}), a low black hole 
occupation fraction is expected in today's dwarf galaxies if direct collapse was the dominant seeding mechanism.

Alternatively, SMBH seeds can form as remnants of massive Pop III stars. 
The observed properties of ancient dwarf galaxies in the Local Group are all
consistent with the idea that these small systems hosted Pop III stars and that 
were likely among the first star-forming galaxies \citep{salvadori09,salvadori15}. These theoretical expectations has been reinforced over 
the years by chemical abundance studies of ancient stars in the newly discovered 
ultra-faint dwarf galaxies, which revealed the chemical imprint of Pop III stars with 
masses $M_*\approx (10-60)\,\mathrm{M}_{\odot}$ \citep[e.g., more recently][]{spite18}. 
Cosmological simulations for the formation of the first stars show that Pop III stars 
can also be very massive, $M_*\approx 10$--$1\,000\,\mathrm{M}_{\odot}$ \citep{hosokawa11,hirano14,hirano15}, a prediction that is consistent with the Pop III IMF 
constraints inferred with Milky Way halo stars \citep{debennassuti17}. 
In conclusion, black hole seeds from Pop III stars likely have masses $M_{BH}\approx10$--$1\,000\,\mathrm{M}_{\odot}$ and their living relics may be found in present-day dwarf galaxies \citep{salvadori09,salvadori15,hartwig15,magg18,magg19}. 
Since Pop III stars are more common than direct collapse BHs, these lighter seeds 
should yield a higher black hole occupation fraction in present-day dwarf galaxies \citep{Volonteri08,vanWassenhove10}.

Several studies have therefore focused on finding IMBHs actively accreting as AGN in dwarf galaxies. A few hundred of such low-mass AGN, with black hole masses $\lesssim10^{6}$ M$_{\odot}$, have been found using either emission line diagnostics or the width of broad emission lines from the gas, assumed to be virialized, around them  \citep[e.g.,][]{2004ApJ...610..722G,2007ApJ...667..131G,Dong2012b,Moran2014,Reines2013,Sartori2015,Chilingarian18, Nguyen_2019}. The AGN nature of some of these sources has been reinforced by the detection of hard X-ray emission \citep{Dong2012a,Baldassare2015,Baldassare2017}. The use of deep X-ray surveys such as \textit{Chandra} COSMOS Legacy \citep{Civano2016} has provided a few more candidates \citep[e.g.,][]{Schramm2013,Pardo2016,Mezcua2018} as well as observational evidence that a population of IMBHs must exist in dwarf galaxies, at least out to z$\sim$1.5 \citep{Mezcua2016}. Their detection is challenging, however, due to their low luminosity and possibly mild obscuration \citep{Mezcua2016}. The redshift record-holder of a low-mass AGN in a dwarf galaxy is cid\_1192, at z=2.39 \citep{Mezcua2018}, which reinforces the scenario in which dwarf galaxies host the relic seed BHs of the early Universe. Such deep X-ray surveys have additionally allowed us to derive the AGN fraction in dwarf galaxies, taken as a lower limit to the local black hole occupation fraction \citep{Pardo2016,Mezcua2018}.

Even when correcting for completeness, the low AGN fraction found \citep[i.e., 0.4\% for $z\leq$0.3;][]{Mezcua2018} and its decrease with stellar mass \citep{Aird2018,Mezcua2018} seem to favor the direct collapse formation scenario for seed BHs \citep{Volonteri08,vanWassenhove10,Bellovary2018}. This scenario is also favoured by the finding that the black hole mass scaling relations flatten in the low-mass regime \citep{Greene2008,Jiang2011,2013ApJ...764..151G,2015ApJ...798...54G,Mezcua17,2018ApJ...855L..20M,Shankar_2019}, as expected from simulations \citep{Volonteri08,vanWassenhove10}. Whether this flattening is an observational bias \citep{Mezcua17}, the result of dwarf galaxies being dominated by supernova feedback \citep{2018ApJ...855L..20M}, or the existence of a bimodality in the accretion efficiency of BHs \citep{Pacucci_2018} remains unclear. 

In addition to dwarf galaxies, some IMBH candidates have also been found in ultraluminous X-ray sources \citep{Farrell2009,Mezcua2011,Webb12,Mezcua2013,Mezcua2015,2018MNRAS.480L..74M}, possibly as the remnant core of a dwarf galaxy that was stripped in the course of a minor merger. We note that mergers of dwarf galaxies are found to be very frequent and could lead to BH coalescence and rapid BH growth of the primordial seed \citep[e.g.,][]{2014ApJ...794..115D, 2019NatAs...3....6M}. The effects of AGN feedback could also significantly impact BH growth, by either shutting down or triggering star formation and thus the BH fodder \citep[e.g.,][]{2018MNRAS.473.5698D, 2018MNRAS.480L..74M, 2018MNRAS.476..979P, 2018arXiv181104953R, 2019NatAs...3....6M}. If IMBHs in dwarf galaxies have significantly grown through mergers and feedback, then they should not be considered as the fossils of the seed BHs of the early Universe \citep{2019NatAs...3....6M}. This has important implications for seed BH formation studies that are based on local IMBHs. Evidence for IMBHs could also be found from tidal disruption events \citep{LinNatAs17}, high-velocity clouds \citep{Oka16,Oka17} and in globular clusters \citep{Gebhardt2005,Luetzgendorf13,Kiziltan17}, though in these cases they could have formed locally hence their relation with the high-$z$ seed BHs is dubious \citep[see][for a review]{Mezcua17}. Finally, we note that virtually all mass measurements \citep[e.g.,][]{2004ApJ...610..722G,2007ApJ...667..131G,Dong2012b,Moran2014,Reines2013,Chilingarian18} rely on virial mass estimates based on empirical line/continuum relations for AGN in the optical part of the spectrum, under the assumption that the emission lines of interest (e.g., H${\alpha}$ or H${\beta}$ lines) originate in the vicinity of an active nucleus (i.e., broad or narrow line AGN). Validating and improving these calibrations is an area of ongoing study \citep{Shankar_2019}, which we discuss in the following subsection.

\subsection{Searching for IMBHs in low luminosity AGN in the multimessenger era}

The discovery of hundreds of SMBHs in the $\lesssim10^{6}$\,M$_{\odot}$ range in the centers of nearby low-mass AGN (see previous section) has set low-luminosity AGN (LLAGN) as prime targets in the search for IMBHs. With the vast majority of candidate IMBHs in nearby low-mass AGN lying in the $10^{5}-10^{6}$\,M$_{\odot}$ range (see recent review by \citealt{Mezcua17}), these measurements tentatively suggest a prevalence of the direct collapse scenario, indicated by the presence of primarily heavy seeds. However, it is very likely that these findings reflect the limitations of our observational capabilities, as well as a bias towards more luminous and therefore more massive candidates in the LLAGN sample. In addition to the selection bias of more luminous host galaxies targeted for spectroscopy in the SDSS sample (see e.g.~discussions by \citealt{2004ApJ...610..722G} and \citealt{2008AJ....136.1179B}), the spectral and spatial resolution of telescopes (like the 5\,m telescope used in the Palomar survey) lies right at the threshold of detecting the signatures of a slowly accreting black hole with a mass on the order of a few $10^{5}$\,M$_{\odot}$. This is in terms of both the size of the broad line region and the broadening of emission lines (see also \citealt{2017mbhe.confE..51K} and references therein).

To investigate the fainter end of LLAGN and dwarf galaxies, one can draw from the correlations between large-scale properties of galaxies and the mass of their central SMBH. One of the most well known scaling relations is that between black hole mass and stellar bulge velocity dispersion ($M_\mathrm{BH}-{\sigma}$: e.g., \citealt{2000ApJ...539L...9F,2000ApJ...539L..13G}). While this relation exhibits a small scatter and is very well established, it is luminosity biased (as more luminous and massive galaxies are selected to populate the relation), and requires the spectroscopic analysis of the host bulge emission. Furthermore, there are considerable doubts on the applicability of the $M_{\rm{BH}}$-$\sigma$ relation in the low mass regime. Most models that aim to reproduce the $M_\mathrm{BH}-{\sigma}$ relation,  through the interaction of massive outflows (during episodes of super-Eddington accretion) with the host galaxy, predict a size for the region of influence that for $M_\mathrm{BH}{<}10^{5}$ \msolar is substantially smaller than what can be resolved by current or near-future optical telescopes \citep[e.g.][]{1998A&A...331L...1S,2003ApJ...596L..27K}.

To bypass some of these drawbacks, and also access fainter sources, the scaling relation between $M_\mathrm{BH}$ and luminosity of the host bulge ($L$) can be considered. Actually preceding the discovery of the $M-{\sigma}$ relation, the $M-L$ relation  \citep[e.g.][]{1989IAUS..134..217D,1993nag..conf.....B} can be applied to fainter sources and spectroscopy is not required. While the $M-L$ relation is also biased towards higher mass/luminosity, recent works have attempted to better constrain the low mass regime, discovering strong indications of a steeper slope in the $M-L$ relation, at  $M_\mathrm{BH}{<}10^{6}\,\mathrm{M}_\odot$ \citep{2013ApJ...764..151G}. Using these improved scaling relations, \cite{2013ApJ...764..151G} presented 40 lower luminosity spheroids that contain AGN which appear to have a central black hole mass that falls in the IMBH range. Other scaling relations between host galaxy and central black hole include a correlation between $M_{\rm{BH}}$ and the galactic S\'ersic index ($n_{sph}$, a measure of bulge concentration 
see e.g., \citealt{2007ApJ...655...77G,2016ApJ...817...21S}), and a correlation between the morphology of the spiral arms and the mass of the central black hole in disk galaxies \citep{2008ApJ...678L..93S,2012ApJS..199...33D} formulated as a relation between $M_\mathrm{BH}$ and the spiral pitch angle (PA, a measure of the tightness of spiral arms). 

The different scaling relations each have their benefits and drawbacks, which may be more notable in the under-explored low mass regime \citep{Shankar_2019}. However, combining the different methods provides a more robust average prediction. Recently, \cite{2017A&A...601A..20K} combined IR, optical and FUV observation of LLAGN and the $M-L$, $M-PA$, $M-n_{sph}$ relations, along with X-ray and radio observations of the accreting central black hole and employing a well known relation between the radio luminosity ($L_{r}$, 5\,GHz), the X-ray luminosity ($L_{x}$, 0.5-10\,keV) and the black hole mass (the "fundamental plane of black hole activity":~\citealt{2003MNRAS.345.1057M}), to estimate the SMBH mass of candidate LLAGN and probe the consistency between the relations in the low mass regime. The authors demonstrated that the different methods yielded consistent results in the low mass regime. Prompted by the promising results of this pilot study, an endeavour has begun to extend it to a complete census of the mass of the central SMBHs in nearby AGN (within 150\,Mpc). The census cross correlates the Third Reference Catalogue of Bright Galaxies (RC3; \citealt{1991rc3..book.....D}),  supplemented with the Catalogue of Neighbouring Galaxies (CNG; \citealt{2004AJ....127.2031K}) and the AGN sample with measured central velocity dispersions (\citealt{2009ApJS..183....1H}; \citealt{2011ApJ...739...28X}), from the Palomar survey (Ho et al.~1997) and the 4th Data Release of the {\it Sloan Digital Sky Survey} \citep{2007ApJ...667..131G}, with the {\it XMM-Newton} and {\it Chandra} catalog and the {\it VLA} and e-Merlin radio surveys of nearby AGN, in search of candidates with multi-wavelength signatures of the accreting central black hole. So far, more than fifty candidates -- that have available X-ray, radio, FUV, optical and IR observations from a slew of major telescopes (e.g,. {\it XMM-Newton}, {\it Chandra}, {\it VLA},{\it HST}, {\it Spitzer},{\it GALEX}, {\it SDSS}, {\it KPNO}, etc.) -- have been identified (Koliopanos et al.~in prep.). When completed this endeavor promises to yield one of the most robust and comprehensive mass measurements of IMBHs in LLAGN to date.

In the era of multi-messenger astronomy, this type of multi-wavelength study is a necessary step towards the ongoing search for IMBHs and the seeds of SMBHs. By scrutinizing the remarkable wealth of observational data that are already available in numerous astronomical catalogs, we will also provide the target sample for the next generation of observatories such as the {\it JWST}, {\it Athena} X-ray telescope and the Square Kilometre Array ({\it SKA}) project. The identification of the seeds to SMBHs is already within reach and promises to be one of the great discoveries of the next decade.

\subsection{Direct Detection of Supermassive Stars with Future Observatories}

\subsubsection{Can We Observe Supermassive Stars Themselves?}

As discussed above, rapidly-accreting SMSs may be the precursors to the first quasars in the Universe. Could this initial stage of early SMBH formation be detected in the NIR by upcoming wide-field surveys such as {\it Euclid} and the Wide-Field Infrared Survey Telescope ({\it WFIRST}), or in narrower but deeper surveys by {\it JWST} and the coming generation of 30 - 40m telescopes on the ground? SMSs typically have luminosities of $\sim$ 10$^6$ \Lsolc\ at the onset of central hydrogen burning that later rise to $\sim$ 10$^{10}$ \Lsolc\ by the time the star collapses to a black hole. There are two main challenges to their detection today.

First, although they are extremely luminous they are also relatively cool, with surface temperatures of $5\,000$--$10\,000\,\mathrm{K}$, set by H$^-$ opacity in their outer layers. At these temperatures, they are not intrinsically bright sources of UV radiation that would be redshifted into the NIR today. However, a few SMSs accreting at rates below $\sim0.005 \mathrm{M}_\odot\,\mathrm{yr}^{-1}$ \citep{haemmerle18a} or intermittently, in bursts \citep{sak15}, may evolve along hotter (bluer) tracks at times and be brighter today.

\begin{figure*} 
\begin{center}
\begin{tabular}{cc}
\includegraphics[width=0.48\textwidth]{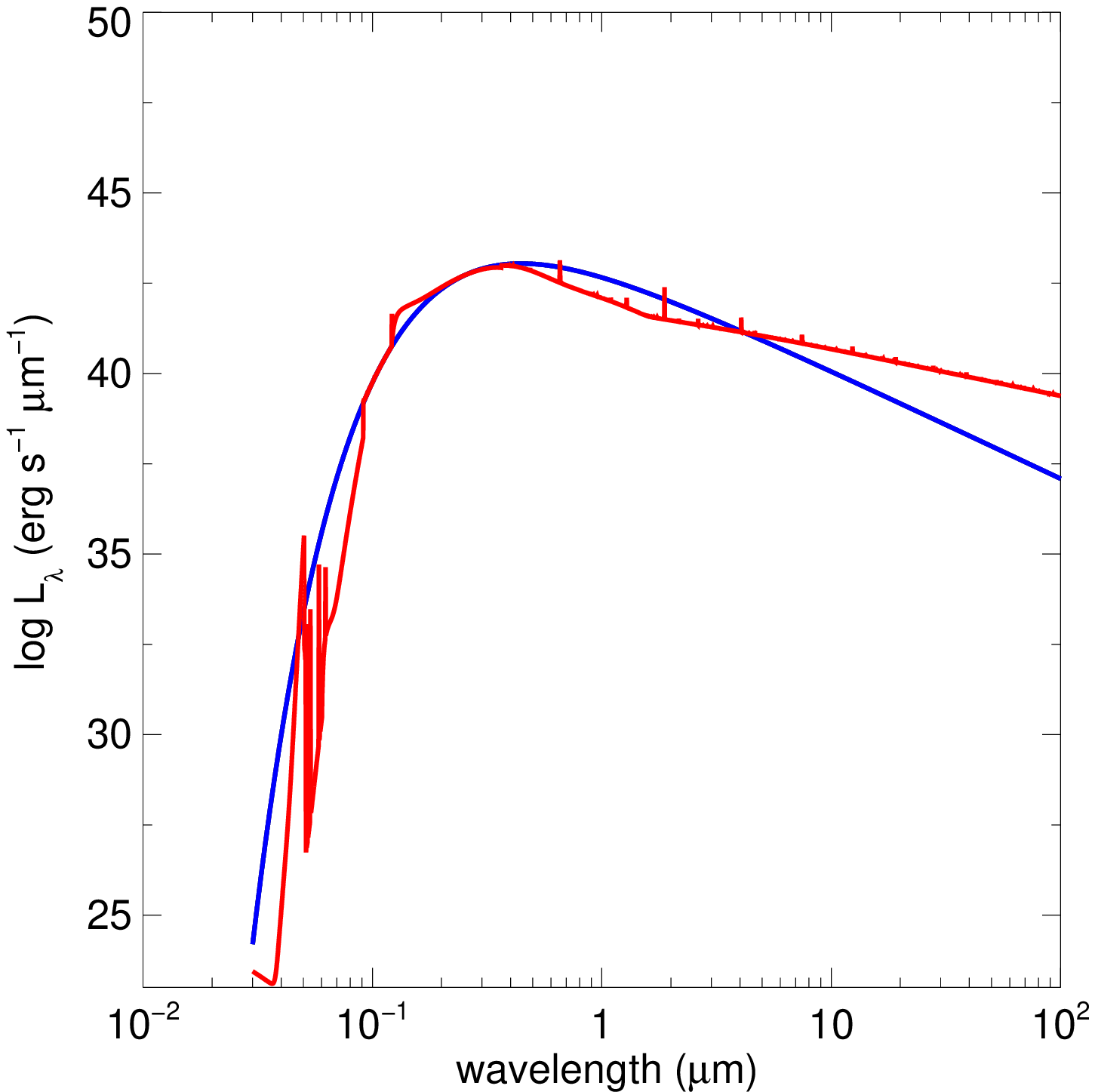}  &
\includegraphics[width=0.48\textwidth]{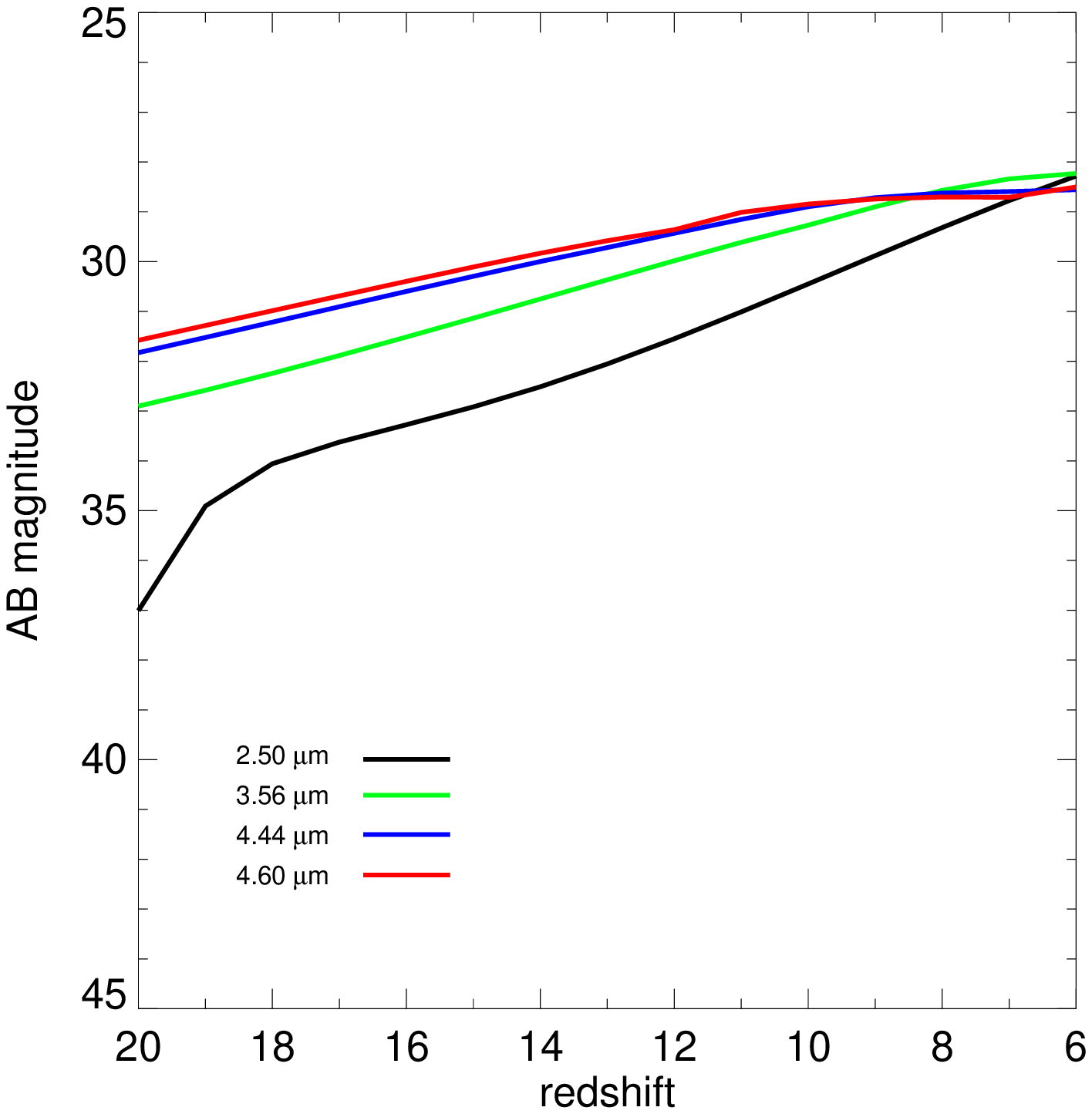}  
\end{tabular}
\end{center}
\caption{Left panel: spectra for an SMS accreting at 1.0 \msolarc\ yr$^{-1}$ at 100,000 yr \citep{haemmerle18a}.  Blue: spectrum of the star itself; red: spectrum of the star in the dense accretion envelope that creates it. Right panel: AB magnitudes for the SMS in its envelope in the JWST
F250 (2.5 $\mu$m, black), JWST F356 (3.56 $\mu$m , green), JWST F444 (4.44 $\mu$m, blue) and JWST F460 (4.60 $\mu$m, red) filters.}
\label{fig:sms_spec} 
\end{figure*}

Second, these stars are only bright for relatively short durations of 10$^5$ - 10$^6$ yr, and do not appear as transients. However, although current evolution models of SMSs cannot resolve pulsations, they are in fact expected to exhibit them, and this might enhance prospects for their detection in future surveys. Depending on the frequency of pulsations in the rest frame, variations in the luminosity of the star might be manifest over the cadences of surveys currently planned for {\it Euclid}, {\it WFIRST}, {\it JWST}, as well as ground-based extremely large telescopes (ELTs) in the near future. 

Source frame spectra for a SMS accreting at 1 \msolarc\ yr$^{-1}$ before and after reprocessing by its dense accretion envelope are shown in the left panel of Fig. \ref{fig:sms_spec} \citep{sur18a}.  Absorption by the envelope at the H and He ionization edges is evident, as is Ly$\alpha$ emission. Continuum absorption due to H$^-$ opacity is also visible at wavelengths shorter than 1.65 $\mu$m. The reprocessing of shorter wavelength photons by the envelope of the star into Ly$\alpha$ enhances prospects for their detection, since at $z \sim$ 10 these photons are redshifted in the NIR.  

We show NIR AB magnitudes for the SMS in the right panel of Figure \ref{fig:sms_spec}. 
Unfortunately, predicted SMS H band AB magnitudes are well below the sensitivities of both {\it Euclid} and {\it WFIRST} at $z \sim 10 - 20$. SMS could, however, still appear in {\it JWST} and ELT surveys out to $z \sim 20$ in the 3.56 and 4.44 $\mu$m NIRCam filters. The likelihood for detection of SMSs is likely greatest at $z \sim 10$, as their numbers are thought to peak at this epoch and they are more easily seen at lower redshifts.

\subsubsection{Do Supermassive Stars Explode?}

\begin{figure}
\includegraphics[width=0.48\textwidth]{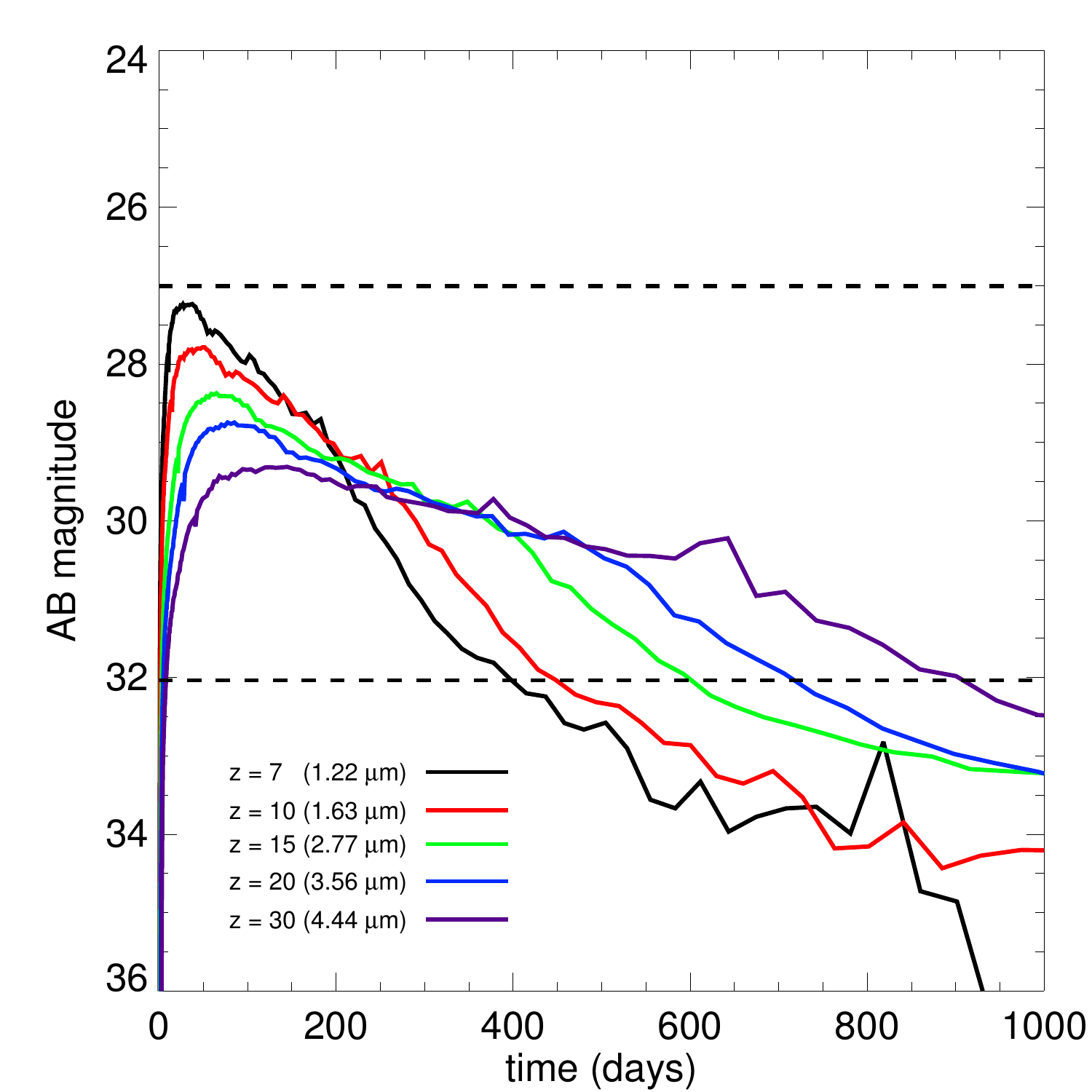}
\caption{Lower limits on the NIR light curves of the thermonuclear explosion of a 55,500 \msolarc\ star.  Each light curve is shown in the filter in which the explosion is brightest at the given redshift.  Note that the wavelength of the filter limits the redshift at which a transient can be detected in it because the IGM absorbs all flux blueward of the Lyman limit in the rest frame of the event prior to the end of cosmological reionization.  Thus, {\it Euclid} and {\it WFIRST} can only detect transients in the H band out to $z \sim$ 15 while {\it JWST} and the ELTs could detect events at $z \gtrsim 20$.}\label{fig:sms_lc}
\end{figure}

To date, no model of rapidly-accreting, zero-metallicity SMSs has produced a SN, however it has recently been suggested that monolithically-formed SMSs within a relatively narrow (5,000 \msolarc) window around 55,500 \msolarc\ \citep{chen14} do end their lives in extremely energetic ($\sim$~10$^{55}$ erg) thermonuclear explosions, capable of destroying the protogalaxies in which they occur \citep{wet13a,wet13b,jet13a}. Numerical studies have found that such explosions would be visible to Euclid and WFIRST at the epochs at which they would likely occur \citep[Figure~\ref{fig:sms_lc};][]{wet12d}. Furthermore, some models of rotating stars above 10$^6$ \msolarc\ \citep{montero12} have been found to explode, as have metal-enriched (0.005 Z$_{\odot}$) non-accreting $\sim$500,000 \msolarc\ stars \citep{fuller86}. While it is unclear at the moment how such massive (55,000--500,000 \msolarc) stars may form monolithically, the abundance patterns predicted for their explosions \cite[e.g.,][]{chen14} suggest that these must be rare events; in particular, they are inconsistent with observations of the Milky Way and DLAs \citep[][see also discussion in $\S$3.4]{kobayashi}. Interestingly, however, mass return and chemical enrichment from a SMS undergoing putative eruptive mass loss during its lifetime on the hydrogen-burning main sequence may help to explain the unusual abundances observed in globular clusters \citep{denissenkov14, gieles18}, although a number of other scenarios exist.

\citet{mats15,mats16} examined the observational signatures of highly energetic but directional gamma-ray bursts (GRBs) driven by the collapse of SMSs.  Such events would be characterized by central engines with unusually long durations of 10$^4$ - 10$^5$ s or more because of the times required for the outer layers of the star to collapse into the black hole accretion disk system (BHAD) at its center, and for its jet to pierce these massive, infalling layers.   \citet{mats16} find that the gamma-ray signal of SMS GRBs would be detected by Swift and its afterglow would be visible to {\it Euclid}, {\it WFIRST} and {\it JWST} in the NIR.  However, to drive a GRB the core of the SMS would have to be spun up to very high angular momenta prior to collapse, in order to produce a BHAD capable of driving a jet for the times required to break out of the star.  The high angular velocities required to form a BHAD in the core approach the breakup limit in stars, and are thought to be why the rate of GRBs is only 10$^{-3}$ - 10$^{-4}$ of all core collapse events today. The SMS could not be born with such angular velocities because the $\Omega\Gamma$ instability would have broken up the star before it could become supermassive and no known mechanism could spin it up at late times prior to its collapse \citep{haemmerle18b}. In any event, the SMS GRB rate is expected to be only a small fraction of the rate of DCBH formation, and the detection rate would be smaller by roughly a factor of 1$/4\pi$ because they are beamed events, making their detection even by wide-field NIR surveys problematic.

If the earliest quasars also formed from low mass black hole seeds via super- or hyper-Eddington accretion, their birth might have been signalled by less massive but still highly energetic Pop III GRBs.  The gamma-ray signal from Pop III star GRBs could in principle be found by {\it Swift} even at $z \gtrsim 20$ \citep[e.g.,][]{nat05,bl06a,nom10,mr10,nsi12}, but the time dilation of the burst might not immediately identify it to be a high redshift event because of the large range of central engine times that have now been found for GRBs in the local Universe.  Prompt spectroscopy of the NIR afterglows of these events in future missions such as EXIST \citep{gou04,wet08c,toma11,wang12,met12a,mes13a} or by radio observations from the ground \citep{im05} could pinpoint their redshifts.  Their event rate, while low, would still be expected to be much higher than those of SMS GRBs \citep[e.g.,][]{bl02,ds11}.

Finally, we note that fragmentation in the vicinity of a DCBH deep in the center of a massive accretion disk of an atomically-cooled halo could lead to energetic transients known as tidal disruption events (TDEs).  TDEs occur when stars forming from such fragments are torn apart by tidal forces just before crossing the event horizon of the DCBH.  In this picture, X-rays from the nascent black hole may also trigger the formation of Pop III stars in its vicinity within 5 - 10 Myr of birth \citep[e.g.,][]{mba03,aycin14}. Statistically, some of these stars would then be scattered onto the loss cone of the black hole within $\sim$ 1 Myr. Numerical studies indicate that the jet emitted in this event could be detected by Swift and eROSITA even at  $z \sim$ 20 \citep{ki16}. The X-ray signal from TDEs could be distinguished from those of DCBH GRBs by their much longer durations, $10^{5 - 6} (1+z)$ s. Detections of the X-ray signal could then be followed up by radio or NIR observations of the afterglow with the extended Very Large Array ({\it eVLA}) and the Square Kilometer Array ({\it SKA}) or {\it JWST} to pinpoint its redshift of origin. The rate of TDEs would be on par with that of DCBH formation itself, much higher than the rate of SMS SNe or GRBs, and would therefore be the likeliest electromagnetic transient associated with DCBH birth.

\subsection{Observability of Direct Collapse Black Holes}
To begin this section, it is important to point out that DCBHs are still black holes: as such, we can expect them to have the typical spectral signatures of black holes. If this is true, how can we possibly distinguish them from normal high-$z$ AGN? In fact, once formed, a black hole does not carry a label indicating which formation channel it followed. It is through the \textit{interaction of the black hole with its galactic environment} that we might be able to understand if an accreting object possibly formed as a DCBH. Differently from other formation mechanisms, DCBHs are predicted to form in a pristine environment (e.g., \citealt{BL03, Agarwal14, Volonteri_2016}). The smoking gun for a DCBH would, thus, be the observation of a black hole spectrum with the absence of any metal line. As the case of CR7 (described in Sec. \ref{subsec:learned_CR7}) taught us, however, the detection of very feeble metal lines does not rule out the DCBH interpretation \citep{Agarwal16_CR7, Pacucci17}. In fact, while the conventional model for the \textit{formation} of a DCBH requires a metal-free gas, once the black hole is in place the gas can certainly become polluted with metals resulting from star formation. This would influence the resulting spectrum of the DCBH. As pointed out in \cite{Pacucci17}, the DCBH would still be the preferred formation model to explain the spectrum of a high-$z$ black hole seed if the metal content of the host galaxy is $\lesssim 10^{-4} \, \mathrm{Z_{\odot}}$, where $\mathrm{Z_{\odot}}$ indicates the solar metallicity.

\begin{figure}
\begin{center}
\includegraphics[width=1.07\columnwidth]{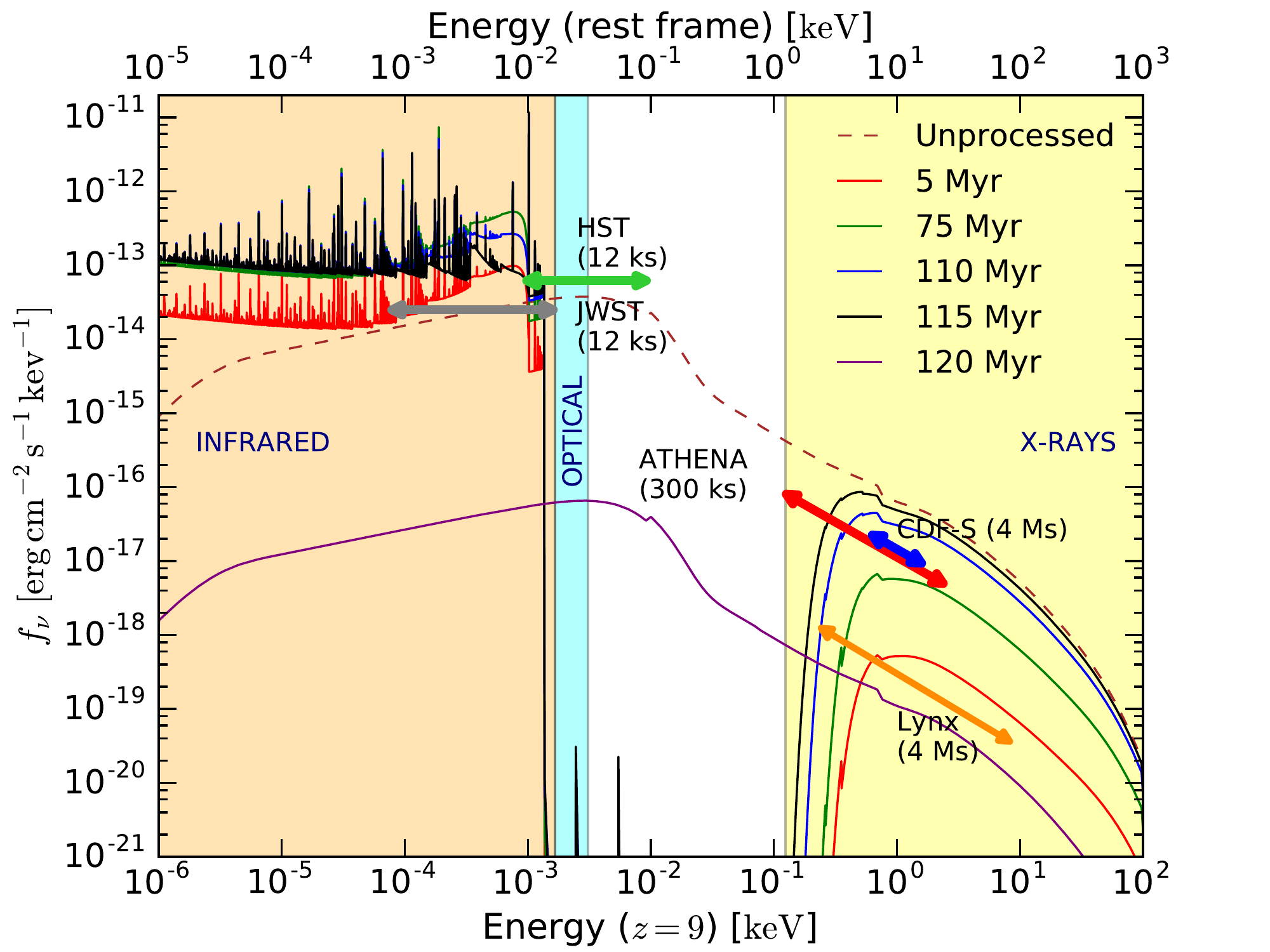}
\caption{Time evolution of the spectrum of a classic DCBH with initial mass $10^5 \, \mathrm{M}_{\odot}$ in a pristine ($Z=0$) environment. The infrared,
optical and X-ray bands are highlighted with shaded regions, while the unprocessed spectrum is reported, at peak luminosity ($t = 115 \, \mathrm{Myr}$), with a dashed line. The flux limits for future (JWST, ATHENA, Lynx) and current (HST, CDF-S) observatories are also shown. Image from \cite{Pacucci_2015_spectrum}.}
\label{fig:DCBH_spectrum}
\end{center}
\end{figure}

The first studies, both analytical \citep{Yue_2013} and numerical \citep{Pacucci_2015_spectrum}, to predict the spectrum of a DCBH indicated predominant emission in the observed infrared ($\gtrsim 1 \, \mathrm{\mu m}$) and X-ray ($\sim 1 \, \mathrm{keV}$) part of the electromagnetic spectrum. Photons with frequencies higher than Lyman-$\alpha$ are absorbed by photo-electric processes and re-emitted in the infrared.
Fig. \ref{fig:DCBH_spectrum} shows a classic DCBH spectrum as computed in \cite{Pacucci_2015_spectrum}. Accretion on a typical DCBH (initial mass $10^5 \, \mathrm{M}_{\odot}$ in a zero-metallicity environment) lasts $\sim 100 \, \mathrm{Myr}$. The figure shows that most of the accretion process on a typical DCBH should be easily observable by the {\it JWST} (see also \citealt{Natarajan_2017,Volonteri_2017}), with peak magnitudes even reaching $m_{AB} \sim 20$ for a very short period of time ($\ll 1 \, \mathrm{Myr}$).
The {\it HST} should be able to observe their emission up to $z \sim 10$ close to the Lyman-$\alpha$ line. In the X-ray part of the spectrum, the CDF-S is predicted to observe the emission of a typical DCBH close to the peak of the high-energy emission, for a short period of time ($\sim 10-20 \, \mathrm{Myr}$). The same result should be obtainable by {\it ATHENA} with a much shorter ($\sim 300$ ks) integration time, while Lynx will be able to easily observe most of the accretion process \citep{Vito_2016,Vikhlinin_2018}.

The two main environmental factors that can change the shape of the spectrum for DCBHs are: (i) the presence of metals, and (ii) very large (Compton thick, $N_H \gg 10^{24} \, \mathrm{cm^{-2}}$) absorbing column densities.
The presence of a non-negligible ($Z \gtrsim 10^{-5} Z_{\odot}$) amount of metals in the galactic environment of a DCBH produces a shift of photons from the X-ray to the infrared part of the spectrum, due to Auger's processes on high-energy electron shells (see, e.g., \citealt{Pacucci17}). Increasing the absorbing column density of the gas (even at $Z=0$) has a similar effect, with a decrease of the X-ray emission followed by an increase of infrared re-emission. The effects of an increase in metallicity/column density are very similar, but not degenerate since the presence of metals also produces metal lines in the spectrum \citep{Agarwal16_CR7,Pacucci17}. The effect of an increase of the column density above Compton-thick levels is particularly interesting for DCBH studies, since most of the high-$z$ sources are predicted to be heavily obscured (e.g., \citealt{Comastri_2015}). The spectral predictions for heavily-obscured DCBHs \citep{Pacucci_2015_spectrum,Valiante_2018} indicate that the detection of these sources could be challenging in high-energy bands, while in the infrared {\it JWST} will be able to observe them. Thus, the {\it JWST} will certainly be the main observatory to unravel highly-obscured sources in the high-$z$ Universe \citep{Natarajan_2017,Valiante_2018}. Very large values of the absorbing column density may also lead to photon trapping effects, with a reduced radiative efficiency of the accretion disk and, consequently, very large accretion rates, possibly above the Eddington limit \citep{Sadowski_2009,Begelman_2017}. As pointed out in \cite{Pacucci_2015_spectrum}, this would reduce the accretion time scale on to the DCBH by a factor $\sim 10$, with typical evolutionary times as low as $\sim 10 \, \mathrm{Myr}$. This effect would certainly add to the challenge of detecting highly-obscured sources in the early Universe.

Even if DCBHs are observable in the early Universe, possibly up to $z \sim 20-25$ with the {\it JWST} \citep{Natarajan_2017}, another interesting question is how we can separate them from other high-$z$ sources. 
The availability of high-resolution spectra, fundamental to confirm the detection of a DCBH (see the discussion in \citealt{Gallerani_2017, Pacucci17}), is limited only to a small fraction of the sources observed, mostly the brighter ones. DCBHs, instead, are expected to be found in the high-$z$ Universe, where the number of sources with spectroscopic information is of course limited.
Deep, wide-field surveys (e.g., the CANDELS survey, see \citealt{CANDELS_1,CANDELS_2}) contain photometric information about hundreds of thousands of sources, thus offering an invaluable amount of information to find sources whose photometry is at least compatible with the one theoretically predicted for DCBHs. Once a source is selected as a DCBH candidate, follow-up spectroscopic observations (if possible) would eventually determine their actual nature.
The first photometric method to identify black hole seed candidates in deep multi-wavelength surveys was developed in \cite{pacucci16}. Supported by numerical simulations, this work predicted that DCBHs are characterized by a steep spectrum in the infrared ($1.6-4.5 \, \mathrm{\mu m}$), i.e., by very red colours. A comparison with other high-$z$ categories of sources clearly shows that the infrared spectrum of DCBHs is predicted to be significantly steeper.
The method selected the only $2$, X-ray detected, objects found in the CANDELS/GOODS-S survey with a photometric redshift $\gtrsim 6$. To date, the selected objects represent the most promising black hole seed candidates, possibly formed via the DCBH scenario, with predicted mass $>10^5 \, \mathrm{M}_{\odot}$. 

Several works have complemented these selection criteria \citep{Natarajan_2017, Volonteri_2017, Valiante_2018} adding, for instance, constraints on the ratio of X-ray flux to rest-frame optical flux.  In particular, upon their birth, DCBHs have BH mass to halo mass ratios much higher than expected for BH remnants of Pop III stars which have grown to the same mass ($\sim 10^6\,\mathrm{M}_\odot$).  A future X-ray mission, such as Lynx, combined with infrared observations should be able to distinguish high-redshift DCBHs from smaller black hole seeds, due to the correspondingly high BH mass to stellar mass ratios of the faintest observed quasars, with
inferred BH masses below $\sim 10^6\,\mathrm{M}_\odot$ \citep{VisbalHaiman2018}.

While the development of improved selection criteria will definitely help in developing a reliable catalogue of DCBH candidates, only the advent of the next generation of telescopes (e.g., JWST, ATHENA, Lynx) will likely lead to the detection of the first, confirmed, DCBH.

\begin{figure*}[t]
\begin{center}
\includegraphics[width=0.75\textwidth]{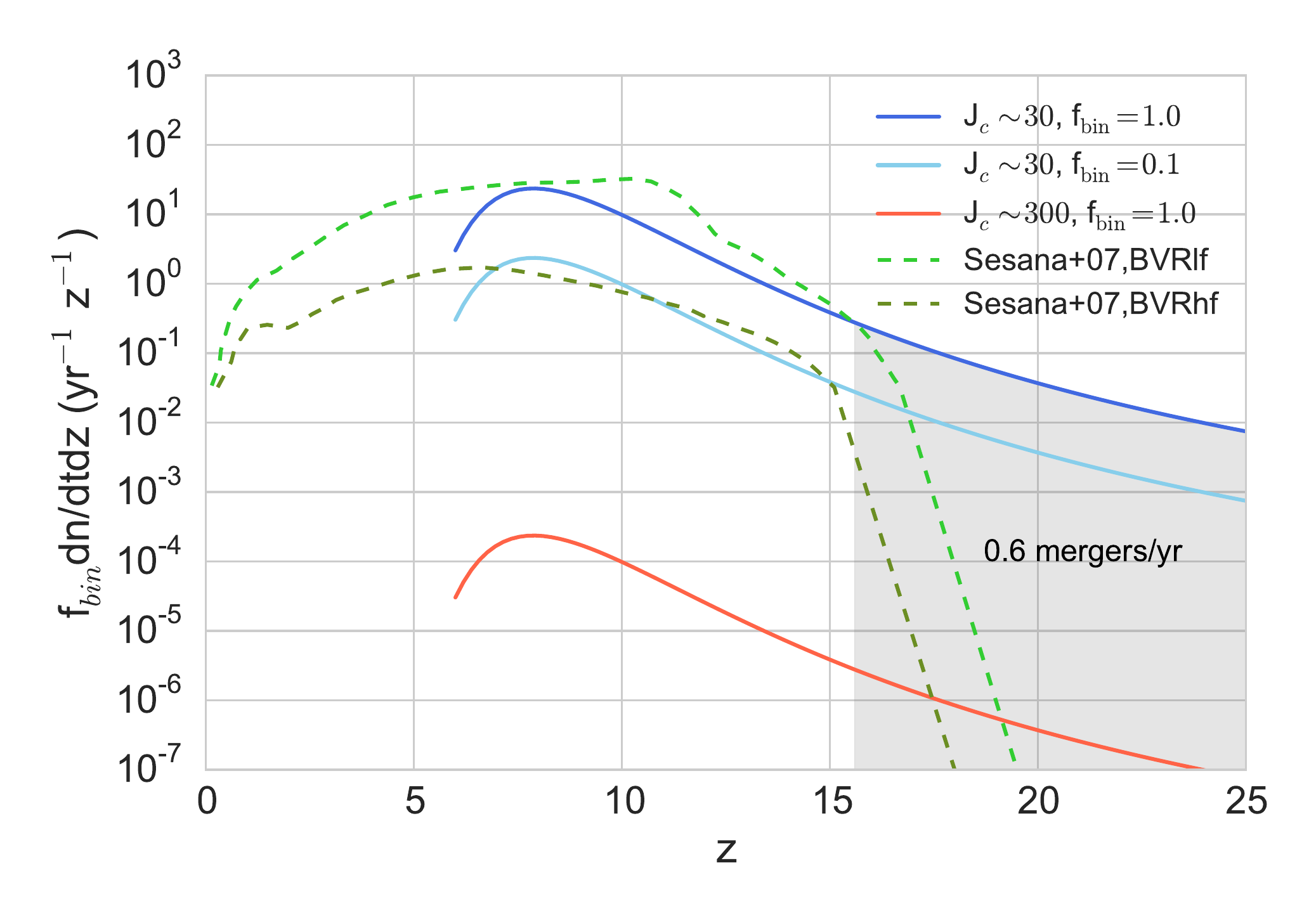}
\end{center}
\caption{Rates for the merging of black hole binaries (BHB) from supermassive stellar binaries as a function of redshift compared to the models by \citet{sesana07}. Only the optimistic scenario with $f_\mathrm{bin}=1$ and $J_{\rm{crit}}=30$ can produce a population of BHB mergers at $z \gtrsim 15$ that are clearly distinguishable from other channels of BHB formation. The total rate of such uniquely identifiable BHB mergers is $\sim 0.6$ per year, highlighted in grey. Adapted from \citet{hartwig18}.}
\label{fig:hartwig18}
\end{figure*}

\subsection{Future Insights from Gravitational Waves}
The gravitational wave detector LIGO has proven its capabilities by detecting binary black hole mergers with total masses of a few times $10\,\mathrm{M}_\odot$ \citep{GW170608}. In the near term, future observations hold substantial promise for furthering our understanding of not only stellar-mass black holes, but their more massive cousins as well. At design sensitivity, LIGO may yet detect IMBH binaries (either IMBH+stellar-mass black hole or of IMBH-IMBH) of $\leq 2 \times 10^3\,\mathrm{M}_\odot$ \citep[e.g.,][]{Fragione18}. Future interferometers are designed to detect even more massive black hole mergers, such as KAGRA \citep{kagra} or LISA \citep{babak17} at lower frequencies. The expected design of LISA has the highest sensitivity in the mass range $10^4 - 10^6\,\mathrm{M}_\odot$, for which black hole mergers are detectable out to $z>20$. This is the same mass range for remnants of SMS that may form in binaries, as recent high-resolution simulations suggest \citep{latif13,Becerra_15,Latif_16,Chon_2018,Regan18,Regan18b}. Depending on the stellar evolution of supermassive binaries and the initial separation of their compact remnants, these objects will merge and can be observed using LISA.

Recently, \citet{hartwig18} have calculated the detection rates of such black hole binary mergers and the resulting constraints on the nature of SMS formation. Since there are many unknowns, such as the binarity of SMS, their stellar binary evolution, and the corresponding delay times until coalescence, this ignorance can at present only be parameterized using the scaling factor $f_\mathrm{bin}$, which quantifies the number of SMS remnant black hole binaries per pristine atomic cooling halo that will merge on a sufficiently short time scale. In their most optimistic model with $f_\mathrm{bin}=1$ and $J_{\rm{crit}}=30$, \citet{hartwig18} show that we can expect up to $\sim 80$ mergers per year of the remnants of supermassive binary stars. However, a major challenge will be to discriminate these events from other formation channels, such as black hole mergers as a consequence of galaxy mergers. To break this degeneracy, \cite{hartwig18} compare their anticipated detection rates to the results of \citet{sesana07}, who derive the rates of black hole mergers in the same mass range as a consequence of galaxy mergers. \citet{hartwig18} demonstrate that black hole mergers from supermassive stellar binaries dominate the LISA detection rates at $z \gtrsim 15$, with about $0.6$ detections per year in their optimistic scenario (see Figure \ref{fig:hartwig18}).

These results also allow one to determine what constraints on the abundance of supermassive stellar binaries can be drawn from a non-detection of such black hole mergers. In \cite{hartwig18}, their optimistic model ($f_\mathrm{bin}=1$, $J_{\rm{crit}}=30$) can already be excluded with $95\%$ certainty after 5 years of non-detections at $z>15$. After 10 years of non-detections, an upper limit on the binarity of supermassive stars of $f_\mathrm{bin}<0.5$ ($J_{\rm{crit}}=30$) can be derived at the $2\sigma$ level.

Further refinements of this work, as well as predictions for other observables (e.g., X-ray and UV emission), await a more refined theory of the formation and evolution of SMS multiples. Given the expected prevalence of fragmentation based on recent numerical simulations \citep[e.g.,][]{Chon_2018,Regan18,Regan18b}, understanding binary interactions in the supermassive regime is a priority. 

\section{Conclusions}

The discovery of massive high-z quasars has provided invaluable insight, and a great challenge, to the study of SMBH formation. Given the difficulty in reaching such masses so early in cosmic history, their existence has motivated a rapidly-growing effort to understand the formation of much more massive ``seed'' black holes than could be produced by ``typical'' Pop III stars. A viable mechanism for the formation of truly supermassive ($>10^5\,\mathrm{M}_{\odot}$) seeds has emerged in the synchronised pairs model for atomically-cooled halos, with an increasingly mature theory for the conditions under which they may arise. A number of alternative models, however, continue to challenge this scenario. 

All of these models, in turn, provide realistic initial conditions for detailed stellar evolutionary calculations, permitting a clearer picture than ever before of the nature and fate of supermassive stars. Many questions remain, however, regarding the mechanism by which SMSs may shed angular momentum, and their fate at the highest plausible accretion rates. 

The state of the field has advanced, however, to the point that one can begin to consider ``subgrid'' models for SMSs. The statistics of SMS formation can also now be addressed, however the essential conditions for atomically-cooled halos to form massive black holes must be robustly determined before the viability of the heavy seed model can be clearly evaluated in this way. In the meantime, observational efforts to determine the seeds of SMBHs continue, from the search for IMBHs, to deep surveys of the distant Universe. Recent efforts have demonstrated that discriminating between light and massive seed models is feasible using multi-wavelength electromagnetic data from the planned and upcoming JWST, WFIRST and Athena missions, as well as anticipated gravitational wave observations of high-z black hole mergers to be detected by the proposed LISA mission \citep{2018MNRAS.481.3278R,Pezzulli17,Volonteri_2017}. In the near future, the next generation of electromagnetic and gravitational wave observatories promises to revolutionize our understanding of the first stars and quasars, and will likely reveal the origin of these titans of the early universe.

\begin{acknowledgements}

TEW, AH, LH, and RSK would like to thank the Joint Institute for Nuclear Astrophysics Center for the Evolution of the Elements (NSF Grant No. PHY-1430152), Monash University, and the European Research Council (ERC Advanced Grant, \textsl{STARLIGHT: Formation of the First Stars} project number 339177) for supporting the workshop ``Titans of the Early Universe,'' as well as the Monash Prato Centre in Prato, Italy, for their hospitality during the meeting, and Cheryl Woynarski for poster and website design. AH was supported by an ARC Future Fellowship (FT120100363), by a Larkins Fellowship from Monash University, by TDLI though a grant from the Science and Technology Commission of Shanghai Municipality (Grants No.16DZ2260200) and National Natural Science Foundation of China (Grants No.11655002). DJW was supported by STFC New Applicant Grant ST/P000509/1. CK was supported by STFC ST/M000958/1. JAR acknowledges the support of the EU Commission through the Marie Sk\l{}odowska-Curie Grant - ``SMARTSTARS'' - grant number 699941.  LH was sponsored by the Swiss National Science Foundation (project number 200020-172505). MM acknowledges support from the Spanish Juan de la Cierva program (IJCI-2015-23944). TH is a JSPS International Research Fellow. FP acknowledges support from the NASA Chandra award No. AR8-19021A. KI was supported by the National Key R\&D Program of China (2016YFA0400702) and the National Science Foundation of China (11721303). ZH acknowledges support from NSF grant 1715661 and NASA grants NNX15AB19G and NNX17AL82G. AF is partially supported for this project by the SNS Internal Grant FI-2017.  YS is supported by Grant-in-Aid for JSPS Overseas Research Fellowships.  SC was financially supported by the Grants-in-Aid for Basic Research by the Ministry of Education, Science and Culture of Japan (17H01102). SS acknowledges funding from the Italian Ministry and Education (MIUR) through a Rita Levi Montalcini Fellowship.  SH acknowledges support from JSPS Research Fellow and JSPS KAKENHI (18J01296). TH has been supported in part by MEXT/JSPS KAKENHI Grant Number 15H00776 and 16H05996. The material is based upon a workshop supported by the National Science Foundation under Grant No. PHY-1430152 (JINA Center for the Evolution of the Elements).  RSK and SCOG acknowledges funding from the European Research Council via the ERC Advanced Grant \textsl{STARLIGHT: Formation of the First Stars} (project number 339177).

\end{acknowledgements}
\vskip0.5cm
\affil{$^{1}$Monash Centre for Astrophysics, School of Physics and Astronomy, Monash University, VIC 3800, Australia}
\affil{$^{2}$Birmingham Institute for Gravitational Wave Astronomy and School of Physics and Astronomy, University of Birmingham, Birmingham, B15 2TT, UK}
\affil{$^{3}$Uni. Heidelberg, Zentrum f\"{u}r Astronomie, Inst. f\"{u}r Theoretische Astrophysik, Albert-Ueberle-Str.\ 2, 69120 Heidelberg, Germany}
\affil{$^{4}$Department of Astronomy, University of Texas at Austin, 2511 Speedway, Austin, TX 78712, USA.}
\affil{$^{5}$Department of Physics, University of Oxford, Keble Road, Oxford, OX1 3RH, U.K.}
\affil{$^{6}$Affiliate Member, Kavli IPMU (WPI), Todai Institutes for Advanced Study, The University of Tokyo, 5-1-5 Kashiwanoha, Kashiwa, Japan 277-8583}
\affil{$^{7}$Institute of Astronomy and Astrophysics, Academia Sinica,  Taipei 10617, Taiwan}
\affil{$^{8}$Astronomical Institute, Tohoku University, Sendai 980-8578, Japan}
\affil{$^{9}$Scuola Normale Superiore Piazza dei Cavalieri 7, 56126 Pisa, Italy}
\affil{$^{10}$School of Physics, Georgia Institute of Technology, Atlanta, GA 30332, USA}
\affil{$^{11}$Observatoire de Gen\`eve, Universit\'e de Gen\`eve, chemin des Maillettes 51, CH-1290 Sauverny, Switzerland}
\affil{$^{12}$Department of Astronomy, Columbia University, New York, NY 10027, USA}
\affil{$^{13}$Department of Physics, School of Science, University of Tokyo, Bunkyo, Tokyo 113-0033, Japan}
\affil{$^{14}$Kavli IPMU (WPI), UTIAS, The University of Tokyo, Kashiwa, Chiba 277-8583, Japan}
\affil{$^{15}$Tsung-Dao Lee Institute, Shanghai 200240, China}
\affil{$^{16}$Department of Earth and Planetary Sciences, Faculty of Sciences, Kyushu University, Fukuoka, Fukuoka 819-0395, Japan}
\affil{$^{17}$Department of Physics, Kyoto University, Sakyo-ku, Kyoto, 606-8502, Japan}
\affil{$^{18}$Kavli Institute for Astronomy and Astrophysics, Peking University, Beijing 100871, China}
\affil{$^{19}$Universit\"{a}t Heidelberg, Interdisziplin\"{a}res Zentrum f\"{u}r Wissenschaftliches Rechnen, INF 205, 69120 Heidelberg, Germany}
\affil{$^{20}$Centre for Astrophysics Research, School of Physics, Astronomy and Mathematics, University of Hertfordshire, Hertfordshire, AL10 9AB, UK}
\affil{$^{21}$CNRS, IRAP, 9 Av. colonel Roche, BP 44346, F-31028 Toulouse Cedex 4, France}
\affil{$^{22}$Universit{\'e} de Toulouse; UPS-OMP; IRAP, Toulouse, France}
\affil{$^{23}$Physics Department, College of Science, United Arab Emirates University, Al-Ain, United Arab Emirates}
\affil{$^{24}$Department of Astronomy and Astrophysics, The Pennsylvania State University,  PA 16802, USA}
\affil{$^{25}$Institute for Cosmology and Gravity, the Pennsylvania State University,  PA 16802, USA}
\affil{$^{26}$Center for Theoretical Astrophysics and Cosmology, University of Zurich,
Winterthurerstrasse 190, 8057, Zurich, Switzerland}
\affil{$^{27}$Institute of Space Sciences (ICE, CSIC), Campus UAB, Carrer de Magrans, 08193 Barcelona, Spain}
\affil{$^{28}$Institut d'Estudis Espacials de Catalunya (IEEC), Carrer Gran Capit\`{a}, 08034 Barcelona, Spain}
\affil{$^{29}$Department of Astronomy, Yale University, PO Box 208101, New Haven, CT 06520, USA.}
\affil{$^{30}$Department of Physics, Yale University, P.O. Box 208121, New Haven, CT 06520, USA}
\affil{$^{31}$Institute of Astronomy, University of Cambridge, Madingley Road, Cambridge CB3 0HA, United Kingdom}
\affil{$^{32}$Centre for Astrophysics and Relativity, School of Mathematical Sciences, Dublin City University, Glasnevin, Ireland}
\affil{$^{33}$Dipartimento di Fisica e Astronomia, Universit\'a di Firenze, Via G. Sansone 1, Sesto Fiorentino, Italy}
\affil{$^{34}$INAF/Osservatorio Astrofisico di Arcetri, Largo E. Fermi 5, Firenze, Italy}
\affil{$^{35}$Observatoire de Paris, PSL University, CNRS, GEPI, Place Jules Janssen, 92195 Meudon, France}
\affil{$^{36}$Dipartimento di Fisica, Sapienza Universit{\'a} di Roma, Piazzale A. Moro 2, 00185 Roma, Italy}
\affil{$^{37}$Institute of Cosmology and Gravitation, University of Portsmouth, Portsmouth PO1 3FX, United Kingdom}
\affil{$^{38}$Independent Scholar}

\bibliographystyle{pasa-mnras}

\end{document}